\documentstyle[aps,epsf,amssymb]{revtex}

\newcommand{\be}{\begin{equation}}
\newcommand{\ee}{\end{equation}}
\newcommand{\bi}{\begin{itemize}}
\newcommand{\ei}{\end{itemize}}

\newcommand{\disp}{\displaystyle}

\newcommand{\leqsim}{\; {\scriptstyle {< \atop ^\sim}} \;}
\newcommand{\geqsim}{\; {\scriptstyle {> \atop ^\sim}} \;}

\renewcommand{\Im}{{\rm Im}}

\begin{document}

\draft

\title{Resonance-Assisted Tunneling}

\author{Olivier  Brodier,$^{(1)}$  Peter  Schlagheck,$^{(1,2)}$ and  Denis
Ullmo$^{(1)}$}

\address{$^{(1)}$Laboratoire 
de Physique Th{\'e}orique et Mod{\`e}les Statistiques (LPTMS), 
91405 Orsay Cedex, France \\
$^{(2)}$Institut f{\"u}r Theoretische Physik,
Universit{\"a}t Regensburg,
93040 Regensburg, Germany }

\date{\today}

\maketitle


\begin{abstract}
We  present evidence  that  tunneling  processes in  near-integrable
systems are  enhanced due to the manifestation  of nonlinear resonances
and their  respective island chains  in phase space.   A semiclassical
description of this ``resonance-assisted'' mechanism is given, which is
based on a local perturbative description of the dynamics in the vicinity of 
the resonances. As underlying picture, we obtain that the quantum state is
coupled, via a succession of classically forbidden transitions across 
nonlinear resonances, to high excitations within the well, from where
tunneling occurs with a rather large rate.
The connection between  this description and the complex classical structure 
of the underlying integrable dynamics is furthermore studied,  
giving ground  to  the  general   coherence  of  the
description as well  as guidelines for the  identification of the
dominant tunneling paths.
The validity of this mechanism is demonstrated within the kicked Harper model,
where good agreement between quantum and semiclassical (resonance-assisted)
tunneling rates is found.
\end{abstract}

\pacs{ \vspace{0.5cm}}













\section{Introduction}

Since  the  early  days  of  quantum  mechanics,  tunneling  has  been
recognized  as  one  of  the   hallmarks  of  the  wave  character  of
microscopic  physics.   The  possibility  of  a  quantum  particle  to
penetrate  an energetic  barrier --  or, more  generally,  a dynamical
barrier  in   phase  space,  as  termed   by  ``dynamical  tunneling''
\cite{davi81}  -- represents  certainly  one of  the most  spectacular
implications of quantum theory and has lead to various applications in
atomic and  molecular physics  as well as  in mesoscopic  science (for
instance  in the  context  of spintronics  devices \cite{barnas98}  to
mention  just  one recent  example).   Despite  its genuinely  quantal
nature,  however, tunneling  is strongly  influenced, if  not entirely
governed  by the structure  of the  underlying classical  phase space.
Tuning  the classical  dynamics from  an integrable  to a  chaotic one
substantially modifies the tunnel coupling - not only if the chaos
affects the classically allowed motion (a case that was studied in
detail  in  \cite{creagh96_99}),  but   also  if  the  destruction  of
invariant tori is restricted to  phase space domains {\em deep inside}
the classically forbidden regions.

Tunneling processes take place in  a variety of contexts.  One may for
instance be  interested in the coupling  of a metastable state to a
continuum,  as, e.g.,   in  the  case   of the $\alpha$-particle decay. 
The tunneling process manifests then in form of a finite life time of the
state, or equivalently, in a finite width of the corresponding spectral
resonance. Another situation concerns systems that possess a discrete symmetry
$P$, in such a way that Rabi oscillations  can take  place  between states  
that are localized in  symmetric regions of  the phase  space.  
The effectiveness  of  tunneling is
then related to the period  of the Rabi oscillations, or equivalently
to   the   splittings   between   the  corresponding   symmetric   and
antisymmetric eigenmodes.

Here, for  the sake  of definiteness, we  shall restrict ourself to this
latter  configuration, though  our study  could be  applied  with minor
modification to  other contexts such  as the computation  of resonance
widths.  To  be more  specific, we shall  consider  systems the
classical dynamics  of which exhibits,  due to the  discrete symmetry,
two congruent but  separate regions of regular, bound  motion in phase
space.   Semiclassical  Einstein-Brillouin-Keller  (EBK)  quantization
then  provides a  prescription how  to construct  local  eigenmodes on
quantized  tori  within each  of  the  regions.  These  ``quasimodes''
fulfill  the  Schr{\"o}dinger  equation,  possibly  to  all  powers  in
$\hbar$,  but  do   not  take  into  account  the   behavior  of  the
wavefunction in the forbidden  domain, and are therefore characterized
by equal energies for any pair of symmetry related quantized tori.  It
is the  nonclassical coupling between  the quasimodes that  lifts this
degeneracy  and selects  the  symmetric and  the antisymmetric  linear
combination of these states as true eigenstates of the quantum system.
Provided  the classical  dynamics is  integrable, the  tunnel coupling
that produces this  level splitting can be reproduced  by the analytic
continuation of the torus structure into complex domain.
In case of a one-dimensional  motion generated by the usual $p^2/2 + V(x)$
Hamiltonian, this amounts to calculating  an imaginary action
integral  of  the type  $\int\sqrt{2(V(x)-E)}dx$  in between the  two
wells.

Such a  dynamical tunneling process also takes  place in nonintegrable
systems, but with  substantially different effectiveness.  By applying
a nonintegrable  perturbation on  the dynamics in  such a way  that an
appreciable  chaotic layer is  introduced in  between the  two regular
regions, the  tunnel coupling between the  quasimodes is significantly
enhanced     with     respect     to     the     integrable     system
\cite{lin90,boh93,tom94}.  Moreover, the  tunneling rates do no longer
exhibit a well-defined scaling  with Planck's constant (which would be
an exponential  decrease with $1/\hbar$ in an  integrable system), but
may  undergo huge  quasierratic fluctuations  for small  variations of
$\hbar$ \cite{lin90,boh93,tom94}.

These  phenomena are  traced back  to the  specific role  that chaotic
states play  in such  systems \cite{boh93,tom94,doron} In  contrast to
integrable dynamics,  the tunnel doublets of  the localized quasimodes
are  in  a  mixed system  no  longer  isolated  in the  spectrum,  but
resonantly interact  with states that are associated  with the chaotic
part of  phase space.  Due  to their delocalized nature,  such chaotic
states  typically exhibit  a  considerable overlap  with the  boundary
regions  of  both  regular  wells.   They  may  therefore  provide  an
efficient coupling  mechanism between the quasimodes  -- which becomes
particularly  effective at values  of $\hbar$  (or any  other external
parameter) at  which one  of the chaotic  level is shifted  exactly on
resonance with the tunnel  doublet.  This interpretation was basically
confirmed by a  statistical analysis of the fluctuations  of the level
splittings,  which  were  shown  to  be in  good  agreement  with  the
corresponding  prediction that a  random matrix  model of  the chaotic
part of phase space  would yield \cite{tom94,leyvraz96}.  Evidence for
chaos-assisted tunneling was found  in theoretical calculations of the
life times of nondispersive  wave packets in microwave driven hydrogen
\cite{zakrzewski98}  and,  most   recently,  also  in  experiments  on
ultracold atoms that were stored within periodically modulated optical
lattices   \cite{hensinger01,steck01,mouchet01}.   Also  for
electromagnetic  systems,  in  particular within  microwave  billiards
\cite{dembowski}  and  optical  microcavities \cite{noeckel},  it  was
verified  that  (ray  optical)  chaos  has a  strong  impact  on  wave
tunneling.

Apart from this qualitative insight and its quantitative statistical modeling,
however, little is known about  chaos-assisted tunneling from the semiclassical
point   of  view.    In  particular,   there  exists   no  practicable
semiclassical tool up to now for estimating the tunnel coupling from a
torus within a regular island  to the chaotic sea (which would provide
valuable information about the magnitude around which the tunnel rates
fluctuate).  This is essentially due to the fact that in nonintegrable
systems  invariant tori  are generally  characterized by  a restricted
domain of  analyticity and can  therefore not be continued  far enough
into   the   complex  domain   to   reproduce   the  tunnel   coupling
\cite{greene81,creagh_in_steve}.   As  has  been  shown by  Shudo  and
coworkers \cite{shudo}, it is in principle possible to semiclassically
reconstruct the tunneling  tail of the time evolution  of some initial
wavefunctions within a mixed system by summing up the contributions of
all  complex classical trajectories  that fulfill  appropriate boundary
conditions.   This  requires the  analysis  of  a highly  complicated
manifold  of initial  conditions  in complex  phase  space, where  the
Stokes  phenomenon   plays  a  crucial  role  for   the  selection  of
contributing  trajectories \cite{shudo}.   Though successful  from the
quantitative point of  view, this method is limited  to initial values
problems,  and it  is not  evident how  much insight it may
provide for an intuitive understanding of the properties of eigenfunctions.

The present contribution can in  some sense be seen as a complementary
approach with respect  to the work of Shudo  et al.\ \cite{shudo}.  We
shall,  however, not tackle  directly mixed  regular-chaotic dynamics,
but consider a conceptually  simpler situation, namely the {\em nearly
integrable} regime in which chaos is not yet fully developed.  In this
context,   it  has   been  pointed   out  by   Bonci   and  co-workers
\cite{bonci98} that  {\em classical resonances} should  play a crucial
role in  such systems.  As nonlinear resonances  between two classical
eigenmodes arise whenever their  frequencies are related by a rational
multiple, they generally lead  to quasi-degeneracies of the associated
levels  in  the  spectrum.   This  may  therefore  induce  significant
couplings  to higher  states  within the  regular  region, from  which
tunneling occurs with a much faster rate than from the original state.
Besides  the  aspect  of  near-degeneracies  --  which  alone  is  not
sufficient for a quantitative understanding of the role of resonances,
as  level   crossings  arise   also  in  integrable   systems  without
introducing  any modification  -- it  is also  important to  take into
account the magnitude of the effective matrix element that couples the
near-degenerate  states.  This  quantity  is directly  related to  the
structure  of  the  underlying  classical  phase  space.   Within  the
action-angle  variable representation  of  the unperturbed  integrable
dynamics,  a nonlinear  resonance  generally manifests  in  form of  a
pendulum-like  perturbation, and  the associated  modification  of the
torus  structure in  its vicinity  is then  responsible  for couplings
across the resonance, the strength of the coupling being determined by
the   extension   of   the   pendulum   structure   in   phase   space
\cite{UzeNoiMar83JCP,Ozo84JPC,FarUze86JCP,BreHol91AP}.

Our previous publication \cite{prlbsu}  has shown that the combination
of  these two  aspects, namely  the near-degeneracy  and  the coupling
induced by the  resonance, form the basis of  a mechanism that governs
tunneling  in the  near-integrable  limit.  This  study was  performed
within the ``kicked  Harper'' model, a one dimensional time periodic
system governed by the Hamiltonian
\begin{equation}  \label{eq:kh_H}   
    H(p,q,t)  =  \cos p  +  
  \tau \sum_{n=-\infty}^{\infty}  \delta(t-n\tau)  \cos q
\end{equation}
the dynamics of which is equivalent to the symplectic map 
\begin{eqnarray}
p' & = & p + \tau \sin{q} \nonumber \\
q' & = & q - \tau \sin{p'} \;
\label{eq:kh_map}
\end{eqnarray}
which describes how the phase space variables $(p,q)$ evolve from time $t =
n\tau$ (more precisely, from the time immediately before the kick) to time 
$t = (n+1)\tau$.
This model has proven its  usefulness in the context of many different
aspects           related          to           quantum          chaos
\cite{Leboeuf90,LimShe91PRL,GeiKetPet91PRL}  (including also dynamical
tunneling \cite{RonO94PRL}).
Our study was restricted to a relatively small value $\tau = 1$ of the 
perturbation parameter, for which the classical dynamics in nearly integrable.
The quantum tunneling rates that are obtained at this perturbation strength 
are shown in Fig.\ref{fig:split1}.
We see that, despite a  seemingly ``regular'' phase  space (shown in
Fig.~\ref{fig:khmap}), they are nontrivial and 
exhibit similar features  as in the  case of a  truly mixed
regular-chaotic  system:   Even for  rather  small deviations  from
integrability, the  tunneling rates may, in  the semiclassical regime,
become appreciably enhanced with respect to the integrable limit (by a
factor that may  reach up to ten orders of magnitude  in the case that
we have  considered in \cite{prlbsu})  and do not follow  a monotonous
exponential scaling  with $1/\hbar$.  

As key  ingredient to understand such a behavior, we have  introduced an 
integrable approximation of the
kicked Harper map (in an analogous way as in \cite{Sok86SJTMP}) which,
even  in the  case of  moderate perturbations,  provides  a reasonable
description  of the nearly  integrable motion  on the  invariant tori.
Expanding the kicked Harper eigenfunctions within the eigenbasis of this
integrable approximation allowed to unambiguously identify resonances as the 
source of  modifications in the tunneling tail  of the eigenfunctions.
A  quantitative reproduction  of  the tunneling  rates  in the  kicked
Harper, the accuracy of  which is visible on Fig.~\ref{fig:split1}, was
then  achieved through  a quantum  perturbative treatment  of  a local
effective   Hamiltonian,  which  is   formally  derived   via  secular
perturbation  theory of  the classical  motion \cite{lichtenberg:book}
and  was  in  practice  obtained  via  the  Fourier  analysis  of  the
separatrix structure associated with the resonance \cite{prlbsu}.

The combination of these ``tools'' has evidently proven successful for
the  identification of  the  underlying  mechanism as  well  as for  a
low-cost  (with  respect to  computer  memory)  calculation of  tunnel
splittings  that would  otherwise be  accessible only  through  a full
quantum treatment  of the problem.  However, the  justification of the
resonance-assisted mechanism  presented in  \cite{prlbsu}  was mainly
based on the demonstration of its quantitative predictive power for
reasonably small values of $\hbar$ within the kicked Harper system.  
In  this paper,  we  would like  to go    further in  the
understanding of the tunneling process in the nearly integrable regime.
A central question that we shall address is to which extent
the  resonance-assisted tunneling
mechanism we propose should in general be the dominant one, and what
modifications  are to be expected as  the system  is pushed  deep  in the
semiclassical regime.  
A general issue that underlies these interrogations is the fact that the
approach we  propose is  based on a  
combination of  perturbative 
techniques
(both classical and quantum) and
semiclassical
concepts, and therefore
involves 
essentially
two small parameters:
the perturbation strength which enters in a purely algebraic way into the
coupling terms, and 
the quantum coarse graining $\hbar$ on which these terms depend both
algebraically and exponentially.
Although  we  obviously do  not
intend to attain  anything like mathematical rigor, our  goal in this
paper is to give evidence that the global picture 
that underlies resonance-assisted tunneling
``makes sense'' and,
on  a  more  practical  tone,  
may lead to
guiding  rules for the
identification of dominating terms in the tunneling mechanism.  

To reach this objective,
we shall  see that  it is  useful to provide  a more  {\em geometric}
vision  of  resonance-assisted  tunneling.   
This means on the one hand that
we  shall
emphasize the  connection between  the coefficients that  describe the
strength  of the  coupling  and  the {\em  complex  structure} of  the
underlying  integrable  approximation.   
On the other hand, we shall see how 
the coupling via a  nonlinear resonance
can be considered as
{\em dynamical tunneling}  process, in very much the same
spirit  as   the  coupling  between   quasimodes  on  symmetry-related
invariant tori.   However, the effective topology of  complex tori that
the  quantum  system encounters  in  order  to  undergo the  tunneling
transition {\em  sensitively} depends on the  quantum coarse graining.
For   rather  large   $\hbar$,   a  direct   connection  between   the
quasidegenerate  tori of  the two  wells  is ``seen''  by the  quantum
system.  Deeper in the  semiclassical regime,  the tori rather
appear as  being connected, via  one or several resonances,  to higher
excitations  within  the well,  from  where  a  transition across  the
separatrix is associated with a rather low imaginary action.

Our study will be  restricted to one-degree-of-freedom systems subject
to  a  time-periodic  perturbation  with period  $\tau$  or  frequency
$\omega  =  2\pi/\tau$.  We  denote  by $\hat{H}(\hat{p},\hat{q},t)  =
\hat{H}(\hat{p},\hat{q},t  +  \tau)$ the  quantum  Hamiltonian and  by
$H(p,q,t)$  its  classical limit.   The  classical  phase is  most
conveniently visualized by means  of a Poincar{\'e} surface of section
in time domain -- i.e., by the area-preserving map
\begin{equation}
  {\mathcal T}:(p_t,q_t) \longmapsto (p_{t+\tau},q_{t+\tau})
\end{equation}   
that describes the evolution of the phase space variables $(p,q)$ from
time  $t$ to  time  $t +  \tau$.   Quantum mechanically  we shall,  in
analogy, consider the quantum propagator
\begin{equation}\label{eq:hatU}
  \hat U = \exp \left( \frac{i}{\hbar} \int_0^\tau \hat H (t) dt \right) \; ,
\end{equation}
and   study  its   eigenfunctions  $|\psi_k\rangle$   and  eigenphases
$\phi_k$, defined by
\begin{equation} \label{quantizemap}
  \hat U |\psi_k\rangle = e^{i\phi_k}|\psi_k\rangle \; .
\end{equation}
Whenever an  illustrative example  appears appropriate, we  shall make
use    of    the    kicked   Harper    Hamiltonian    \cite{Leboeuf90}
Eq.~(\ref{eq:kh_H})  the  Poincar{\'e}  map   of  which  is  given  by
Eq.~(\ref{eq:kh_map}).  We shall, however,  try to keep the discussion
as general as possible in order  to allow an application also to other
time-periodic  tunneling  problems  such  as the  driven  double  well
\cite{lin90,GroO91PRL}  or the effective  Hamiltonian \cite{mouchet01}
that was  employed in  the context of  the recent  dynamical tunneling
experiments in cold atoms \cite{hensinger01,steck01}.

To lay firm foundations, we  begin in Sec.~\ref{secII} with a brief
review  of  what  we  like  to name  ``regular  tunneling''  
-- i.e., 
the  attempt  to  semiclassically  describe
tunneling by a direct analytic continuation of the invariant tori into
the complex phase  space.  We shall argue, however,  that this concept
is,  strictly  speaking, limited  to  exactly  integrable systems  and
breaks  down when a  small nonintegrable  perturbation is  applied.
This naturally leads  to the question
of  how  nonlinear  resonances  influence tunneling,  which  we  shall
discuss in  Sec.~\ref{sec:resonances}.  We shall  begin, in
Sec.~\ref{sec:secular}, with a formal description of the classical dynamics
in  the   vicinity  of  a   nonlinear  resonance,  based   on  secular
perturbation  theory,  and use  then,  in Sec.~\ref{sec:pendulum},
quantum  perturbation theory as  well as  semiclassical WKB  theory to
study transitions across the  resonance.  The practical calculation of
the  coupling coefficients  that parametrize  this description,  and a
discussion of the general properties  of their scaling, is given in
Sec.~\ref{sec:spt}.  Plugging  these basic  elements  together, we
then   obtain,   in   Sec.~\ref{sec:mechanism},   a   satisfactory
semiclassical picture of how tunneling  proceeds in presence of one or
several resonances at given value  of the quantum coarse graining.  To
demonstrate  its feasibility as  well as  to verify  basic assumptions
that  have been  made  in the  course  of its  derivation, we  finally
return, in Sec.~\ref{sec:kh}, to  the particular case of the kicked
Harper Hamiltonian, in a parameter regime where its classical dynamics
is nearly integrable.

\section{``Regular'' tunneling}

\label{secII}

\subsection{Tunneling in integrable systems} 

\label{sec:int}

For  one-dimensional  time-periodic systems, integrability  can be defined 
by the  existence of a
function  ${\cal F}(p,q)$ that is  conserved by  the  Poincar{\'e} map
${\cal T}$ describing the evolution of $(p,q)$ from time $t$ to time $t+\tau$. 
This  can be  shown  to  be equivalent  to  the  existence of  a
$\tau$-periodic     canonical     transformation    $(p,q)     \mapsto
(\tilde{p}(p,q,t),\tilde{q}(p,q,t))$ such that the Hamiltonian in the new
coordinates is  {\em time  independent}  \cite{moser:lecture} --
in  which case  the
conserved  quantity   is  simply   the  energy.   
Without loss of generality, therefore, we discuss in this subsection the
properties of time-independent Hamiltonians $H = H(p,q)$.

Integrability   quite    naturally   yields   a    great   number   of
simplifications.  
Due to the existence of a constant of motion, 
the iterates by the Poincar{\'e} map of
a given point  in phase space lie on an  invariant curve $\Gamma$ 
(see, e.g, Fig.~\ref{fig:Htildemap})
which we call,
in analogy to higher dimensional systems, a ``torus'' throughout this paper.
It will be  convenient to use the {\em action-angle} variables $(I,\theta)$
associated with $H$.
For a given phase space point $(p,q)$ on the invariant torus $\Gamma$,
the action is defined by
\begin{equation}
  I = \frac{1}{2\pi} \oint_\Gamma pdq
\end{equation}
and corresponds, up to the factor $2\pi$, to the area that is enclosed
by  the  torus  in  phase  space.  The  angle  $\theta  \in  [0,2\pi[$
represents the  conjugate variable and corresponds  to the propagation
time that elapses  from a given reference point on  $\Gamma$ up to the
point $(p,q)$ (normalized in such a way that $\theta = 2\pi$ after one
full round-trip).   Expressed in these new  variables, the Hamiltonian
is, by construction, a function of the action only:
\begin{equation}
  {H}(I,\theta) = {H}(I) \; .
\end{equation}
(In order not to overload the notation, we shall use the same symbol $H$ for
the Hamiltonian in the original phase space variables $(p,q)$ and in the
action-angle variables $(I,\theta)$).

Quantum mechanically, the time-invariance  of the Hamiltonian $\hat H$
implies that the propagator $\hat U$ of the wavefunction from time $t$
to  time $t+\tau$  (Eq.~\ref{eq:hatU}) is  simply given  by $\hat  U =
\exp(-i \hat{H} \tau /  \hbar)$.  Its eigenfunctions $\psi_k$ are then
also  the  eigenfunctions  of  the  Hamiltonian,  and  the  associated
eigenphases $\phi_k$ are related via  $\phi_k \equiv E_k \tau / \hbar$
to the eigenenergies $E_k$ of $\hat{H}$.
They can, moreover, be  semiclassically constructed using standard EBK
theory.   More  precisely,  the  semiclassical  wavefunction  that  is
associated with an invariant curve $\Gamma$ is defined by
\begin{equation}
  \psi^{(qs)}(q) = \sum_{ {x_\alpha=(p_\alpha,q_\alpha) 
  \in \Gamma} \; ; \; {q_\alpha=q}}
  \frac{1}{\sqrt{2\pi\left({\partial q_\alpha}/
    {\partial \theta}\right)_I}}
  \exp{\left(\disp \frac{i}{\hbar}
  {\mathop{\int_{x_0}^{x_\alpha}}_{\Gamma}
  p dq'} + i \frac{\pi}{2} \nu(x_0,x_\alpha) \right)}
\label{semic_eigenf}
\end{equation}
where  $\nu(x_0,x_\alpha)$ is the [algebraic]  number of  vertical tangents
that are encountered by $\Gamma$ between the phase space points $x_0$ and $x_\alpha$
\cite{maslov:book}.   $\psi(q)$  can  be  properly defined  (i.e.\  is
mono-valued)  if  and only  if  the  action  enclosed by  the  curve
$\Gamma$ fulfills the quantization condition
\begin{equation} \label{eq:ebk}
  I = \frac{1}{2\pi}\mathop{\oint}_{\Gamma} pdq 
  = \hbar\left(k+\frac{1}{2}\right) = I_k
\end{equation}
for  some  integer $k$.   In  that  case, the semiclassical energy
$E_k^{(qs)}  =  H(I_k)$  is a  good
approximation  of the  true  eigenenergy $E_k$, and the associated
semiclassical eigenfunction $\psi^{(qs)} \equiv \psi_k^{(qs)}$ fulfills
\begin{equation} \label{eq:qmode}
  \hat U \psi_k^{(qs)} = \exp{\left(- i E_k^{(qs)} \tau /\hbar\right)} 
  \, \psi_k^{(qs)} \; + O(\hbar^2) \; .
\end{equation}

It  is,  at  least  in   principle,  possible  to  improve  the  above
approximation  to an  arbitrary  order in  $\hbar$.  Nevertheless,  it
should be born in  mind that Eq.~(\ref{eq:qmode}) does not necessarily
imply   that  $\psi_k^{(qs)}$   is  an   approximation  of   the  true
eigenfunction  $\psi_k$  of $\hat  U$  (or  $\hat  H$).  This  becomes
particularly relevant for systems
that are  invariant under some  discrete symmetry --- say,  e.g.\, the
inversion ${P} : q \mapsto -q$ ---
which  is  such  that  the  invariant  curve  $\Gamma_k$  obeying  the
quantization  condition Eq.~(\ref{eq:ebk})  and its  symmetric partner
$\Gamma_k' =  {P} \Gamma_k$ are distinct.  In  such circumstances, the
semiclassical    wavefunctions    $\psi_k^{'(qs)}$   constructed    on
$\Gamma'_k$ will be the  symmetric equivalents of $\psi_k^{(qs)}$, and
the    corresponding   semiclassical    energies    $E_k^{(qs)}$   and
$E^{'(qs)}_k$ will be exactly degenerate.

Since ${P}$  admits only representations  of dimension one,  there is,
however,  a  priori no  reason  that  the  two exact  eigenergies  are
degenerate.   Classically  forbidden  processes, that  we  generically
refer  to as  tunneling events  even  when no  potential barriers  are
explicitly  involved, will  generally give  rise to  an exponentially
small   (in   $\hbar$)   coupling  matrix   element   $\alpha=|\langle
\psi_k^{(qs)}|\hat H  | \psi^{'(qs)}_k \rangle|$.   Using standard WKB
methods, this  matrix element  can be evaluated  semiclassically. For
instance in  the case  considered above where  ${P}$ is  the inversion
symmetry relating two invariant  curve $\Gamma_k$ and $\Gamma_k'$, one
obtains \cite{landau:book}
\begin{equation}
  \alpha = \frac{\hbar}{T} \exp \left(-\frac{\sigma}{\hbar} \right) \;
\label{semicsplit}
\end{equation}
where $T$ is the classical period on the torus $\Gamma_k$ and
\begin{equation}
  \sigma = \Im\left[\int_{\Gamma_k \to \Gamma_k'} p \, dq\right]
\end{equation}
is  the  imaginary part of the action  integral  taken  on a  path  joining
$\Gamma_k$  and 
$\Gamma_k'$ on their analytical  continuation in the complex phase space
(see in this context also \cite{Cre94JPA}).
This is illustrated in Fig.~\ref{fig:integrable_cont} where we plot the
analytic continuation of an invariant torus and its symmetric counterpart
in the Harper model $H = \cos p + \cos q$.

The   projection  of   $\hat   H$  on   the   subspace  generated   by
$\psi_k^{(qs)}$  and  $\psi^{'(qs)}_k$  then  reads (with  the  proper
choice of their phases)
\begin{equation}   
 	\disp
	\hat H \left(\begin{array}{c}  \psi_k^{(qs)} 
	\\ \psi^{'(qs)}_k \end{array} \right)
	= \left(\begin{array}{cc}
	E_k^{(qs)} & \alpha \\
	\alpha  & E_k^{(qs)}
	\end{array} \right) 
	\left(\begin{array}{c}  \psi_k^{(qs)} \\ 
	\psi^{'(qs)}_k \end{array} \right) \; .
\label{2matrix}
\end{equation} 
Therefore,  although the  eigenenergies are  only slightly  shifted with
respect to $E_k^{(qs)}$, yielding a splitting $\delta E_k = 2 \alpha$,
(and   thus   an  eigenphase   splitting   $\delta   \phi_k  =   2\tau
\alpha/\hbar$),  the  true  eigenstates  are not  $\psi_k^{(qs)}$  and
$\psi^{'(qs)}_k$   but  their   symmetric  and   antisymmetric  linear
combinations.  Arnold  \cite{arnold:book} has  suggested  to call  the
semiclassical     wavefunctions     (Eq.~(\ref{semic_eigenf}))    {\em
quasi-modes}   to  stress   that,  although   they  may   fulfill  the
Schr{\"o}dinger's equation  up to an arbitrary order  in $\hbar$, they
are  not  necessarily  an   approximation  of  the  true  eigenstates.
Intuitively, this can  be seen from the propagation  of a wavefunction
that is  initially prepared on  one of the tori  $\Gamma_k$.  Although
Eq.~(\ref{eq:qmode}) is fulfilled for  a single iteration of $\hat U$,
the population of the wavefunction will,  after a very long time (or a
large  number of  iterations), be  fully encountered  on  the symmetric
torus   $\Gamma'_k$,  and   oscillates  between   $\psi_k^{(qs)}$  and
$\psi^{'(qs)}_k$  with an  exponentially long  period $2  \pi  \hbar /
\delta E_k$.

In the quasi-integrable regime we consider in the following, quasimodes
can  again be  defined, and  one can  still observe  tunneling between
symmetry  related   quasimodes  which   are  degenerate  at   the  EBK
approximation.   We shall  see, however, that  the way  the  tunneling
mechanism takes place  is sensibly  more complicated than the  two-level
process sketched above in the integrable case.

\subsection{From integrability to quasi-integrability}
\label{sec:quasiint}

We consider from now on a system with a Hamiltonian which depends on a
small  parameter  $\epsilon$  in  such  a way  that  the  dynamics  is
integrable  for  $\epsilon =  0$  and  non-integrable otherwise.   For
sufficiently small  but finite values  of the perturbation,  the system
will  display a quasi-integrable  dynamics, which  more or  less means
that  the classical  motion  is visibly  not  distinguishable from  an
integrable  one.  As  stated by  the  Kolmogorov-Arnold-Moser (K.A.M.)
theorem  (cf.   \cite{arnold:book}),  the   phase  space  of   such  a
near-integrable  system  is still  characterized  by  dense layers  of
invariant tori -- so-called K.A.M. tori -- which are slightly deformed
with respect to the integrable limit.

This  modification of  the  phase space  structure  can be  explicitly
reconstructed by  means of  classical perturbation theory.   Using for
instance  the  Lie  transformation method  \cite{lichtenberg:book},  a
(time    dependent)    canonical    transformation   $(p,q)    \mapsto
(\tilde{p}(p,q,t),\tilde{q}(p,q,t))$ of  the phase space variables  
$(p,q)$ can be
defined  in  such  a way  that  
the Hamiltonian  is  effectively   time-independent
in  these  new coordinates.
This  procedure  is
described in detail in appendix  \ref{app:lie} for the special case of
rapidly driven systems (where $\epsilon$ is given by the period of the
driving).  Generally, it yields the  new Hamiltonian as a power series
in  the  perturbation  parameter  $\epsilon$,  which  in  practice  is
iteratively calculated up to some maximum order $n$:
\begin{equation}
H_{\epsilon} \simeq \widetilde{H}_{\epsilon}^{(n)}(\tilde{p},\tilde{q})
= \sum_{l=0}^{n}\widetilde{H}_l(\tilde{p},\tilde{q}) \epsilon^l.
\label{seriesH}
\end{equation}
The convergence of this series is in general of asymptotic nature, which 
means that for any finite $\epsilon$ the development converges up to some 
optimal order $n_0$ and starts diverging beyond.

As is well known and as was first emphasized by the Poincar{\'e} Birkhoff 
theorem (cf. \cite{arnold:book}), 
the  development  (\ref{seriesH}) diverges  particularly  fast in  the
vicinity of nonlinear resonances.   
If the frequency of the oscillation generated by $\widetilde{H}_{\epsilon}^{(n)}$ --
given by $\Omega(I) = \partial \widetilde{H}_{\epsilon}^{(n)} / \partial I$ in the action angle
variables $(I,\theta)$ of $\widetilde{H}_{\epsilon}^{(n)}$ -- is a rational multiple of
the frequency $\omega = 2\pi/\tau$ that characterizes the time-periodic perturbation,
then even a small
strength of the
perturbation  causes a  substantial  modification of  the phase  space
structure.  Except  for a stable  and an unstable periodic  orbit, the
resonant torus and the tori  in its immediate vicinity are broken.  At
their place, a new regular  substructure is appearing which is winding
around  the stable  orbit and  which manifests  within  the Poincar{\'e}
surface  of  section  in  form  of a  chain  of  eye-like  structures,
so-called ``resonance islands'' (we use this terminology in analogy to
mixed regular-chaotic systems where  they may appear as ``islands'' of
regular  motion embedded into  a ``sea''  of chaotic  dynamics).  This
island  chain is  separated from  the  remaining set  of the  unbroken
K.A.M.\  tori  by a  tiny  chaotic  layer  which originates  from  the
separatrix  structure  associated   with  the  unstable  fixed  point.
Compared to the size of  the resonances, the extension of such chaotic
layers is  practically negligible if the perturbation  is rather small
and  if  overlaps   of  different  resonances  do  not   play  a  role
\cite{chirikov}.

As a typical example,  Fig.~\ref{fig:khmap} shows the phase space
portrait of the kicked Harper map in the near-integrable regime ($\tau
= 1$).  In comparison  with Fig.~\ref{fig:Htildemap}, we see that the
phase  space   structure  does  not  substantially   differ  from  the
corresponding integrable limit.   The most significant modification is
in fact the appearance of island chains which are induced by nonlinear
resonances between the kick periodicity and the free oscillation.

However, despite  the overall  regularity of the  phase space  at that
strength  of  the  perturbation,  the  tunneling  process  is  already
substantially modified with respect  to the integrable case.  This was
already  discussed   in  the  Introduction.   It   is  illustrated  in
Fig.~\ref{fig:split1} where we show the scaling of a typical tunneling
rate with the  quantum coarse graining.  As will  be explained in more
detail in section~\ref{sec:kh}, we plot here the level splittings (or,
more   precisely,   the  difference   of   the  evolution   operator's
eigenphases)   between   the   symmetric  and   antisymmetric   states
constructed on  the tori shown  on Fig.~\ref{fig:khmap}.  We  see that
the tunneling rates  do not follow the smooth  and monotonous decrease
with $1/\hbar$ that was  predicted for integrable systems, but exhibit
rather significant fluctuations.   Moreover, the tunnel splittings are
by many orders  of magnitude larger than the  ones calculated from the
integrable approximation (\ref{seriesH}) (dashed line) which otherwise
reproduces the near-integrable phase space structure quite well.

These  findings are in  accordance with  the fact  that the  method of
analytic continuation of the phase space tori to complex domain, which
essentially provided  the basis  for the semiclassical  description of
tunneling in  integrable systems, does  not work in  the nonintegrable
case.  It  is obvious  that the two  equivalent tori between  which we
consider  tunneling do  no longer  form  a single  smooth manifold  in
complex phase space if the dynamics is not integrable (since such
a  manifold would  imply the  existence of  an additional  constant of
motion).    This  alone,   however,  does   not   necessarily  disable
continuation   methods   of   the   kind   that   was   described   in
section~\ref{sec:int}.  If  the   manifolds  that  correspond  to  the
analytic continuation  of the two equivalent tori  happen to intersect
under some  finite angle  somewhere in complex  phase space,  then the
respective semiclassical wavefunctions Eq.~(\ref{semic_eigenf}) can be
continued  until that intersection  line, and  their splitting  can be
evaluated  by  means of  their  overlap at  that  line.   As has  been
demonstrated by Wilkinson  \cite{wilkinson86}, this yields essentially
the same  exponential decrease of  the splitting with $1/\hbar$  as in
integrable  dynamics, but  with a  different power  of $\hbar$  in the
algebraic prefactor.

In reality,  however, the analytic  continuations of the tori  do {\em
not}  meet each  other,  but  are interrupted  at  their {\em  natural
boundaries},  consisting of  lines of  singularities in  complex phase
space.  This  phenomenon has  been discussed in  detail by  Greene and
Percival \cite{greene81} for the case of the standard map: by means of
the Fourier representation  of the K.A.M.\ torus as  a function of the
angle  variable, the location  and nature  of these  singularity lines
were  analyzed, and  it  was found  that  the complex  tori acquire  a
fractal-like structure in their vicinity.

This behavior  is qualitatively confirmed for the  kicked Harper map.
By   means  of  an   optimization  program   which  is   described  in
Appendix~\ref{app:cont}, we are able to identify the complex invariant
manifold  that corresponds  to  the continuation  of  a given  K.A.M.\
torus.   A   typical  example   of  such  a   manifold  is   shown  in
Fig.~\ref{fig:khcont}.  Although  the   dynamics  is  rather  close  to
integrability,  the K.A.M.\ torus  cannot be  continued far  away into
imaginary domain.   In fact,  the projection of  the complex  torus to
real phase space is restricted to regions {\em far inside} the regular
K.A.M.\ tori regime -- i.e., far away from the chaos border around the
separatrix.  
At this stage of our investigation, we therefore note that
the concept of  analytic continuations does not
seem  to represent  the  appropriate framework  for the  semiclassical
study of near-integrable tunneling  phenomena -- which again indicates
that in  near-integrable systems tunneling  proceeds in a way  that is
very different from integrable ones.

\section{Nonlinear resonances and tunneling}

\label{sec:resonances}

\subsection{Effective Hamiltonian in the vicinity of resonances}

\label{sec:secular}

As we have seen in Fig.~\ref{fig:khmap} and Fig.~\ref{fig:Htildemap}, the
major modification of phase space that arises  when going from integrable to
near-integrable 
dynamics  is   the  appearance  of  substructures   due  to  nonlinear
resonances.   It  is  therefore  natural  to ask  what  would  be  the
influence of  these substructures on the tunneling  process.  In fact,
the effect  of resonances  on semiclassically quantized  energy levels
and  eigenstates  in  a   quasi-integrable  system  has  already  been
discussed  under a  variety  of aspects,  mostly  within the  chemical
physics  literature  \cite{UzeNoiMar83JCP,Ozo84JPC,FarUze86JCP,BreHol91AP}.
The  approach
that we are adopting follows more  or less the lines of the derivation
undertaken  by Ozorio  de Almeida  \cite{Ozo84JPC}~:  we shall
introduce an effective integrable  Hamiltonian for the dynamics in the
vicinity of the resonance and  then discuss, in the following section,
how   this  Hamiltonian  may   induce  couplings   between  integrable
eigenmodes.

Most conveniently, the effective integrable Hamiltonian that generates
the dynamics in  the vicinity of a nonlinear  resonance is constructed
by means of secular perturbation theory \cite{lichtenberg:book}.  This
procedure  is  illustrated hereafter  for  the  particular  case of  a
periodically driven  one-degree-of-freedom system.  For  this purpose,
we write the Hamiltonian of our weakly perturbed system in the form
\begin{equation}
  H(I,\theta,t) = H_0(I) + \epsilon' V(I,\theta,t) \label{Haa}
\end{equation}
where  $H_0$  represents the  unperturbed  Hamiltonian  or a  suitable
integrable approximation of $H$,  obtained e.g.\ by standard classical
perturbation theory as a series  of the form (\ref{seriesH}),
and $(I,\theta)$ are the  action-angle variables associated with $H_0$
(which  implies  that   $H(I,\theta+2\pi,t)  =  H(I,\theta,t)$).   The
perturbative  term   $\epsilon' V$  contains  then  ``all  the   rest''  of  the
Hamiltonian $H$ and is simply  given by $H(I,\theta,t) - H_0(I)$ for a
particular choice of $H_0$.

A  resonance  condition  arises   whenever  the  frequency  $\omega  =
2\pi/\tau$ of the  external driving equals a rational  multiple of the
internal oscillation  frequency $\Omega(I) = dH_0/dI$  of the system
-- i.e.,
\begin{equation}
  r \Omega_{r:s} = s \omega
\end{equation}
where $r$, $s$ are coprime positive integers and $\Omega_{r:s} = \Omega(I_{r:s})$ is
the oscillation frequency at the action $I_{r:s}$ at resonance.  
In the vicinity of such
a $r$:$s$  resonance, standard classical  perturbation theory diverges
rather quickly due  to small denominators.  To avoid  this problem, it
is convenient to perform a  canonical transformation to the frame that
co-rotates with  the angle variable $\theta$ on the  resonance.  This is
done by introducing the new angle variable
\begin{equation}
  \Theta = \theta - \Omega_{r:s} t \label{thetarot}
\end{equation}
which remains constant, under the  time evolution generated by $H_0$, on
the $r$:$s$  resonance, and varies  slowly in its vicinity.   After the
corresponding transformation $H \mapsto H^{(r:s)}  = H - \Omega_{r:s} I$
of  the  Hamiltonian  (which  is necessary  since  the  transformation
(\ref{thetarot}) depends  explicitly on time), we obtain
\begin{equation}
H^{(r:s)}(I,\Theta,t) = H_0(I) - \Omega_{r:s} I + \epsilon' V^{(r:s)}(I,\Theta,t)
\end{equation}
as new Hamiltonian that describes the time evolution of the new phase space 
variables $(I,\Theta)$, with the perturbation term
\begin{equation}
V^{(r:s)}(I,\Theta,t) = V \left( I,\Theta 
+ \Omega_{r:s} t,t \right). \label{tildev}
\end{equation}

Since $\Theta$ varies on a time  scale that is rather long compared to
the periodicity  $2 \pi/\omega$  of the external  driving, we  can now
apply    adiabatic   perturbation    theory    to   the    Hamiltonian
\cite{lichtenberg:book} and eliminate  the explicit time dependence by
a canonical transformation $(I,\Theta) \mapsto (\bar{I},\bar{\Theta})$
to     new,     slightly     shifted     phase     space     variables
$(\bar{I},\bar{\Theta})$, which  is accompanied by  the transformation
$H^{(r:s)}  \mapsto \bar{H}^{(r:s)}$  of the  Hamiltonian.   In lowest
order in  the perturbation, this  amounts to replacing  $V^{(r:s)}$ by
its time average  over $r$ driving periods (note  that $V^{(r:s)}$, as
defined in  Eq.~(\ref{tildev}), is $2 \pi r/\omega$  periodic in $t$).
We thereby obtain the effective time-independent Hamiltonian
\begin{equation}
  {\bar{H}^{(r:s)}}(\bar{I},\bar{\Theta}) = 
  H_0^{(r:s)}(\bar{I}) +  
  \bar{V}^{(r:s)}(\bar{I},\bar{\Theta}) \; , 
\label{Hres}
\end{equation}
with
\begin{eqnarray}
  H_0^{(r:s)}(\bar{I}) & := & 
  H_0(\bar{I}) - \Omega_{r:s} \bar{I} \; , 
  \label{H0rs}\\
  \bar{V}^{(r:s)}(\bar{I},\bar{\Theta}) & := &
   \frac{1}{r\tau} 
  \int_0^{r \tau} \epsilon'V^{(r:s)}(\bar{I},\bar{\Theta},t) dt. 
  \label{vtilde}
\end{eqnarray}
The new phase space variables $(\bar{I},\bar{\Theta})$ are given by
\begin{eqnarray}
  I & = & \bar{I} + \frac{\partial G}{\partial \Theta}
  (\bar{I}, \Theta,t), \\
  \bar{\Theta} & = & \Theta + \frac{\partial G}{\partial \bar{I}}
  (\bar{I}, \Theta,t)
\end{eqnarray}
where $G$ is, in first order in the perturbation, evaluated as
\begin{equation}
  G(\bar{I},\Theta,t) = - \int_0^t (\epsilon'V^{(r:s)}(\bar{I},\Theta,t') -
  \bar{V}^{(r:s)} (\bar{I},\Theta) ) dt'.
\end{equation}

Further  insight  into the  properties  of  the effective  Hamiltonian
Eq.~(\ref{Hres}) is obtained by the Fourier series ansatz
\begin{equation}
  \epsilon'V(I,\theta,t) = \sum_{k,l=-\infty}^{\infty} {V}_{k,l}(I) 
  \exp (i(k \theta + l \omega t)) \label{Vfourier}
\end{equation}
for     the     perturbation     term     in   Eq.~(\ref{Haa})     (with
$V_{-k,-l}=V_{k,l}^*$).    This    yields   according   to
Eq.~(\ref{vtilde})
\begin{equation}
  \bar{V}^{(r:s)}(\bar{I},\bar{\Theta}) = V_0( \bar{I} ) + 
  \sum_{m=1}^{\infty} V_{r.m}(\bar{I}) \cos(m r \bar{\Theta} + \varphi_m) \; ,
\end{equation}
where  the  real  expansion  coefficients  $V_{r.m}$  and  their
associated phases $\varphi_m$ are introduced via
\begin{equation}
  {V}_{mr,-ms} \equiv \frac{1}{2} \, V_{r.m} e^{i\varphi_m} \; . 
  \label{Vfour}
\end{equation}
We  note that,  in lowest  order  in the  perturbation, the  effective
Hamiltonian  Eq.~(\ref{Hres})  corresponds  to  a $2  \pi/r$  periodic
function in $\bar{\Theta}$.

In general, the magnitude  of the Fourier coefficients $V_{r.m}$ decreases
rather rapidly with  $m$.  More precisely, assuming the  perturbation
$V$ to be an analytic function, the decrease  of $|{V}_{k,l}|$
with  $k$  would be  exponential, i.e.
\begin{equation}
    \lim_{m \to \infty} 
  \frac{\log \left (V_{r.m}/V_{r.1} \right)}{m} 
  =  \vartheta \; .
  \label{eq:exp-decay}
\end{equation}
with  the  exponent  $\vartheta$  governed  by  the  location  of  the
singularities of  $V$.  We  shall come back  in section~\ref{sec:spt},
and in more detail in  section~\ref{sec:kh} in the particular case of
the kicked Harper model, to the descriptions of these singularities.

Expanding $H_0$ up to second  order around the action $I_{r:s}$ of the
$r$:$s$ resonance, we then obtain
\begin{equation}
   \bar{H}^{(r:s)} = H_0(I_{r:s}) + V_0(I_{r:s}) + 
  \frac{(\bar{I}-I_{r:s})^2}{2\mu} + \sum_{m=1}^\infty V_{r.m}(I_{r:s})
   \cos{\left(mr\bar{\Theta}+\varphi_m\right)} \; ,
\label{localapprox}
\end{equation}
as lowest  order expression for the integrable  Hamiltonian, with $\mu
\equiv (d^2H_0/dI^2)^{-1}(I_{r:s})$  the inverse change  of frequency with
action  at   the  resonance  (note  that  the   linear  dependence  on
$(\bar{I}-I_{r:s})$ is canceled  by the $\Omega_{r:s} I$  term in (\ref{H0rs})).  
As the first
$m=1$ term dominates the Fourier expansion of the $\bar{\Theta}$ dependent
part,  the  dynamics generated  by  Eq.~(\ref{localapprox}) is  nearly
equivalent  to  the  one   of  a  generalized  mathematical  pendulum,
containing $r$ regions  of bound motion within $0\leq\bar{\Theta}\leq2\pi$
instead of one.
At large deviation from resonance ($(\bar{I} - I_{r:s})^2/(2\mu) \gg \max_\Theta 
\bar{V}^{(r:s)} \simeq V_{r.1}$), the system performs a free rotation in $\Theta$ that is only
marginally perturbed by the presence of $\bar{V}^{(r:s)}$,
while in the immediate vicinity of $(\bar{I} \!  = \!  I_{r:s})$, librational
motion around the (co-rotating) angles $\Theta \!  =  \!   2l\pi/r$ ($l\in{\mathbb Z}$)
is also possible.

This is illustrated in Fig.~\ref{fig:res_10} where
we plot the  phase space of the kicked  Harper map at $\tau =  1$ as a
function of the action-angle variables $(I,\theta)$ associated with an
integrable approximation of type Eq.~(\ref{seriesH}).  Clearly, we see
that the structure of the  10:1 resonance chain strongly resembles the
one of a pendulum with 10 islands.

\subsection{Transitions in the generalized pendulum}

\label{sec:pendulum}

The most straightforward way  now to quantitatively analyze the effect
of  the resonance  onto the  unperturbed  eigenstate is  given by  the
framework  of quantum  perturbation  theory, directly  applied to  the
effective  pendulum  Hamiltonian  (\ref{localapprox}).  This  requires
that quantities like energies, matrix elements, transition rates etc.\
remain  invariant under  the succession  of  canonical transformations
that  leads  from  $(p,q)$  to  $(\bar{I},  \bar{\Theta})$,  which  is
generally  fulfilled  in   the  semiclassical  regime.   We  therefore
consider now the quantum Hamiltonian
\begin{equation}
  \widehat{H} = \widehat{H}_0 + \widehat{V} \; .
\end{equation}
The unperturbed part
\begin{equation}
  \widehat{H}_0 = \frac{(\widehat{I}-I_{r:s})^2}{2\mu}
\end{equation}
is  more or  less  equivalent to  the  initial integrable  Hamiltonian
$H_0(I)$  (see   (\ref{Haa}))  near  $I  =  I_{r:s}$   with  the  term
$\Omega_{r:s}  I$  being   substracted,  and  the  perturbation  which
contains the effect of the resonance is given by
\begin{equation}
  \widehat{V} = \sum_{m=1}^\infty V_{r.m}
  \cos{\left(mr\widehat{\Theta}+\varphi_m\right)} 
\label{pendpert}
\end{equation}
(constant terms are omitted throughout).  $\bar{I}$ and $\bar{\Theta}$
being canonically conjugate variables,  the action operator is defined
by
\begin{equation}
  \widehat{I} = \frac{\hbar}{i} \frac{\partial}{\partial \bar{\Theta}} \; ,
\label{quac}
\end{equation}
(with anti  periodic boundary conditions in  $\bar{\Theta}$ to account
for the Maslov indices in the original variable \cite{Ozo84JPC}).  The
unperturbed  modes $|\widetilde{\psi}_k\rangle$,  which  correspond to
the  eigenfunctions of $H_0$,  are then  given by  plane waves  in the
angle $\bar{\Theta}$
\begin{equation}
  \disp \langle\bar{\Theta}|\widetilde{\psi}_k\rangle = 
  \frac{1}{\sqrt{2\pi}}\exp{\left(i(k+1/2)\bar{\Theta}\right)},
\label{unpertmode}
\end{equation}
with the  quantized actions $I_k  = \hbar (k+1/2)$.   Their associated
energies (with respect to $\widehat{H}_0$) read
\begin{equation}
\bar{E}_k = \frac{\hbar^2}{2 \mu} (k - \kappa_{r:s})^2 \label{enpend}
\end{equation}
with $\kappa_{r:s}+1/2 \equiv I_{r:s} / \hbar$.

The  matrix  elements   of  the  perturbation  operator  $\widehat{V}$
Eq.~(\ref{pendpert}) within the unperturbed basis are evaluated as
\begin{equation}
      \langle \widetilde{\psi}_{k'}|\widehat{V}|\widetilde{\psi}_k\rangle 
  = \frac{1}{2\pi} \int_0^{2\pi}
  \bar{V}(\bar{\Theta}) e^{i(k-k')\bar{\Theta}} d\Theta =
  \sum_{m=1}^{\infty}\frac{V_{r.m}}{2} \delta_{|k-k'|,mr} \; .
\end{equation}
Hence, within a perturbative approach,  the modification  of the
eigenmode $|\widetilde{\psi}_k\rangle$ reads
\begin{equation}
       |\psi_k\rangle = 
  |\widetilde{\psi}_k\rangle + \sum_{m \neq 0} {\mathcal A}_{k,m}
        |\widetilde{\psi}_{k+mr}\rangle
\label{pertmode}
\end{equation}
where in first order approximation, the transition amplitudes are given by
\begin{equation}
  {\mathcal A}^{(1)}_{k,m} = \frac{V_{|r.m|} / 2}
    {\bar{E}_k - \bar{E}_{k+mr}} \; .
  \label{eq:first_order}
\end{equation}
Second and higher order corrections contain sums over products of type
$(V_{r.m} / 2) / (\bar{E}_k - \bar{E}_{k+mr})$ $(V_{r.m'} / 2) /
(\bar{E}_k   -   \bar{E}_{k+(m+m')r})$  $|\widetilde{\psi}_{k+(m+m')r}
\rangle$.  As  a consequence, a $r$:$s$ resonance  couples, as expressed  by
Eq.~(\ref{pertmode}), only those  unperturbed modes to  the state
$|\widetilde{\psi}_k\rangle$ the quantum  numbers of which differ from
$k$ by integer multiples of $r$.

The  perturbative   expansion  converges   rather  fast  as   long  as
$|V_{r.1}|  \ll |\bar{E}_k  - \bar{E}_{k\pm  r}|$ --  that  is, with
$\bar{\Omega}_k=(I_k-I_{r:s})/\mu$,
\begin{equation} \label{eq:pert_cond}
  r \hbar |\bar{\Omega}_k| \gg |V_{r.1}| \; ,
\end{equation}
which is well fulfilled as long as the action range spanned by the 
librational islands is small in front of $r\hbar$.
Due to  the exponential scaling  of $V_{r.m}$
with  $m$  (see  Eq.~(\ref{eq:exp-decay})),  the  resulting  overlap  matrix
elements $|\langle \widetilde{\psi}_{k+rm}|\psi_k\rangle|$ decrease in
general  rapidly  with  $m$.   Significant  admixtures,  however,  are
induced  from  states  $|\widetilde{\psi}_{k'}\rangle$  the  quantized
actions $I_{k'}$  of which  are located on  the {\em other}  side with
respect  to the  pendulum center  at $I_{r:s}$  and lie  close  to the
symmetric equivalent of $I_k$.  In this case, $(k+k')/2 \simeq \kappa_{r:s}$,
or equivalently, $k' = k + r m$ with $m \neq 0$ such that
\begin{equation}
  r m \simeq 2 (\kappa_{r:s} - k)\; ,
  \label{eq:deg_cond}
\end{equation}
and  the energies  (\ref{enpend}) of  the  states $|\widetilde{\psi}_k
\rangle$ and  $|\widetilde{\psi}_{k'} \rangle$ become near-degenerate,
which strongly enhances their  coupling with respect to the neighbors
$|\widetilde{\psi}_{k+(m\pm 1)r}\rangle$.  Though relatively weak as
compared    for   instance   with    the   $|\widetilde{\psi}_k\rangle
\leftrightarrow  |\widetilde{\psi}_{k\pm  r}\rangle$ couplings,  these
transitions  across  the island  chain  play  a  crucial role  in  the
tunneling  process.   

This  makes  it necessary,  however,  to  consider
somewhat further the  perturbation expansion.  Indeed, the exponential
behavior Eq.~(\ref{eq:exp-decay}) of the coefficient $V_{r.m}$ makes
it  a priori  not  obvious to  decide  whether, in  the evaluation  of
${\mathcal  A}_{k,m}  = \langle  \widetilde{\psi}_{k  +  mr} |  \psi_k
\rangle$, the  first order  contribution in $V_{r.m}$  dominates the
$m$'th  order  contribution  in   $V_{r.1}$,  since  this  latter  is
proportional  to   $(V_{r.1})^m$,  and   therefore  both  terms   have  an
exponential part $\exp(-m \vartheta)$.  As  we shall see, it turns out
that  the amplitudes  $\langle  \widetilde{\psi}_{ k  +  mr} |  \psi_k
\rangle$   are  dominated   by   the  first   order  term   ${\mathcal
A}^{(1)}_{k,m_k}$  in  the  limit  of small  perturbations  (at  fixed
$\hbar$), while for more strongly  perturbed systems (or deeper in the
semiclassical  regime at  fixed strength  of the  perturbation) higher
order coupling terms may become dominant.

For  this purpose,  it  is useful  to  consider in  more detail  the
special case of the {\em exact} pendulum dynamics
\begin{equation}
    \widehat{H} = 
  \frac{(\widehat{I}-I_{r:s})^2}{2\mu}+V_{r.1}\cos{r\theta} \; ,
\label{Hpend}
\end{equation}
with $V_{r.m} = 0$ for $m > 1$.
In this  case, the
coupling          from         $|\widetilde{\psi}_k\rangle$         to
$|\widetilde{\psi}_{k+rm}\rangle$   is  described   by  perturbation
theory of order $m$, which can be straightforwardly evaluated due to
the tight-binding  structure of the  Hamiltonian matrix.  As  shown in
appendix \ref{app:tunnpend}, one has for $k < \kappa_{r:s}$
\begin{equation} \label{eq:from_appendix4_1}
  \langle \widetilde{\psi}_{k+m r}|\psi_{k}\rangle \simeq 
  \left(\frac{\mu V_{r.1}}{\hbar^2 r^2}\right)^{m}
  \frac{\disp\Gamma(\delta_k)}{\disp m! \Gamma (m+\delta_k)} 
\end{equation}
in the limit of large $m \simeq 2 ( \kappa_{r:s} - k)$.  Here we introduce
\begin{equation}
  \delta_k \equiv 2 \, \frac{\kappa_{r:s} - k}{r} - m 
  = s_k/(\hbar\pi) - m\; , 
\label{delta}
\end{equation}
the equivalent of the energy  denominator in terms of quantum numbers,
where
\begin{equation}
  s_k \equiv 2\pi\hbar \, \frac{\kappa_{r:s} - k}{r} \label{eta}
\end{equation}
represents the phase space area that is enclosed between the quantized
torus $I_k$ and the center  $I_{r:s}$ of the pendulum within the angle
range $0 \leq \bar{\Theta} \leq 2 \pi / r$.

From   the  semiclassical   point   of  view,   the  transition   from
$|\widetilde{\psi}_k  \rangle$ to  its counterpart  on the  other side
with respect  to the pendulum  center corresponds to a  {\em dynamical
tunneling} process.   Unless $\kappa_{r:s}$ is integer  or half-integer, this
tunneling process is, as in the case of a non-symmetric double well, a
{\em non-resonant} one, which means that the states that are connected
by tunneling are not quasi-degenerate, but well separated in energy --
or, alternatively formulated,  that quantized tori on one  side of the
barrier are connected to non-quantized  ones on the other side.  Under
such  circumstances, only  a tiny  fraction of  the population  may be
encountered on the forbidden side of the barrier.

Based on  this point of  view, we can  derive, by means of  WKB theory
\cite{maslov:book,langer37},   a  semiclassical  expression   for  the
wavefunction  within  the  generalized  pendulum, which  includes  the
tunneling component beyond the  pendulum center.  This construction is
shown  in detail  in appendix  \ref{app:tunnpend} in  the case  of the
exact   pendulum    Eq.~(\ref{Hpend}),   and   can    be   generalized
straightforwardly in  the more general  case Eq.~(\ref{pendpert}).  It
yields
\begin{equation}
  \psi^{\rm sc}_k(\bar{\Theta}) = \frac{1}{\sqrt{T_k |\dot{\bar{\Theta}}_k|}}
         \left[ \exp \left( \frac{i}{\hbar} S_{k}(\bar{\Theta}) \right) + 
         {\mathcal A}_T \sum_{\scriptstyle k'=k+mr \atop 
          \scriptstyle k'\simeq 2\kappa_{r:s} -k}
         \frac{\sin\left(\pi \delta e_{kk'}\right)}{\pi \delta e_{kk'}}
         \exp \left(\frac{i}{\hbar} S_{k'}(\bar{\Theta}) \right) \right]
\label{eq:from_appendix4_2}
\end{equation}
as semiclassical eigenfunction of the state $| \psi_k \rangle$, with
\begin{equation}
  S_{k}(\bar{\Theta})  =   \int_0^{\bar{\Theta}}
  {\mathcal I}_k(\bar{\Theta}') d\bar{\Theta}' \; 
\end{equation}
the action integral along the torus and
\begin{equation} \label{eq:ekk}
  \delta e_{kk'} = (2\kappa_{r:s} - (k+k')) = 
  \frac{\bar{E}_k - \bar{E}_{k'}}{r\hbar\bar{\Omega}_k} \; .
\end{equation}
Here,  $I = {\mathcal  I}_k(\bar{\Theta})$ parametrizes  the quantized
torus  associated with  the  excitation $k$  (which naturally  implies
$S_{k}(2\pi) = 2 \pi \hbar  ( k + 1/2)$), $\dot{\bar{\Theta}}_k \equiv
\dot{\bar{\Theta}}_k(\bar{\Theta})$ symbolizes  the time-derivative of
the angle  variable along the  quantized torus, and $T_k$  denotes its
period,  i.e., the  classical  propagation time  that elapses  between
$\bar{\Theta} = 0$ and  $\bar{\Theta} = 2\pi$.  The coupling amplitude
${\mathcal A}_T$ is given by
\begin{equation}
  {\mathcal A}_T = 
  \frac{\exp(-\bar \sigma_k/\hbar)}{2\sin(s_k/\hbar)} \; ,
  \label{eq:semiclassical_A}
\end{equation}
where $\bar \sigma_k$  denotes the imaginary  part of the action  along the
complex classical manifold that  connects the quantized torus with its
symmetric   counterpart,   and   $s_k$   is  given   by   (\ref{eta}).
Interestingly, these two actions  $\bar \sigma_k$, $s_k$ and their relation
to $\hbar$ fully determine the transition rate across the resonance in
the semiclassical limit.

The semiclassical expression (\ref{eq:from_appendix4_2}) is explicitly
derived  in appendix  \ref{app:tunnpend} for  the special  case  of an
exact pendulum dynamics (\ref{Hpend}).  As it is based on the topology
of the phase  space structure rather than on the  explicit form of the
potential, we expect its validity also in the presence of nonvanishing
(but comparatively weak) higher  harmonics.  The case (\ref{Hpend}) is
nevertheless instructive, as it  permits an analytic evaluation of the
parameters   that    enter   into   (\ref{eq:from_appendix4_2}).    If
$\bar{V}(\bar{\Theta}) = V_{r.1} \cos{r\bar{\Theta}}$, we have
\begin{equation}
  \bar \sigma_k = \sqrt{2\mu V_{r.1}}\int_1^{\bar{E}_k/V_{r.1}}
  \sqrt{\frac{\bar{E}_k/V_{r.1}-x}{x^2-1}}~dx.
\end{equation}
Assuming $V_{r.1}  \ll \bar{E}_k$,  i.e.\ that the  quantizing torus
$k$  is far away  from the librational islands 
of  the resonance,  one can  use that
$I(\bar{\Theta})      \simeq      I_k      +      (V_{r.1}/\bar{\Omega}_k)
\cos(r\bar{\Theta})$.   If   furthermore  the  perturbative  condition
Eq.~(\ref{eq:pert_cond})   applies,  we have $\exp(i  S_k(\bar{\Theta})/\hbar)
\simeq  \exp(i k  \bar{\Theta}) (1  - i  V_{r.1} \sin(r\bar{\Theta})
/r\hbar \bar{\Omega}_k)$ and
\begin{equation} \label{eq:plane_wave_exp}
  \frac{1}{\sqrt{T_k |\dot{\bar{\Theta}}_k|}}
  \exp \left( \frac{i}{\hbar} S_{k}(\bar{\Theta}) \right)
  \simeq  \widetilde{\psi}_k(\bar{\Theta})
  - \frac{V_{r.1}}{2r\hbar\bar{\Omega}_k}(\widetilde{\psi}_{k+r}(\bar{\Theta}) - 
  \widetilde{\psi}_{k-r}(\bar{\Theta})) \; .
\end{equation}
In the  regime $\kappa_{r:s}-k  \gg r$, this  readily gives  the first
order  perturbation  Eq.~(\ref{eq:first_order})  (with only  $V_{r.1}$
non-zero).   Moreover, we  verify in  appendix~\ref{app:tunnpend} that
the  insertion  of  (\ref{eq:plane_wave_exp}) into  the  semiclassical
expression (\ref{eq:from_appendix4_2}) of  the $k'$ component recovers
the  quantum transition  amplitude (\ref{eq:from_appendix4_1})  in the
limit $|\delta_k| \ll 1$ (or,  more precisely, in the limit $|V_{r.1}|
\ll |\bar{E}_k - \bar{E}_{k'}|  \ll |\bar{E}_k - \bar{E}_{k'\pm r}|$),
where  only one  quantum state  from the  other side  is significantly
coupled   \cite{validity}.   The   semiclassical   expression  becomes
particularly  useful  when   the  condition  (\ref{eq:pert_cond})  for
quantum perturbation theory does not hold any more.

\subsection{Determination of the coupling strength}
\label{sec:spt}

The description of the local  dynamics near a $r$:$s$ resonance by the
Hamiltonian Eq.~(\ref{localapprox}) gives rise to a mechanism by which
the quasimodes located on opposite sides of the resonance are coupled.
This  will constitute  the basic  ingredient to  the  global tunneling
mechanism which we shall develop in the next subsection.  To allow for
a quantitative  prediction of the  associated transition rates,  it is
necessary,  however,  to  specify  how the  parameters  $V_{r.m}$  and
$\varphi_m$  that  enter  into  the  expression of  $H_{r:s}$  can  be
computed in practice.   The purpose of this subsection  is to show how
this can  be done  from the classical  motion near the  resonance.  We
shall  furthermore  discuss   some  qualitative  properties  of  these
quantities, in  particular the  asymptotic behavior of  the $V_{r.m}$
for large $m$.
 
The only slight technical difficulty  we shall need to address here is
due to the fact that we consider maps.  More precisely, the integrable
Hamiltonian  $\widetilde  H^{(n)}(p,q)$  Eq.~(\ref{seriesH}) has  been
introduced in  such a  way that the  map $\widetilde {\mathcal  T}$ it
generates  is  the same  as  ${\mathcal  T}$  up to  $O(\epsilon^{n})$
corrections.   In  other  words  $g^{\widetilde{H}^{(n)}}_\tau  \simeq
g^{H}_{\tau}$,  where  $g^{\mathcal  H}_t$  is  the  hamiltonian  flow
generated  by  ${\mathcal  H}=H$ or  $\widetilde{H}^{(n)}$.   However,
nothing  imposes  a priori  that  for  intermediate times  $0<t<\tau$,
$g^{\widetilde{H}^{(n)}}_t(q,p) \simeq g^{H}_t (q,p) $ up to order $n$
corrections.  As   a  consequence,  the  original   Hamiltonian  $H  =
H(I,\theta,t)$  is  not  well  approximated  by  the  time-independent
expression $\widetilde{H}^{(n)}(I)$.

Starting   from   the   action-angle  coordinate   $X=(I,\theta)$   of
$\widetilde{H}^{(n)}(p,q)$, we shall  therefore first need to define
a   periodically   time-dependent   coordinate  system   $X^b(X,t)   =
(I^b,\theta^b)$ such that in  these new coordinates, the kicked Harper
Hamiltonian is  well approximated by  $\widetilde{H}^{(n)}(I^b)$ for
all times, up to small corrections that we can then deal with by using
the    standard    secular    perturbation   theory    described    in
section~\ref{sec:secular}.   We are  thus looking  for  a periodically
time  dependent canonical  coordinates  transformation $\gamma^b_t:  X
\mapsto \gamma^b_t(X) = X^b(X,t)$ such that
\begin{eqnarray*} 
	X^b(X,\tau) & = & X^b(X,0) \, = \, X \; , \\ 
	\mbox{and} \quad 
	H^b(I^b,\theta^b,t) & = & \widetilde{H}(I^b) + O(\epsilon^{n_0}) \; . 
\end{eqnarray*} 
A way to fulfill these constraints is to define $\gamma^b_t$ as 
\begin{equation} 
    \gamma^b_t = g^{\widetilde{H}}_t\circ g^{H}_{-t} \; ,  
    \quad \mbox{for} \quad 0 \leq t<\tau \; , 
\end{equation} 
for $0  \leq t < \tau$,  and by periodicity  for the rest of  the real
time axis,  where $g^{\mathcal  H}_t$ symbolizes the  Hamiltonian flow
over  time  $t$  generated  by  the Hamiltonian  ${\mathcal  H}=H$  or
$\widetilde{H}^{(n)}$.
The following scheme 
\begin{equation}
\begin{array}{ccc} 
    X & \disp\mathop{\longrightarrow}^H &  
    g^H_{t} (X) \\ 
    \gamma^b_0\downarrow &  & \uparrow (\gamma^b_t)^{-1} \\ 
    X^b(X,0) = X & \disp\mathop{\longrightarrow}_{\widetilde{H}} & 
    g^{\widetilde{H}}_{t} (X^b(X,0)) 
\end{array}
\end{equation} 
illustrates why  the motion  under the  Hamiltonian $H$  in  the original
variables $X$ is equivalent to the one generated by $\widetilde H^{(n)}$ in
the  $X_b$ variables,  for   $t  \in  [0,\tau[$.   

The transformation $\gamma^b_t$ is, by explicit construction, periodic
in time.   However, it is  in general {\em  not} continuous at $t  = n
\tau$, as  a consequence  of the fact  that $\widetilde  {\mathcal T}$
does   not  perfectly  approximate   ${\mathcal  T}$.    The  complete
definition  of  the  new   Hamiltonian  $H^b$  requires  therefore  to
introduce  a perturbation  term $\epsilon'  V^b(I^b,\theta^b,t)$ which
becomes active only  at $t = n \tau$ and  which accomplishes the final
``jump'' from $X^b(X,t\to\tau)$ to $X^b(X,\tau)$.
One
therefore has
\begin{equation} 
   H^b(I^b,\theta^b,t) = \widetilde{H}(I^b) +  
  \epsilon' V^b(I^b,\theta^b,t), 
\end{equation} 
with 
\begin{equation} 
  \epsilon' V^b(I^b,\theta^b,t) = \sum_{n=-\infty}^{\infty} 
  \delta^{(-)}(t-n\tau)R(I^b,\theta^b,t). 
\end{equation} 
$\epsilon'$   is  the   strength  of   the  perturbation   induced  by
$V^b(I^b,\theta^b,t)$   and  corresponds  to   the  accuracy   of  the
integrable approximation of $H$. From  a strictly formal point of view
this  strength is 
of order ${\mathcal  O}(\epsilon^n)$.  
This scaling, however, applies only to contributions that are analytic in $\epsilon$
(e.g., a global deformation of the K.A.M.\ tori) and does not take into
account non-analytical contributions (e.g., of  the form  $C_1  \exp{ \left(
-C_2/\epsilon  \right) }$) which result from the vicinity of nonlinear resonances.
$\delta^{(-)}$  is a  Dirac distribution
that, for consistency, we need  to consider as being smeared on the interval
$[\tau^-,\tau]$,  with   $\tau^-  \to\tau$.  (In   practice,  we  take
$\delta^{(-)}(t) =  1/(\tau-\tau^-)$ if $t  \in [\tau^{(-)},\tau]$, an
zero elsewhere.)  $R(I^b,\theta^b,t)$ is  a time-periodic function with period $\tau$.

A  natural  interpretation  of  what  $R(I^b,\theta^b,t)$  is  can  be
obtained by integrating Hamilton's equations of motion associated with
$H^b(X^b,t)$ from $\tau^-$ to $\tau$. This yields
\begin{equation}  
  g^H_{\tau}(X,\tau) - g^{\widetilde{H}}_{\tau}(X,\tau) =   
  \frac{1}{\tau-\tau^-}\int_{\tau^-}^{\tau} {\mathcal I}  
  \nabla R(X^b(t),t)~dt 
\label{propagVp}  
\end{equation}  
where ${\mathcal I} = \; \scriptstyle \left({ 0 \; \; 1 \atop -1 \; 0}
\right)$  and $X^b(t)$  is the  path that relates $g^H_{\tau}(X,\tau)$ to
$g^{\widetilde{H}}_{\tau}(X,\tau)$.  Notice that these equations would
be inconsistent  without a  time dependence for  $R$. However,  as the
path  $X^b(t)$  is  of  typical  size  $\epsilon'$,  one  can  rewrite
perturbatively (\ref{propagVp}) as
\begin{eqnarray}
  I^{H}_{\tau}(X) - I^{\widetilde{H}}_{\tau}(X) & = &  
  -\disp\frac{\partial \langle R\rangle}{\partial \theta} 
  (I^{H}_{\tau}(X),\theta^{\widetilde{H}}_{\tau}(X))  
  + {\mathcal O}((\epsilon')^2) \; ,\\ 
  \theta^{H}_{\tau}(X) - \theta^{\widetilde{H}}_{\tau}(X) & = &  
  \disp\frac{\partial \langle R\rangle}{\partial I} 
  (I^{H}_{\tau}(X),\theta^{\widetilde{H}}_{\tau}(X)) 
  + {\mathcal O}((\epsilon')^2) \; ,
\end{eqnarray} 
where $\langle  R\rangle$ denotes the 
time average  of $R$ between  $\tau^-$ and  $\tau$.
We recognize  that $\langle  R\rangle$ is, in first  order in  $\epsilon'$,  the
{\em generating function} of the canonical transformation
\begin{equation} 
  g^{\widetilde{H}}_{\tau}(X,\tau) \mapsto  g^H_{\tau}(X,\tau)
\end{equation} 
-- that  is, of  the difference  between  the map  and the  motion of  its
integrable approximation during a time $\tau$.

Within  the  $X^b$  variables,  we  can now apply the standard  secular
perturbation  theory described in Sec.~\ref{sec:secular}.  We
obtain in this way
\begin{equation}  
    H_{r:s}(\bar{I},\bar{\Theta},t) = {H}_0^{(r:s)}(\bar{I})  
    + \bar{V}^{(r:s)} (\bar{I},\bar{\Theta})  
    + {\mathcal O}(\epsilon'^2). 
\end{equation}  
with  
\begin{equation}  
    \bar{V}^{(r:s)} (\bar{I},\bar{\Theta}) =  
  \frac{1}{r\tau}\sum_{n=0}^{r-1}\langle R\rangle 
  (\bar{I}, \bar{\Theta} + \Omega_{r:s}n\tau) \; .
\end{equation}  
The Fourier coefficients of the averaged perturbing potential
\begin{equation}  
   \bar{V}^{(r:s)}(\bar{I},\bar{\Theta}) =  
    \sum_{m=0}^{\infty} V_{r.m} \cos(rm\bar{\Theta} + \varphi_m) 
\end{equation}  
can, with  $\bar{\Theta} \simeq \Theta$ and  $\bar{I} \simeq I \simeq I_{r:s}$,
then be written as
\begin{equation}  
    V_{r.m} e^{i\varphi_m}  = \frac{1}{\pi \tau}\int_0^{2\pi}\exp(-irm\theta) 
    \langle R\rangle (I_{r:s},\theta ) d\theta \; .
\end{equation}  
This transforms 
after integration by parts into  
\begin{equation}  
    V_{r.m} e^{i\varphi_m}   =  
  \frac{1}{i\pi mr\tau}\int_0^{2\pi}\exp(-irm\theta) 
     \delta I_{r:s}(\theta)  d\theta \; .
\label{couplTPS} 
\end{equation}  
Here, $\delta I_{r:s}(\theta)$ is defined by
\begin{equation}
  \delta I_{r:s}(\theta) = I^{(-1)}(I_{r:s},\theta) - I_{r:s}
\end{equation}  
where $I^{(-1)}(I,\theta)$ symbolizes the action variable that is obtained by
applying the
inverse Poincar{\'e} map ${\mathcal T}^{-1}$ to $(I,\theta)$ (or alternatively, the
backward propagation with $H$ from time $t=\tau$ to $t=0$).
Eq.~(\ref{couplTPS})  therefore provides a convenient way
to obtain the  numerical value of the coefficients  $V_{r.m}$, which is based
only on the propagation of classical trajectories.

The  effect of  averaging out  the time  dependence on  the integrable
contributions of  $V$ leads to the  $\theta$--independent coefficient
$V_0$ which is  of order $\epsilon^{n}$. On the  other hand, the other
coefficients $V_{r.m}$ with $m>0$ correspond  to the non-integrable effect of
the resonances, and  therefore their magnitude is  not simply proportional
to  $\epsilon^{n}$   (we should actually expect them to be
essentially independent of  $n$, in some range near  the optimal value
$n_0$).  As the $V_{r.m}$ result from the Fourier integrals of $\delta
I_{r:s}(\theta)$,  their scaling  with $m$  can be  inferred  from the
analytical  structure of  $\delta I_{r:s}(\theta)$.   Assuming $\delta
I_{r:s}(\theta)$ to be  an analytic function in $\theta$,  the line of
integration in Eq.~(\ref{couplTPS}) can be displaced into the negative
imaginary  direction  of  $\theta$  (for  $m>0$),  where  it  gives  a
vanishing contribution due to the exponentially small prefactor.  As a
consequence,  the Fourier  integral  Eq.~(\ref{couplTPS}) is  entirely
described by the {\em singularities} $\theta^\xi = \Omega_{r:s} t^\xi$
of $\delta I_{r:s}(\theta)$ in  the complex $\theta$ domain, and will,
for large  $m$, be  dominated by the  contribution of  the singularity
that  is  closest  to  the   real  axis  (see  in  this  context  also
\cite{bogomolny84}).


The calculation of $\delta  I_{r:s}(\theta)$ involves in practice three
steps. The first one is to determine the coordinate  $(p,q)$ of the point
$(I_{r:s},\theta)$ under  consideration.  The  second one is to  apply the
map  ${\mathcal T}^{-1}$  to  $(p,q)$,  and  the last one is to  determine
the  action coordinate $I$ of the resulting point.  
In general, these two latter steps 
should not involve any singularity:  the map ${\mathcal T}$, the function
$\tilde H^{(n)} (p,q)$, as  well as the function $I(\tilde  H)$ will usually
be  analytical.   As  a  consequence,  the  singularities  of  $\delta
I_{r:s}(\theta)$  should  be  the  one  of  the torus  $\Gamma_{r:s}$,   that   is
the   complex   angles   $\theta$  such   that 
$(p,q) = (p(I_{r:s},\theta),q(I_{r:s},\theta))$ lies at infinity.  This corresponds to
trajectories 
which, starting from  $(I_{r:s},\theta \! = \! 0)$  on the real torus,
go  to infinity  in a  finite complex  time under  the  dynamics of
$\tilde H^{(n)}$.

One can therefore write in the asymptotic regime
\begin{equation}  
  V_{r.m} \varpropto 
   (rm)^{\gamma} V_{\xi_0} e^{-(\Omega_{r:s} t_{r:s}^{\xi_0}) rm}, m>0,
  \label{varVnsecular} 
\end{equation}   
with $t_{r:s}^{\xi_0}$  the imaginary  part of the  time to  reach the
closest singularity,  and where $V_{\xi_0}$  and $\gamma$ characterize
the  behavior  of $\langle  R  \rangle$  near  the singularity.   (If
$\langle R \rangle$ was a  meromorphic function, $\gamma$ would be the
degree  of the pole,  and $V_{\xi_0}$  the corresponding  residue.)  We
would like to  stress here is that there are  two sources of smallness
in this expression.   One is the exponential dependence  in $m$, which
is entirely controlled by  the dynamics of the integrable approximation
($t^{\xi_0}$ is determined by $\tilde H^{(n)}$).  In the semiclassical
limit, this  will give rise  to an exponential dependence  in $\hbar$,
since  one  should use  $m  = \Delta  I/r\hbar$  to  connect two  tori
differing   by   an  action   $\Delta   I$.    The  other   parameters
characterizing  the  asymptotic  behavior  of the  $V_{r.m}$,  namely
$\gamma$  and $V_{\xi_0}$,  depend  on the  complete  dynamics of  the
perturbed system and contain  in particular the perturbation parameter
$\epsilon$.   To  have a  well  defined  {\em classical}  perturbation
expansion, and in particular  for the first order secular perturbation
approximation we have used to be valid, the corresponding $(rm)^\gamma
V_{\xi_0}$ terms should be small {\em on the classical scale},
although not  exponentially. We  shall always assume  the perturbation
parameter $\epsilon$ to be small enough for this property to hold.

In addition,  The general scaling  behavior Eq.\ (\ref{varVnsecular})
has  consequences  for the  {\em  quantum}  perturbative expansion  to
evaluate  the  overlap  $  \langle  \widetilde{\psi}_{k+rm}  |\psi_{k}
\rangle$, and  determines up to  which order this expansion  should be
done.
To  illustrate
this, let us consider for a particular $r$:$s$ resonance the coupling
between two states that are symmetrically located on opposite sides 
with respect to the
resonance (i.e.\ such that Eq.~(\ref{eq:deg_cond}) holds).
The second order correction to Eq.~(\ref{eq:first_order}) reads
\begin{equation}
  \langle \widetilde{\psi}_{k+mr} 
  | \psi_k\rangle = \frac{V_{r.m}}{2(\bar{E}_k - \bar{E}_{k+mr})}
  +  \sum_{m_1 + m_2 = m} 
    \frac{V_{r.m_1}V_{r.m_2}}{4(\bar{E}_k - \bar{E}_{k+m_1r})
      (\bar{E}_k - \bar{E}_{k+mr})}
    \; .
\label{pert_two}
\end{equation}
If Eq.~(\ref{varVnsecular}) applies, we see that the condition for the
second order  term to  be smaller  than the first  order one  does not
involve the exponential, but that for  each $m_1$ in the sum $(m_1 m_2
r/m)^\gamma V_{\xi_0} /(\bar{E}_k  - \bar{E}_{k+m_1r}) \ll 1$.  For  a given value
of the perturbation parameter $\epsilon$  and at fixed $\hbar$, such a
condition may  very well be fulfilled.  However,  in the semiclassical
limit  $\hbar \to 0$  with {\em  fixed} $m$ -- and in practice $m$ will always be
more or less fixed when the system undergoes the transition over a particular
$r$:$s$ resonance (as will be discussed in the following subsection) --
  the denominator  $(\bar{E}_k -
\bar{E}_{k+m_1r})$  goes to  zero
(being  bounded by  $E_k  - E_{r:s}  \simeq \hbar^2  (m^2  r^2/8\mu))$.  
Therefore,  as  one  goes  deeper in  the
semiclassical regime,  the second order term  will eventually dominate
over the first one.

In the  same way, one can see  that assuming Eq.~(\ref{varVnsecular}),
the      condition      for      the     $m^{th}$      order      term
Eq.~(\ref{eq:from_appendix4_1}) to be larger  than the first order one
is that
\begin{equation}  
  \frac{\mu e^2 r^\gamma V_{\xi_0}}{\hbar^2 r^2 m^2} \gg 1 \; .
  \label{valordre1_raf} 
\end{equation}  
In  the semiclassical  limit  (with fixed  $m$),  this condition  will
eventually be reached at one point.

As a consequence, we see that, assuming $V_{\xi_0}$ to be small on
a classical scale, a first order quantum perturbative treatment will
be  valid for  moderately small  values  of $\hbar$,  but higher  order
should be taken into account as $\hbar \to 0$.  
Note that this is not
incompatible with  the fact that the  quantum perturbation development
is   convergent,   since   the condition Eq.~(\ref{eq:pert_cond}) for its
validity 
involves   the  exponential  term  $\exp(-r \Omega_{r:s} t_\xi)$ which can  be
extremely small, especially for high-order resonances with $r \gg 1$.  
Very far  in the semiclassical regime (or for small $r$), 
quantum perturbation theory might nevertheless fail at some point, in
which case  the it would  become necessary to resort  to semiclassical
expressions such as Eq.~(\ref{eq:from_appendix4_2}).

We  see that  considering  the analytical  structure  of the  function
$\delta  I_{r:s}(\theta)$, is  important to  decide what  term  in the
perturbation expansion will  be the dominating one, as  well as, as we
will see in the next section,  what is the dominating mechanism in the
tunneling `process.  This should be  reconciled with the fact that the
analytical structure  of the invariant tori may  sensitively depend on
the  precise choice  of $\tilde  H^{(n)}$,  and in  particular on  the
degree $n$  of the  integrable approximation.  We  shall come  back on
this issue in section~\ref{sec:kh}.

\subsection{Mechanism of resonance assisted tunneling}

\label{sec:mechanism}

In the  previous sections, we have examined in detail the characteristics of
couplings that are locally induced by the presence of  a nonlinear resonance.
We shall now see how these couplings  can be combined at a larger scale to
form a global mechanism of  tunneling for quasi-integrable  systems.
Furthermore, we analyze  why,  and  under  which  condition,  this  mechanism
is  the dominating one.

As we have seen in Sec.~\ref{sec:secular}, the dynamics near a $r$:$s$
resonance is locally described by a Hamiltonian of the form
\begin{equation}
   H^{(r:s)} = \frac{(I-I_{r:s})^2}{2\mu} + \sum_{m=1}^{\infty} 
  V^{r:s}_{r.m} \cos{\left(mr\theta+\varphi_m\right)} \; ,
\label{localapprox2}
\end{equation}
where  the  parameters $V^{r:s}_{r.m}$  and  $\varphi_m$  can be  computed with
Eq.~(\ref{couplTPS})
through    the   propagation    of    classical   trajectories.
Furthermore,  when   discussing  the  order  of
magnitude of  the various terms,  we shall assume that  the asymptotic
expression derived in the last section can be used, and thus that
\begin{equation}
  V^{r:s}_{r.m} \thicksim (mr)^{\gamma}
  V^{r:s}_{\xi_0} \exp(-\Omega_{r:s}t_{r:s}^{\xi_0} mr) \; ,
\label{varV}
\end{equation}
where  $\Omega_{r:s}=s\omega/r $  is  the angular  frequency of  the
integrable torus at the $r$:$s$ resonance, $t_{r:s}^{\xi_0} > 0$ is the
imaginary part of  the  classical  time  to  reach the  closest  singularity
of  the analytic  continuation  of $\Gamma_{r:s}$ into  complex
phase   space,  and  $\gamma$ and $V^{r:s}_{\xi_0}$ characterize the
generating function $\langle R \rangle$ near the singularity.

Let  us,   to  start   with,  consider  the   unperturbed  Hamiltonian
$\hat{H}(\epsilon=0)$,  and  one of  its  quasi-modes  $\psi_k^{(qs)}$  built on  an
invariant  torus  $\Gamma_k$.   As  discussed  in  Sec.~\ref{sec:int},  the
symmetry $P$ of our system is assumed
such that $\Gamma_k$ exhibits a symmetric, but distinct equivalent $\Gamma'_k$ 
on which one can  build another quasi-mode $\psi_k^{'(qs)}$ analogous
to  $\psi_k^{(qs)}$.  $\Gamma_k$ and  $\Gamma'_k$  are separated  in
phase  space by a  separatrix ${\mathcal S}$.  The true  eigenstates of the
evolution operator $\hat U$ correspond  to the  symmetric  and
anti-symmetric linear combination  of $\psi_k^{(qs)}$ and $\psi_k^{'(qs)}$,  the
eigenphases of which differ  by the  splitting $\delta  \phi_k$.  
The  semiclassical expression  of the  splitting is
given   by   $2\alpha\tau/\hbar$,  where   $\alpha$   is  defined   by
Eq.~(\ref{semicsplit}).

Now we increase  $\epsilon$ and follow the adiabatic  evolution of the
eigenmodes  of   $\hat U$,  which   can  be   considered  as
perturbations   of  the   quasi-modes   $\widetilde{\psi}_k^{(qs)}$  associated
with   the integrable approximation  $\widetilde{H}(\epsilon)$.  
At some  point, a resonance  $r$:$s$  grows significantly  as  compared  to
$\hbar$  and couples  $\widetilde{\psi}_k^{(qs)}$ to some
$\widetilde{\psi}_{k+rm}^{(qs)}$, with $m>0$.    
Since the torus $\Gamma_{k+rm}$  is located  closer to the separatrix than $\Gamma_k$, 
$\widetilde{\psi}_{k+rm}^{(qs)}$  exhibits a slower  exponential decrease in the
forbidden domain than $\widetilde{\psi}_k^{(qs)}$.
As a consequence, if the strength of the coupling between
$\widetilde{\psi}_k^{(qs)}$ and $\widetilde{\psi}_{k+rm}^{(qs)}$ is not too small,
the admixture of this latter component will eventually dominate the behavior
of the perturbed quasi-mode in phase space regions close to the separatrix,
and thereby determines the eigenphase splitting $\delta  \phi_k$ between the symmetric
and the antisymmetric linear combinations of the quasi-modes.
One obtains in this way a splitting
\begin{equation}
  \delta \phi_k \simeq 
  \left| {\mathcal A}^{r:s}_{k,m} \right|^2
  \frac{2\tau}{T_{k+rm}}\exp{\left(\disp -\frac{\sigma_{k+rm}}{\hbar}\right)}
  \; , \label{locresonsplit}
\end{equation}
where $T_{k+rm}$ is the classical period of the torus $\Gamma_{k+rm}$,
$\sigma_{k+rm}$ is the imaginary  part of  the classical  action along  a  complex
trajectory relating  $\Gamma_{k+rm}$  to  its symmetric counterpart, and 
${\mathcal  A}^{r:s}_{k,m}$ represents the coupling amplitude (\ref{pertmode})
associated with the resonance.
For sake  of
clarity,   we  shall   consider  below   the  case   where  ${\mathcal
A}^{r:s}_{k,m}$  can be  approximated  by the  first order  expression
${V^{r:s}_{r.m}}/({\bar{E}^{r:s}_k - \bar{E}^{r:s}_{k+rm}})$ with $\bar{E}^{r:s}_k = E_k -
\Omega_{r:s}I_k$.
Our argumentation, however, does
not rely on this precise form.

To  compare  the  relative  effectiveness  of  the  above  ``resonance
assisted''  mechanism  with respect to the  direct  (integrable-like)  one, we
use Eq.~(\ref{varV}) and obtain that
\begin{equation}
  \delta \phi_k \varpropto (rm)^\gamma
  \Big| \frac{V^{r:s}_{\xi_0}}{\bar{E}^{r:s}_k - \bar{E}^{r:s}_{k+rm}}\Big|^2 
  \exp{\left(\disp - \frac{2\Omega_{r:s}t_{r:s}^{\xi_0}(I_{k+rm} - I_k) 
  + \sigma_{k+rm}}{\hbar}\right)} \; .
\label{behavlocresonsplit}
\end{equation}
If, for a moment, we just compare the exponential factors of the above
expression  with  the  one,  $\exp{(-\sigma_k/\hbar)}$,  of  the  direct
tunneling  mechanism, we  see  that the  condition  for the
resonance-assisted one to be dominant would be that
\begin{equation}
  2\Omega_{r:s}t_{r:s}^{\xi_0}(I_{k+rm} - I_k) \leq
  \sigma_{k}- \sigma_{k+rm} \; .
  \label{eq:ram_cond1}
\end{equation}
Now, in the semiclassical regime, one can assume $I_k$, $I_{k+rm}$ and
$I_{r:s}$ classically  close, and thus $\Omega_{r:s}  (I_{k+rm} - I_k)
\simeq  \Delta  E  \equiv  E_{k+rm}   -  E_k$.   In  the   same  way,
$\sigma_{k+rm} -  \sigma_{k} \simeq \Delta E \cdot  (\partial \sigma / \partial  E)_{r:s}$, 
with  $(\partial \sigma  /  \partial E)_{r:s}  = t^\sigma_{r:s}$  the
imaginary  part of  the time  needed to  follow the complex path $\cal  P$ 
from the resonant torus $\Gamma_{r:s}$ to its symmetric counterpart $\Gamma_{r:s}'$, on
which the  action $\sigma$ is computed.
As a consequence Eq.~(\ref{eq:ram_cond1}) reads
\begin{equation}
  t_{r:s}^{\xi_0} \leq  \frac{t^\sigma_{r:s}}{2}\; ,
  \label{eq:ram_cond2}
\end{equation}
or in other words, that the  imaginary part of the time needed to reach
the closest singularity should be smaller than half the imaginary part
of the  time required to go  from one torus to  its symmetric partner.
This condition is necessarily fulfilled, as can be seen from propagating
$\cal   P$   under   $\tilde   H^{(n)}$   for   real   time   $t   \in
[0,t_{r:s}^{\xi_0}]$.   ($g_t^{\tilde H^{(n)}}  {\cal  P}$ necessarily
encounters  at  least  one  singularity  of $\Gamma_{r:s}$,  and  by
symmetry,   one    of   these   singularities    necessarily   fulfill
Eq.~(\ref{eq:ram_cond2}).)  As  a consequence, the  resonance-assisted
mechanism  will {\em always}  dominate   the  ``regular''  tunneling  
process (Sec.~\ref{sec:int}) in the semiclassical limit.

Considering   now   the   prefactor,   the   energy   denominator   in
Eq.~(\ref{locresonsplit})   will   make   it  favorable   to   connect
$\widetilde{\psi}_k^{(qs)}$          to          a          quasi-mode
$\widetilde{\psi}_{k+mr}^{(qs)}$ such that $\bar{E}^{r:s}_k$ is almost
(i.e.  up  to a difference  of order $\Omega_{r:s}  \hbar$) degenerate
with $\bar{E}^{r:s}_{k+rm}$, which  implies that $\Gamma_{r:s}$ should
lie at  mid distance between the tori  $\Gamma_k$ and $\Gamma_{k+mr}$.
Note that this  is the case not only if  the first order approximation
of     ${\mathcal      A}^{r:s}_{k,m}$     is     used,      as     in
Eq.~(\ref{behavlocresonsplit}), but also if  higher order terms of the
perturbation  are   included,  or  if   the  semiclassical  expression
Eq.~(\ref{eq:semiclassical_A}) is used.

For small $\hbar$, $\Gamma_{k+mr}$ will  then a priori not be close to
the  separatrix.   However,  nothing   prevents  from  making  use  of
couplings via other $r'$:$s'$ resonances in order
to  gradually approach the vicinity of the separatrix ${\cal S}$.
In this way, $\widetilde{\psi}_k^{(qs)}$ can eventually be connected to a
quasi-mode $\widetilde{\psi}_{k'}^{(qs)}$ the action $I_{k'}$ of which is only a
few $\hbar$ smaller than the action $I_{\cal S}$ of the separatrix and from
where ``regular'' tunneling takes place with a rather large rate.

Using successively  the resonances $r_1$:$s_1$,  $r_2$:$s_2$, $\ldots$
$r_N$:$s_N$, which we assume to appear in ascending order (i.e., 
$I_{r_1:s_1} < I_{r_2:s_2} < \ldots < I_{r_N:s_N}$),
the resulting expression for the splitting is then
\begin{equation}
  \delta \phi_k \simeq 
  \left| {\mathcal  A}^{r_1:s_1}_{k,m_1} \right|^2 
  \ldots
  \left| {\mathcal  A}^{r_N:s_N}_{k_N,m_N} \right|^2 
  \frac{2\tau}{T_{k'}}
  \exp{\left(\disp -\frac{1}{\hbar}\sigma_{k'}\right)} \; .
\label{globresonsplit}
\end{equation}
Here, $k \equiv  k_1$, $k_2 = k_1 +  m_1 r_1$, $k_3 = k_2  + m_2 r_2$,
\ldots  $k'  = k_N  +  m_N  r_N$ denote  the  quantum  numbers of  the
intermediate quasi-modes that are involved in the coupling scheme.
The   $m_i$   are   always   chosen   such   that   the   denominators
$\bar{E}^{r_i:s_i}_{k_i}   -   \bar{E}^{r_i:s_i}_{k_i+r_i  m_i}$   are
quasi-degenerate,  which  means  that  the tori  $  \Gamma_{k_i}$  and
$\Gamma_{k_i+r_im_i}$  should  be  almost  symmetric with  respect  to
$\Gamma_{r_i:s_i}$.

In   the   particular  case   where   the  semiclassical   expression
Eq.~(\ref{eq:from_appendix4_2})   can  be   used   for  the   amplitudes
${\mathcal A}^{r:s}_{k,m}$, one obtains the expression
\begin{eqnarray}
  \delta \phi_k & \simeq &
  \left |\frac{\sin(\pi \delta e_{kk_1})}{\pi \delta e_{kk_1}}
  \frac{\exp(-\bar \sigma_{k}^{(r_1:s_1)}/\hbar)}
  {2\sin(s_k^{(r_1:s_1)}/\hbar)} \right|^2 
  \ldots
  \left| \frac{\sin(\pi \delta e_{k_{N-1}k_N})}{\pi \delta e_{k_{N-1}k_N}}
  \frac{\exp(-\bar \sigma_{k_N}^{(r_N:s_N)}/\hbar)} 
  {2\sin(s_{k_N}^{(r_N:s_N)}/\hbar)}  \right|^2 \nonumber \\
   && \qquad \times \qquad\frac{2\tau}{T_{k'}} 
  \exp{\left(\disp -\frac{1}{\hbar}\sigma_{k'}\right)},
\label{globresonsplitsemic}
\end{eqnarray}
with  $e_{k_{i-1}k_i}$ defined by Eq.~(\ref{eq:ekk}) and
\begin{equation}
s_{k_i}^{(r_i:s_i)} = 2 \pi \frac{I_{r_i:s_i} - I_{k_i}}{r_i} = 2\pi\hbar \,
\frac{\kappa_{r_i:s_i} - k_I}{r_i}  
\label{eta2}
\end{equation}
(see Eq.~(\ref{eta}))  the phase space  area that is  enclosed between
$\Gamma_{k_i}$ and  $\Gamma_{r_i:s_i}$ within the angle  range $0 \leq
\theta  \leq  2  \pi  / r_i$.   Here  $\bar  \sigma_{k_i}^{(r_i:s_i)}$
represents the imaginary action  along the complex path that connects,
within the  effective pendulum Hamiltonian  Eq.\ (\ref{localapprox2}),
the perturbed torus $\Gamma_k$ with  its counterpart on the other side
of the $r_i$:$s_i$ resonance.
The overall picture is illustrated in Fig.~\ref{fig:saute_moutons}.
Within the  corresponding  secular perturbation  approximation,
each resonance provides a complex path allowing to join a torus on one
side of the  resonance to its symmetric counterpart on the  other side. The
global ``resonance  assisted  tunneling''  mechanism  we propose  consists  in
following this succession of complex paths across a series of resonances
to reach the neighborhood of the separatrix. From there, ``regular'' tunneling
occurs into the symmetric island.


In the above scenario, it remains to decide what is the most effective
chain of  resonances to be used  in the tunneling process.   For a given
value of $\hbar$, one constraint is naturally that a $r$:$s$ resonance
needs to be  taken into account only if the area $2 \pi I_{\cal S}$ enclosed by the
separatrix is  larger than $2 \pi r  \hbar$.  For relatively  large $\hbar$,
this can leave only a  moderate amount of possibilities.  Deep in
the  semiclassical regime however,  a large  number of  resonance will
fulfill this condition, and it is useful to design a guiding principle
on which one to use.

For the sake  of clarity, we shall address this  question again under the
hypothesis      that      the      first      order      approximation
Eq.~(\ref{eq:first_order}) can  be used for  the transition amplitudes
${\mathcal A}^{r:s}_{k,m}$. With Eq.~(\ref{varV}), we can therefore
write ${\mathcal A}^{r:s}_{k,m} \simeq {\mathcal B}^{r:s}_{k,m}
\exp(-\Omega_{r:s} t^{\xi_0}_{r:s} rm)$, with 
\begin{equation}
  {\mathcal B}^{r:s}_{k,m} = (rm)^\gamma 
  \frac{V^{r:s}_{\xi_0}}{\bar{E}^{r:s}_{k} - \bar{E}^{r:s}_{k+rm}} \; .
\end{equation}
With this notation, we have, for a given choice of resonances
$r_1$:$s_1$, $\cdots$, $r_N$:$s_N$,
\begin{equation}
  \delta \phi_k \simeq 
  \left| {\mathcal  B}^{r_1:s_1}_{k,m_1} \right|^2 
  \ldots
  \left| {\mathcal  B}^{r_N:s_N}_{k_N,m_N} \right|^2 
  \frac{2\tau}{T_{k'}}
  \exp{\left(\disp -\frac{1}{\hbar} (J_{kk'} + \sigma_{k'})\right)},
\label{eq:withB}
\end{equation}
with, noting $r_im_i = \Delta I^{(r_i:s_i)}/\hbar$,
\begin{eqnarray}
  J_{kk'} & = & \frac{1}{\hbar} \sum_{i=1}^N \Omega_{r_i:s_i} 
  t^{\xi_0}_{r_i:s_i} \Delta I^{(r_i:s_i)} \\
  & \simeq & \frac{1}{\hbar}  \int_{I_k}^{I_k'} 
    \Omega(I) t^{\xi_0}(I) dI \; .
\end{eqnarray}
The exponential term $\exp{\left( -(J_{kk'} + \sigma_{k'})/\hbar
\right)}$ will therefore not depend too much on the precise choice of
the sequence of resonances used to connect $I_k$ and $I_k'$.  On the other
hand, using a large amount of resonances between $I_k$ and $I_k'$,
each of them inducing only a small change in the action, will have a
tendency to increase the number of ${\mathcal B}^{r:s}_{k,m}$ terms in
the prefactor of $\delta \phi_k$.  Hence, the choice of the dominant path
will in the end depend on the magnitude of these coefficients.

The    energy   denominator    of   ${\mathcal   B}^{r:s}_{k,m}$   is in
general of the order of
$\Delta E_{r:s}  = r \hbar \, \Delta  I^{(r:s)} / 2\mu$.  
Therefore, the condition that it is favorable to introduce
a $r$:$s$ resonance into the coupling path reads 
\begin{equation}
  {\mathcal B}^{r:s}_{k,m} \simeq 
  \frac{(r m)^{\gamma-1} 2\mu V^{r:s}_{\xi_0}}{\hbar^2 r}
  \gg 1
  \label{eq:Bcond}
\end{equation}
Even though the  parameter $V_{\xi_0}$  is  usually small when  the
system is  close to integrability,  this condition will  eventually be
met if one goes high enough  into the semiclassical regime.  This can be
interpreted as an  upper bound for $\hbar$ above  which it is impossible to
``resolve''  the $r$:$s$ resonance.   We  stress  though that  this
criterion is  {\em not} directly related to  the size of the  islands of the
resonance.  Indeed, this  latter quantity is proportional to 
$\sqrt{V^{r:s}_{r.1}}$ and involves therefore   a   factor    $\exp(-   \Omega_{r:s}
t_{r:s}^{\xi_0}  r)$  which can  be extremely  small for  large  $r$.
$V_{\xi_0}^{r:s}$ should,  on the other  hand, smoothly depend on $\Omega / \omega = s/r$
(except  for symmetry considerations, see Sec.~\ref{sec:kh})
and might not be very different between, say, $\Omega / \omega = 1/2$ and $\Omega / \omega  =
5/11$.  Note finally that using higher order terms in the perturbative
calculation of ${\mathcal A}^{r:s}_{k,m}$ can only make the condition
Eq.~(\ref{eq:Bcond}) valid sooner in the semiclassical regime.

In any case, the above consideration implies  that as long as
Eq.~(\ref{eq:Bcond})  applies for  all resonances that are used in the
coupling sequence, and as long as the final torus is kept as close as possible
to the  separatrix, a  large number of  small steps will generally be favored
with respect to  a small number of large ones,  since each extra step
imply  the  multiplication  by  a large  prefactor (which, roughly speaking,
comes from a small energy denominator).
This  justifies a  posteriori the use of  a local description  of the dynamics,
since  the dominant  tunneling mechanism takes into  account the effect  of a
resonance only in  its close neighborhood.


\section{Resonance assisted tunneling in the kicked Harper model}
\label{sec:kh}

After the  general discussion in  the previous sections, we  shall now
consider in more  detail a particular system in  the nearly integrable
regime, namely the  kicked Harper model.  The purpose  of this section
will be to  check, for this particular case, the  accuracy of the final
semiclassical expression Eq.~(\ref{globresonsplit}) -- which gives the
splitting between  the quasi-energy of  a symmetric Floquet  mode and
its antisymmetric counterpart  -- as well as to verify the degree  of validity
of the various hypotheses that were made along the way of its derivation.

The  classical  dynamics of  the  kicked  Harper  is governed  by  the
hamiltonian   Eq.~(\ref{eq:kh_H}),  yielding   the   stroboscopic  map
Eq.~(\ref{eq:kh_map}).  In the limit $\tau \to 0$, this dynamics is equivalent to
the  one  generated  by   the  time-independent  (integrable)  Harper
hamiltonian
\begin{equation} 
    H_0(p,q) = \cos{p} + \cos{q} \; .
\label{eq:harper_ham}
\end{equation}
In Eq.~(\ref{eq:kh_H}), $\tau$ is thus both  the period of the kick and the
perturbation  parameter (i.e.\  $\epsilon \equiv  \tau$).

Quantum mechanically, the map Eq.~(\ref{eq:kh_map}) can be associated
with the evolution operator
\begin{equation}
    \widehat{U} = \exp\left(\frac{i\tau\cos\widehat{p}}{\hbar}\right)
    \exp\left(\frac{i\tau\cos\widehat{q}}{\hbar}\right).
\end{equation}
The $2\pi$ periodicity in $\hat q$ and $\hat p$ makes the quantum treatment of
the kicked Harper particularly easy if
\begin{equation}
\hbar = \frac{2\pi}{N}
\end{equation}
with integer $N > 0$.
For these particular values of $\hbar$, the eigenfunctions $\psi$ of $\widehat{U}$
can be written as Bloch functions in both position and momentum -- i.e.,
\begin{eqnarray*}
  \langle q+2\pi | \psi \rangle 
  & = &\exp(i\varphi_q) \langle q |
  \psi\rangle \; ,\\
  \langle p+2\pi | \psi \rangle & = &\exp(i\varphi_p) \langle p |
  \psi\rangle
\end{eqnarray*}
for some pair of Bloch phases $0 \leq \varphi_q, \varphi_p < 2 \pi$, where $| q \rangle$ and 
$| p \rangle$ denote the eigenfunctions of the position and momentum operator,
respectively.
For each pair $(\varphi_q,\varphi_p)$ of Bloch phases, the corresponding subspace of the
Hilbert space  is {\em finite} dimensional and contains $N$ linearly
independent wave functions, spanned, e.g., by the basis states 
\begin{equation}
|n\rangle = \sum_{l=-\infty}^{\infty}\exp(i(l+n/N)\varphi_q) \ |q \equiv ( 2 \pi n + \varphi_p ) / N + 2 \pi l \rangle
\end{equation}
for $0 \leq n < N$ \cite{Leboeuf90}.
The eigenvectors $|\psi_k\rangle$ of $\widehat{U}$ and their eigenphases $\phi_k$ can
therefore be computed up to numerical (quadruple) precision, by diagonalizing
the $N$$\times$$N$ matrix $\langle n | \widehat{U} | n' \rangle$.
In the following, we  shall consider only the two 
pairs $(0,0)$ and  $(0,\pi)$ of Bloch  phases, corresponding  to periodic
boundary conditions in momentum, and periodic or anti-periodic
boundary conditions in position.  This choice is equivalent
to restricting $p$  to the  interval $[-\pi,\pi]$  and $q$  to  the interval
$[-\pi,3\pi]$ with periodic boundary  conditions (see Fig.~\ref{fig:khmap}), and
to consider the even  and  odd symmetry  classes  with  respect  to the
inversion  $q \to -q$.

The  calculation of the  integrable approximation  $\widetilde{H}$ for
the  kicked  Harper is  performed  straightforwardly  by applying  the
formalism  of Appendix  \ref{app:lie}.   One obtains  for instance  as
zeroth      order     coefficient      the      Harper     Hamiltonian
Eq.~(\ref{eq:harper_ham}), and  Eq.~(\ref{Hint}) for the approximation
of order  three (recall that  $\epsilon \equiv \tau$).   In principle,
one may construct  $\widetilde{H}^{(n)}$ up to orders as  high as $n =
20$  fairly easily  with symbolic  programs such  as {\tt  MAPLE}.  As
mentioned  in   Sec.~\ref{sec:quasiint},  however,  the   series  Eq.\
(\ref{seriesH}) of $\widetilde{H}^{(n)}$ tends to re-diverge beyond an
optimal order $n_0$, which, for $\tau = 1.$, is generally found around
$n_0 \simeq  6$.  This is illustrated  in Fig.~\ref{fig:asympH_n}: For
various  orders  $n$ of  the  integrable  approximation, $40$ randomly
distributed initial  phase  space  points  have  been
propagated during  a given time by  means of the kicked  Harper map as
well as by its integrable approximation $\widetilde{H}^{(n)}$, and the
distance in phase space between  the two resulting sets of final points 
is plotted
as a function of $n$, yielding a minimum at rather moderate values ($n
\simeq 6$ in  this particular example). We shall  therefore mainly use
$\widetilde{H}^{(n)}$ with $n = 6$ in the following.

\subsection{Resonances parameters}

Fig.~\ref{fig:khmap} and Fig.~\ref{fig:Htildemap} compare the phase  space
portraits of the kicked Harper and of  its integrable (6th order) approximation
in the near-integrable regime at $\tau =  1.0$.  In fact, one observes that the
only significant difference between the two Poincar{\'e}  sections  is the
presence of  the  resonances.  One  may further note the  relative importance
of $r$:$1$ resonances with $r = 10$ and $14$ as compared to the $8$:$1$ and
the $12$:$1$ resonances
(the absence of resonances with odd $r$  is an obvious consequence of the
rectangular symmetry  of the kicked  Harper).
As a matter of fact, these latter resonances, with $r$ a multiple of $4$,
are rather weakly developed at $\tau = 1$ and systematically exhibit $2 r$
(instead of $r$) islands in the Poincar{\'e} surface of section.
We conjecture that this behavior is a consequence of the initial square
symmetry of  the Harper Hamiltonian, which is  still relevant for  small
values of $\tau$.
As the period in the center of the regular region is already larger than $6$
and monotonously increases when moving towards the separatrix, $r$:$s$
resonances with $r \leq 6$ do not exist at $\tau = 1$.

To obtain  a quantitative  prediction for the  tunneling rates,  it is
necessary   to  characterize  the   resonances  through   the  Fourier
coefficients $V^{r:s}_{r.m}$.   This is done  in practice by  a direct
application of Eq.~(\ref{couplTPS}),  i.e.\ by Fourier transforming the
function $\delta I_{r:s}(\theta) \equiv \delta I_{r:s}(p(I_{r:s},\theta),q(I_{r:s},\theta))$
where $I_{r:s}$  is the
action of the resonant  torus ${\Gamma}_{r:s}$.  
On this torus, the  angle  variable  is  given by  $\theta  =  \Omega_{r:s}  t$,  with
$\Omega_{r:s} = \partial H / \partial  I (I_{r:s}) = 2\pi s/ (r \tau) $.  
For a given $\theta$, $\delta I_{r:s}$ is computed
through  the following  successive steps:  
i)  Choose once  for all  a
reference point $(p_0,q_0)$ on the resonant torus ${\Gamma}_{r:s}$
of $\widetilde{H}^{(n_0)}$.  
ii)  Propagate $(p_0,q_0)$ under $\widetilde{H}^{(n_0)}$ dynamics during the
time $t =  (r/s) (\theta /2\pi)  \tau$.  
iii) Apply the time reverse of  the Poincar{\'e} map Eq.~(\ref{eq:kh_map}) on the
resulting point.   
iv) Compute the  difference between the action of this iterated point and 
the action $I_{r:s}$ of ${\Gamma}_{r:s}$.  
The values obtained
in   this   way   for   the   $10$:$1$ resonance  are   plotted   on
Fig.~\ref{fig:Vk_prl_tps}, for various orders $n$ of the
integrable approximation,  showing that  for $3 \leq n \leq 6$ the coefficients
$V^{r:s}_{r.m}$ do not depend  sensitively on $n$.  Also shown 
in this  figure are  the values obtained  by the method  introduced in
\cite{prlbsu}, which is based on a Fourier analysis of the (pseudo-)separatrix
structure that is associated with the resonance.

Within  our setting for the kicked Harper,  the tunnel splitting is
defined as the difference
\begin{equation}
      \delta \phi_k = |\phi_k(\varphi_q=0) - \phi_k(\varphi_q=\pi)|.
\end{equation}
As already stated, the exact quantum values of $\delta \phi_k$ can be calculated up to
numerical precision.
Using  the coefficients   $V^{r:s}_{r.m}$ obtained in the above way, as well
as the unperturbed energies  $\tilde{E}_k$, the periods $T_{k}$ and the
tunneling actions $\sigma_{k}$ which are straightforwardly calculated from the
integrable approximation  $\widetilde{H}^{(n)}$ of the  kicked Harper,
these exact splittings can be  compared with the ones derived from our
semiclassical  expression   Eq.~(\ref{globresonsplit})  based  on  the
resonance-assisted tunneling mechanism.

Before performing this comparison, 
let us first verify that  the qualitative description of the tunneling
mechanism we  gave in Sec.~\ref{sec:pendulum}  and \ref{sec:mechanism}
actually applies  in this particular  example.  To start with,  we can
check that  all the  resonance involved in  the tunneling  process are
well within the quantum perturbative regime.  Indeed, for the value of
the perturbation parameter we  consider, $\tau=1$, the largest Fourier
coefficients  for the  resonances coming  into play  are $V^{8:1}_{16}
\simeq 9.0  \cdot 10^{-7}$, (as already stated,  the $8$:$1$ resonance
exhibits  16  islands),  $V^{10:1}_{10}  \simeq  2.5  \cdot  10^{-4}$,
$V^{14:1}_{14} \simeq 9 \cdot 10^{-4}$, while, in the range of $\hbar$
we  consider, the  energy difference  between  quasi-degenerate states
with respect  to the  resonance is typically  of the order  of $\simeq
10^{-2}$.  Furthermore,  taking into account the actual  values of the
$V^{r:s}_{m.r}$ we observe that as $\hbar=2\pi/N$ gets smaller, higher
orders  of the  quantum  perturbation theory  become  dominant in  the
calculation     of     the     transition    amplitudes     ${\mathcal
A}^{r:s}_{k,m}$. This  can be  specifically verified for  the $10$:$1$
resonance: For  this resonance,  the $k \to  k+20$ transitions  are of
order  one --  i.e.,  are dominated  by  the first-order  perturbative
coupling terms -- for $N  \leqsim 38$, but involve perturbation theory
of order two for  $N \geqsim 38$.  Similarly, we find that  the $k \to k+30$
transitions are of order one for  $N \leqsim 38$, of order two for $38
\leqsim N \leqsim  127$ and involve higher terms  beyond ($k \to k+10$
transitions are, of course, always of order one).
Effectively, one finds here the (possibly unusual) situation which will
generally be encountered in  the semiclassical  limit -- namely that the 
lowest order terms of the perturbative expansion (which converges nevertheless
well) are not the dominating ones.

\begin{figure}[h]
\begin{center}
\leavevmode
\epsfxsize9cm
\epsfbox{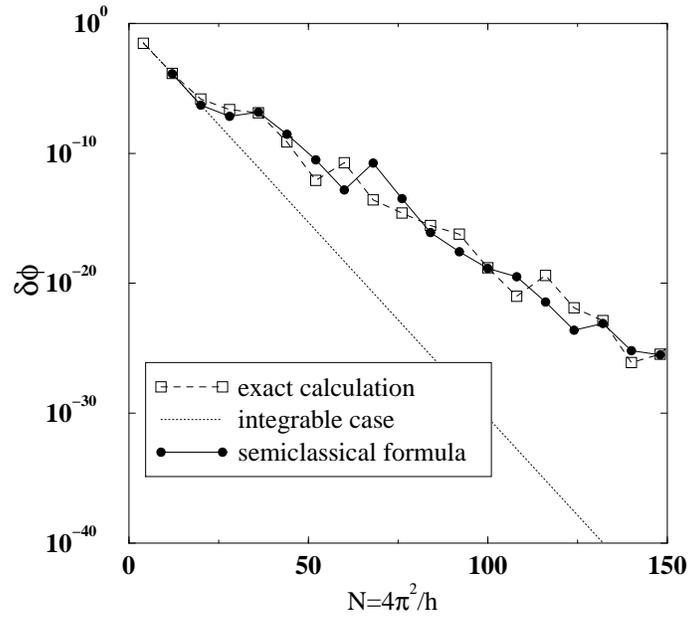}
\end{center}
\caption{Eigenphase  splitting for  the  kicked Harper  system in  the
nearly integrable  regime $(\tau = 1)$.  The  splittings correspond to
the Floquet  mode whose WKB  approximation quantizes on  the classical
torus   with  action  $I_{\pi/4}=\pi/4$,   for  different   values  of
$N=2\pi/\hbar$.   Squares:  exact  numerical calculation.   Filled circles:
semiclassical approximation based  on the resonance assisted tunneling
mechanism.   The  doted line  corresponds  to  the  splitting for  the
integrable approximation.   The solid and dashed lines  are just guides
to the eye.
\label{fig:split1}
}
\end{figure}

\begin{figure}[t]
\begin{center}
\leavevmode
\epsfxsize15cm
\epsfbox{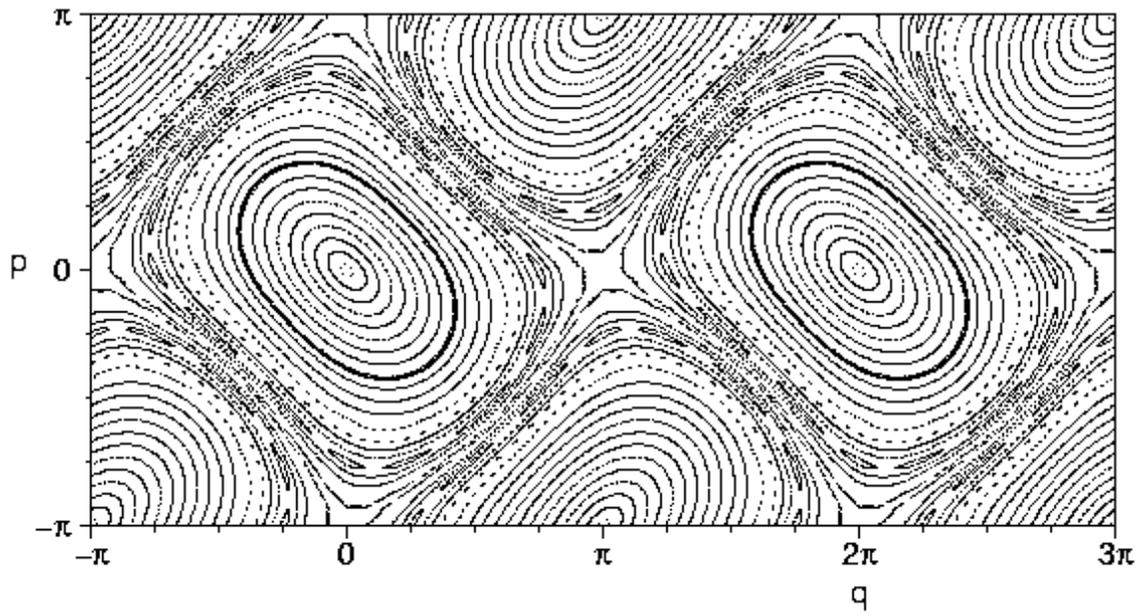}
\end{center}
\caption{Poincar{\'e} section for the kicked Harper map Eq.~({\protect
\ref{eq:kh_map}})  with $\tau=1$.  The  eigenphase splitting  shown on
Fig.~{\protect   \ref{fig:split1}}   correspond   to  the   tunneling
processes between the quasimodes constructed on the tori with action I
= pi/4 marked as thick lines in the left and right wells.
\label{fig:khmap}
}
\end{figure}

\begin{figure}[b]
\begin{center}
\leavevmode
\epsfxsize15cm
\epsfbox{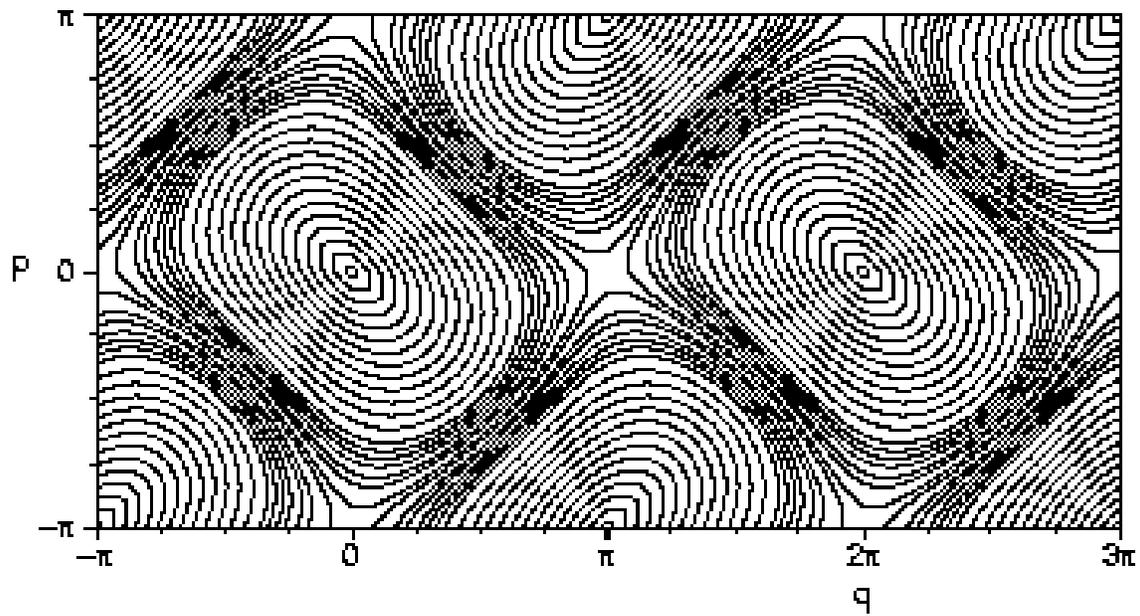}  
\end{center}
\caption{Poincar{\'e}    section   for   the    integrable   approximation
$\widetilde{H}^{(n)}$ with $n = 6$ of the the kicked Harper map, see
Eq.~({\protect \ref{seriesH}})  at  $\tau=1$. Note the good agreement between the
invariant tori of this integrable system and the K.A.M.\ tori of the kicked
Harper at $\tau = 1$ (Fig.~\ref{fig:khmap}).
\label{fig:Htildemap}
}
\end{figure}

\begin{figure}[h]
\begin{center}
\leavevmode

\epsfxsize15cm
\raisebox{8cm}{\epsfbox{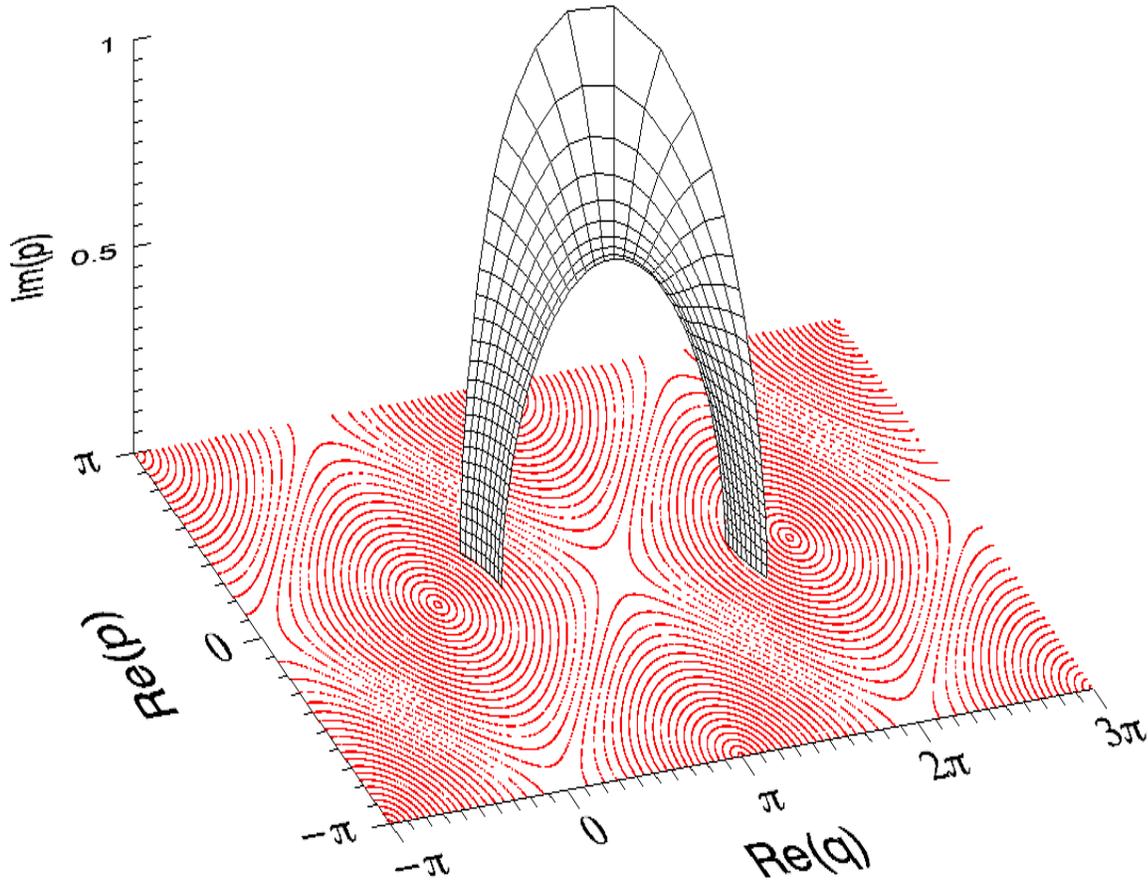}}
\end{center}
\caption{ Part of the analytic  continuation of an  invariant torus
and its  symmetrically shifted  counterpart for the  integrable Harper
hamiltonian.   The  imaginary part  of  momentum  of  the manifold  is
plotted versus the real parts of  position and momentum, on top of the
real phase space.
\label{fig:integrable_cont}}
\end{figure}

Figures~\ref{fig:split1} and  \ref{fig:split2} show for a varying value of
$\hbar$, i.e., a varying total number $N  = 2 \pi / \hbar$ of states, the eigenphase
splittings of the eigenmode of $\widehat{U}$ that corresponds  to a 
fixed classical torus, with action $I=\pi/4$ in Fig.~\ref{fig:split1} and with
action $I=\pi/6$ in Fig.~\ref{fig:split2}.
Evidently, these splittings can be calculated only for particular values of
$N$, namely for $N = 4 ( 2 k + 1 )$ and $N = 6 ( 2 k + 1 )$ with $k = 0,1,\ldots$
in Fig.~\ref{fig:split1} and Fig.~\ref{fig:split2}, respectively, for which
this torus is selected by semiclassical quantization and supports the
$k$th excited quasi-mode.
In both cases, the perturbation parameter  $\tau$ equals $1.0$.  
The resonance involved are the $8$:$1$,  $10$:$1$, and $14$:$1$.   
We observe  that the  agreement  between the
quantum  and  semiclassical  results   is  extremely  nice.   For  the
moderately small values of $\hbar$ that we consider, it is possible to 
try  all the possible  coupling paths  that participate at the tunneling
process,  and  in  Figs.~\ref{fig:split1}  and  \ref{fig:split2},  the
semiclassical   prediction   is   obtained   by  summing up  all   these
contributions.    However,   as  shown   on   the   lower  panel   of
Fig.~\ref{fig:split2},   where   the   action   coordinates   of   the
intermediate states that participate at  the dominant tunneling  path are
displayed,   we  see   here  that,   as  discussed   at  the   end  of
section~\ref{sec:mechanism},  this dominant path  is always  such that
the  number of steps is as large  as possible,  taking into  account the
contraints due to $\hbar$.

Finally we  show on Fig.~\ref{fig:split-ki} a comparison,  for a fixed
value of $\hbar$  and a variable  initial   torus,  between  the  exact  quantum
mechanical  splitting  and  the  one calculated  from  the  expression
corresponding to integrable tunneling, with no resonance coupling.  We
observe on this figure that, although the two curves strongly differ in
the interior of the regular region, they match perfectly as
one gets close  to the separatrix.  This shows  that the presence
of  the separatrix  does not  introduce any  additional effect (e.g.\ from a
small chaotic layer) to the tunneling mechanism.

\subsection{Singularities of the invariant manifold of the integrable
approximation}

In addition to  the  numerical values  of  the  coefficients $V^{r:s}_{r.m}$,
needed to  obtain quantitative prediction  for the tunneling  rates, a
qualitative understanding of their  behaviour, and in particular their
asymptotic  properties for  large  $m$,  is, as  seen  for instance  in
section~\ref{sec:mechanism},  also   required  to  guaranty   that  the
tunneling mechanism we propose  is indeed the dominating one.  Since
the  $V^{r:s}_{r.m}$ are proportional  to the  Fourier coefficients  of the
function  $\delta  I_{r:s}  (\theta)$,  their  asymptotic  behaviour  is
related  to  the  singularities of this function,  for  complex  values  of
the  angle $\theta$.

Let us consider,  more generally, for fixed values  of the energy $E$
and the order $n$ of  the integrable approximation,  the invariant
manifold ${\Gamma}$ of  $\widetilde{H}^{(n)}$, defined by the equation
$\widetilde{H}^{(n)}(p,q) = E$ and characterized by the angular frequency
$\Omega_0$.
Let a function $f(\theta)$ be defined on ${\Gamma}$ as
\begin{equation}
  f(\theta) = \hat f (p(\theta),q(\theta)) \; ,
\end{equation}
where $\hat f(p,q)$ is an entire  function of the phase space variables.
As a consequence,
the  singularities  of
$f(\theta)$  are the ones  of $(p,q)(\theta)$.
What we therefore  need to  study  are the  singularities  of the  analytic
continuation of  $(p,q)(\theta)$ for complex  angles $\theta$.  Due to the
linear relation between $\theta$ and $t$, this analytic continuation
is  straightforwardly constructed  by     propagation (under
$\widetilde{H}^{(n)}$) of some real initial point $(p_0,q_0)$ on ${\Gamma}$, taken
as the origin of the  angle axis, over complex time $t$.
A singularity of $(p,q)(\theta)$  is an angle $\theta^\xi$ such that for the
time $t^\xi  = \theta^\xi / \Omega$ the point $g^{\widetilde H}_{t} (p_0,q_0)$ goes  to
infinity.  Note that  because of the  existence of these 
singularities, $(p,q)(\theta)$  actually depend not only  on the final
time $\theta / \Omega_0$, but  also on the homotopy  class of the
path joining $t=0$ to $\theta / \Omega_0$ in  complex the time plane.  In other words,
$(p,q)(\theta)$ is a priori a multivalued function of $\theta$.

To search for the singularities of $(p,q)(\theta)$, the first step will
consist  in  finding  asymptotic  expression describing  the  manifold
$\widetilde{H}^{(n)} (p,q) = E$ when the imaginary part of $p$ and/or  $q$ goes
to infinity.  For this purpose, we introduce the variables
\begin{eqnarray}
X & = & \exp(ip) \; , \\
Y & = & \exp(iq) \; .
\end{eqnarray}
In these new variables, the integrable approximation of the kicked Harper
Hamiltonian takes the polynomial form
\begin{equation}
   \label{eq:HXY}
   \widetilde{H}^{(n)} = \sum_{i,j=-(n+1)}^{+(n+1)} a_{ij}^{(n)} X^i Y^j \; ,
\end{equation}
with  known   real  coefficient  $a_{ij}^{(n)}$.    For  $H^{(0)}$  for
instance, the non-zero  coefficients are $a_{-10}^{(0)} = a_{10}^{(0)}
= a_{0-1}^{(0)} =  a_{01}^{(0)} = 1/2$.

 The manifold  ${\Gamma}$ is  invariant under the symmetries  $s_O: (X
\mapsto 1/X,  Y \mapsto 1/Y)$, $s_\Delta: (X  \leftrightarrow Y)$, and
$\bar s: (X \mapsto 1/\bar X, Y \mapsto 1/\bar Y)$.  Moreover, one can
check  easily that  if $\widetilde{H}^{(n)}  [\tau] (X,Y)  =  E$, then
$\widetilde{H}^{(n)}  [-\tau] (1/X,Y)  = E$.   We shall  call $s_\tau$
this transformation, although this is not properly speaking a symmetry
of $\Gamma$.  The asymptotic  regions  of ${\Gamma}$ -- i.e., the neighborhood of points
at infinity  on ${\Gamma}$ -- can be obtained by application of one of the above
transformation from one region  such that $\Im  [p] \to +\infty$,  (i.e.~ $X \to 0$  )
and $\Im [q]$ is either bounded  or goes to $+\infty$ (i.e.~ $Y$ bounded).
For such regions, one can assume an asymptotic expression of the form
\begin{equation}
    \label{eq:asymptot}
    Y^\xi(X) = \gamma^\xi_{0} + \gamma^\xi_{1} X + 
    \gamma^\xi_{2}   X^{2} + \cdots \; .
\end{equation}
where $\xi$ label the asymptotic region.
Introducing Eq.~(\ref{eq:asymptot}) in the expression
Eq.~(\ref{eq:HXY}) of the Hamiltonian to solve the equation
$\widetilde{H}^{(n)}(X,Y) = E$ yields a series of polynomial equations
for the coefficients $\gamma^\xi_l$, which can be solved order by order
to determine successively $\gamma^\xi_{0}$,  $\gamma^\xi_{1}$,
$\gamma^\xi_{2}$, etc..  Again for the zeroth order Hamiltonian
$H^{(0)}$, the set of equations obtained in this way are
\begin{eqnarray*}
  \gamma_{0} & = & 0 \\
  \frac{1}{2} \left( \gamma_1 + 1 + \gamma_{0} \right)  
  & = & E \gamma_{0} \\
  \gamma_2 / 2 + \gamma_1 \gamma_{0} + \gamma_{0}/2 
  & = &  E \gamma_1 \\
  \cdots && \cdots \; ,
\end{eqnarray*}
yielding $\gamma_{0} = 0$, $\gamma_{1} = -1$, $\gamma_{2} = -2E$,
$\cdots$.  In other words, for small $X$, the manifold defined by the
implicit expression $\widetilde{H}^{(0)}(X,Y) = E$ admits the explicit
asymptotic expression 
\begin{equation} \label{eq:asym_H0}
Y(X) = -X - 2E X^2 + \cdots \; .
\end{equation}

Using the above equation with $X$  small enough allows to find a point
with    a    large    imaginary    part    for    $p$,    such    that
$\widetilde{H}^{(0)}(p,q)$ is  very close to  $E$.  This point  can be
brought  back to  the  energy $E$  by  following the  gradient of  the
Hamiltonian,    giving     a    point    $(p^\xi,q^\xi)$     on    the
$\widetilde{H}^{(0)}(p,q) =  E$ manifold and in  the asymptotic region
of  large  $\Im  [p]$.   From  this  point,  we  integrate  Hamilton's
equations  of motion  choosing the  path in  the complex  time  in two
different ways: i)  First we take a purely  imaginary direction, until
$t=i  t_I$  such  that   the  trajectory  crosses  the  real  manifold
${\Gamma}_{\mathbb R}  = {\Gamma} \cap {\mathbb  R}^2$.  The imaginary
part of the angle coordinate of $(p^\xi,q^\xi)$ it then given by $-t_I
/ \Omega$.   ii) Then we  start again from $(p^\xi,q^\xi)$  and choose
the  complex phase  of each  time step  $dt$ in  such a  way  that the
imaginary part of $p$ remains constant.  The time $t$ describes then a
small loop  in the complex  time plane that contains  the singularity.
This  gives  the order  of  magnitude  of  the time  distance  between
$(p^\xi,q^\xi)$ and  the singularity,  which is in  practice extremely
small as soon as $\Im(p)$ is taken reasonably large.  For $n=0$, there
is only one independent (i.e.   up to symmetries) singularity, and the
imaginary part of its time  coordinate is just half of $t^\sigma$, the
imaginary  time  required  to   go  from  ${\Gamma}_{\mathbb  R}  \cap
[-\pi,\pi] \times [-\pi,\pi]$  to ${\Gamma}_{\mathbb R} \cap [\pi,3\pi
] \times [-\pi ,\pi]$.

Such  a procedure  can be  reproduced for  various orders  $n$  of the
integrable  Hamiltonian, and we  have performed  it explicitly  up to
$n=3$.   Although the method  we apply  is basically  the same,  a few
important differences may be noticed
\begin{itemize}
\item[i)]  The  number  of  singularities (i.e.   more  precisely,  of
asymptotic regions  of the manifold)  increases with the order  of the
Hamiltonian.  Counting only the  number of  independent singularities,
that  is the ones  that cannot  be deduced  one from  each other  by a
symmetry, there is only one  for $n=0$, but $(2n+2)$ for $n = 1, 2,3$.
\item[ii)] If one starts form  a point $(p^\xi,q^\xi)$ in an asymptotic
region  such as  Eq.~(\ref{eq:asymptot}) and propagates along a time path that
describes  a small closed  loop of infinitesimal  radius around  the
singularity  in time plane, one  can show that the  real part of the
resulting momentum is not Re$[p^\xi]$, but Re$[p^\xi] + 2\pi/\ell^\xi$, 
where $\ell^\xi$
is  an  integer  which depends  on  the  order  $n$ of  the  integrable
approximation   and    on   the   singularity    under   consideration
($\ell^\xi$  is equal  to one for  $H^{(0)}$ and
$H^{(1)}$,  to two  for $H^{(3)}$  and  four of  the singularities  of
$H^{(2)}$, but to three for the two remaining ones). If one identifies
$p$ and $p+  2\pi$, this means that for  $n \neq 0,1$, $(p,q)(\theta)$
are not meromorphic  functions.
Instead, the singularities  are of logarithmic
type.   More precisely,  there are  $\ell^\xi$ distinct  sheets  of the
manifold $(p,q)(\theta)$ around each singularity.
\item[iii)] As a consequence, when one tries to reach the complex torus
from the  neighborhood of  a singularity, one  should specify  on what
sheet  one  places  oneself.   Moreover,  this implies  that  not  all
singularities are ``visible'' from the real torus: assuming the best way
to compute  the Fourier integral Eq.~(\ref{couplTPS}) is  to shift the
integration contour  in the  imaginary direction, the  only singularities
that will be encountered in this way are the ones that can be reached
by  purely imaginary time  propagation from  the real  manifold.  For
$n=3$, only four out of the eight singularities are ``visible'' from the
real torus.
\item[iv)] Starting from the neighborhood of a ``visible'' singularity
and following the Hamiltonian flow,  one may, depending on whether time
runs in  the positive or  negative imaginary direction,  and depending
also on the chosen sheet of the manifold, cross the real manifold
${\Gamma}_R$ in different cell $[(2l- 1) \pi, (2l+1)\pi] \times  [(2l'- 1)\pi,
(2l'+1)\pi]$. Depending on the final cell, the time can be $\pm i
t^\xi$ or $\pm i (t^\sigma - t^\xi)$
\end{itemize}

On  Figs.~\ref{fig:txi_vs_tau:2} and \ref{fig:txi_vs_tau:3},  we plot,
for the resonant torus $10$:$1$  and as a function of the perturbation
parameter $\tau$,  the imaginary  part of the  time coordinate  of the
``visible'' singularities of $\widetilde{H}^{(n)}$ for $n = 2$ and $3$
respectively.  What we  are waiting for is that  the $k$ dependence of
the $V^{r:s}_{k}$ (for $k=rm$ as well as $k \neq rm$) is given by an
expression like Eq.~(\ref{varV}), with $t_{r:s}$ the imaginary part of
the time coordinate of the  singularity closest to the real torus.  On
Fig~\ref{fig:txi_vs_tau:2}  and \ref{fig:txi_vs_tau:3} are  also shown
the  values $\theta_\xi$  obtained  by fitting  the numerically  obtained
$V^{r:s}_{k}$  with  the  expression  Eq.~(\ref{varV}).  We observe that
for $n=2,3$  the variation of  the fitted values follows  nicely the
predicted ones.   For higher $n$,  and up to $n=6$,  the $V^{r:s}_{k}$
are insensitive  to the variation  of the order of  the approximation,
and therefore the fitted values  remains on the curve corresponding to
the $n=3$ closest singularity.


The      data     shown     in      Fig~\ref{fig:txi_vs_tau:2}     and
\ref{fig:txi_vs_tau:3}  give  a  pretty  convincing  picture,  which
justifies  to   use  confidently  the   expression  Eq.~(\ref{varV})  to
describe the behavior of the $V^{r:s}_{r.m}$ coefficients.  Although we
believe this to be true from a practical point  of view, one should, however,
resist the temptation to oversimplify this issue.   Indeed, it is, to start
with, a  priori  not  obvious  to  justify  on a  rigorous  basis  the  form
$(mr)^{\gamma} V^{r:s}_{\xi_0}$ we have written for the prefactor, and
this can only be taken as a sensible parameterization.  Moreover, even
if we did  not extend the analysis of the  location of all singularities
for  $n$  greater than  three  because  the  approach described  above
becomes  somewhat  cumbersome, it  is  still  possible  to locate  the
closest singularity  by a brute  force search in the  complex $\theta$
plane.  Doing this for $n=4,5,6$ for the the torus $10$:$1$ at $\tau =
1$  shows  that  this  closest  singularity  slightly  drifts  as  $n$
increase, and  that its  imaginary part goes  from $0.8$ for  $n=3$ to
$0.6$  for $n=6$, 
in spite of the $V^{r:s}_{r.m}$ being not affected
by this change.   This drift, although moderate, is  still larger than
the numerical precision of our fit.
In accordance, if we take $\hat
f(p,q)  \equiv \cos(p)$  as  done of  Fig.~\ref{fig:fk},  we see  that,
contrary to  $\delta I_{r:s}$,  the Fourier coefficient  of $f(\theta)$
changes with  $n$ even when this  latter is greater than  three, and in
particular  follows  the  asymptotic  slope  $0.6$  for  $n=6$.   This
indicates  that  although  the basic picture  we  gave  to  interpret  the
asymptotic  behavior  of the  $V^{r:s}_{r.m}$  is certainly 
correct,  the complete  description  is presumably  more complicated  and
might involve, for instance, the  link between the kicked Harper map 
${\mathcal T}$ and the  integrable Hamiltonians  $\widetilde{H}_{(n)}$, as
well  as a more  careful  analysis  of  the  different ranges  in  the
asymptotic behavior of the $V^{r:s}_{r.m}$.

\section{Conclusion}

We  have  presented  a semiclassical picture  of  how
tunneling proceeds  in near-integrable systems.   Nonlinear resonances
are identified to  provide a key coupling mechanism  that leads to a
strong  enhancement  of  the   tunneling  rate  with  respect  to  the
corresponding  integrable  system.  In  an  analogous  way  as in  the
generalized pendulum, these resonances induce a periodic modulation of
the  phase  space  structures  in  their vicinity,  which  couples  the
invariant  tori that  are symmetrically  located with  respect  to the
resonance.   Via the  succession of  one or  more such  couplings, the
quantum state may be connected  to higher excitations within the well,
from which it is significantly easier to tunnel outside of the island.

The practical calculation of tunneling rates between separate wells in
near-integrable systems  basically includes the  following four steps.
First, the major resonances, with respect to a given value of $\hbar$,
need to be  identified within the well regions  (which at rather large
$\hbar$ can  be done by  direct visual inspection of  the Poincar{\'e}
surface  of section).  Second,  the coupling  coefficients $V^{r:s}_k$
associated with  the effective pendulum  Hamiltonian of each  of these
resonances  need  to be  calculated.   This  requires  in general  the
construction  of  an  integrable  Hamiltonian which  approximates  the
dynamics with good accuracy.   The low-order coupling coefficients are
most conveniently evaluated by means  of the method that was presented
in section \ref{sec:spt} -- i.e.,  by a Fourier analysis of the action
difference $\delta I_{r:s}(\theta)$ (Eq.~(\ref{couplTPS})) between the
backward propagation of the resonant torus  and  the resonant torus itself  --
whereas  high  orders coefficients  can be deduced  from the  knowledge of
their asymptotic 
behavior.  With these coupling coefficients, the transition amplitudes
across the resonances  can be evaluated by means  of first, or higher,
order quantum  perturbation theory for rather  large, or intermediate,
values  of  $\hbar$,  respectively,   and  are  described  by  a  local
semiclassical tunneling process deep  in the semiclassical regime.  As
a final step, the transition amplitudes associated with the resonances
are  then  combined  to  form  the coupling  scheme  that  the  system
undergoes  to  tunnel  into  the  symmetry-related  well.   For  small
$\hbar$,  many tunneling  paths,  characterized by  the succession  of
resonances  used in  the process,  are contributing  to  the tunneling
rate.  As a general rule, we  state that the dominant one is such that
a  maximum number  of resonance  are implied,  provided  the condition
Eq.~(\ref{eq:Bcond}) is met for all of them.

To  insure that the  tunneling mechanism  we propose  not only  is the
dominant  one,  but  ``makes  sense''  to  start  with,  requires  some
understanding   of  the   qualitative  properties   of   the  coupling
coefficients $V^{r:s}_k$.  This in  addition makes it possible to give
general criteria concerning which  resonances to use in the tunneling
path,  and  how  to deal  locally  with  the  effect of  a  particular
resonance.   Such qualitative  properties  can be  obtained from  the
study of  the analytical  structures, in complex  phase space,  of the
invariant manifold of the integrable Hamiltonian.  We have here made a
first step toward the clarification  of these issues, from which some
confidence in the global coherence of the description can be obtained.
We have however  not tried to address some  more delicate points, such
as the computation  of prefactors in the asymptotic  dependence of the
$V^{r:s}_k$.   Moreover,  although we  have  verified  for the  kicked
Harper model the independence of the $V^{r:s}_k$ on the precise choice
of the integrable  Hamiltonian, and in particular on  the order $n$ of the
approximation, we can only provide an educated guess on the reason why
this property holds.  More research on complex classical dynamics from
a general, mathematical point of  view is certainly required to obtain
a deeper understanding of these issues.

Despite these  remaining questions, we  could show that  the resonance
assisted  tunneling mechanism provides  a quantitative  description of
tunneling  in the near-integrable  kicked Harper  Hamiltonian.  Though
exemplified only  within this particular  model system, the  theory is
described in such  a way that it is  straightforwardly applicable also
to    other   semiclassical    one-degree-of-freedom    systems   with
time-periodic perturbations  (such as the driven  double well dynamics
\cite{lin90,GroO91PRL} for  instance).  A further  generalization that
is straightforwardly achievable is the  one to decay problems -- i.e.,
to the dynamics of quasi-bound  states that are coupled to a continuum
via a tunneling barrier (as is the case e.g.\ in the $\alpha$ decay of
nuclei  or,  to mention  also  a more  recent  example,  in laser-  or
microwave-driven   atoms   \cite{zakrzewski98}).    Here,  the   major
modification with respect to level splitting problems to which we have
restricted our study in this paper concerns the spectral observable of
tunneling --  which would  be given  by a finite  width of  a spectral
resonance instead of a  difference of quasidegenerate eigenlevels.  In
the same way, our theory should apply up to minor modifications to the
case of autonomous two-degrees-of-freedom systems.

The  quantitative  success  of  resonance-assisted  tunneling  in  the
near-integrable kicked  Harper model gives us the  confidence that the
underlying  mechanism  plays   a  key  role  also  in   a  more  mixed
regular-chaotic situation  where macroscopic parts of  the phase space
are occupied by chaos.   Specifically, we believe that the classically
forbidden coupling from  a regular island into the  chaotic sea is now
amenable  to a natural description  by  means  of  the  major
resonances  that characterize  the dynamics  within the  island.  This
still requires  to resolve a  number of open questions,  in particular
related to the (more or  less smooth) transition from regular dynamics
inside to chaotic dynamics  outside the island.  We estimate, though,
that resonance-assisted  tunneling could develop into  a powerful tool
to  quantitatively  understand  semiclassical tunneling  processes  in
mixed systems.

{\bf acknowledgment}

We  thank   E.~Bogomoly,  O.~Bohigas,  P.~Leboeuf,   S.~Tomsovic,  and
A.~M.~Ozorio  de Almeida  for helpful  and inspiring  discussions.  PS
acknowledges  financial support  from Alexander  von Humboldt-Stiftung
and DFG.   The LPTMS  is an ``Unit\'e  de recherche  de l'Universit\'e
Paris 11 associ\'ee au C.N.R.S.''

\appendix

\section{Integrable approximation for rapidly driven Hamiltonians}
\label{app:lie}

The integrable approximation of periodically driven  Hamiltonians, such
as the the Kicked Harper, is most
conveniently constructed by the Lie transformation method.
Following its description in standard textbooks \cite{lichtenberg:book}, the
Lie transformation method introduces a ``pseudo'' Hamiltonian 
$w_{\epsilon} = w_{\epsilon} ({\bf p},{\bf q})$ in order to perform the
canonical transformation from the original (in general multidimensional)
phase space variables $({\bf p},{\bf q})$ to the new variables
$({\bf \tilde{p}},{\bf \tilde{q}})$:
The evolution from $({\bf p},{\bf q})$ to $({\bf \tilde{p}},{\bf \tilde{q}})$
is generated by Hamilton's equations of motion associated with $w_{\epsilon}$,
where the ``evolution time'' $\epsilon$ is given by the perturbation strength
of the actual Hamiltonian ${\cal H} = {\cal H}({\bf p},{\bf q})$ under 
consideration.
The aim is to construct $w_{\epsilon}$ in such a way that ${\cal H}$ acquires, 
for each value of the perturbation parameter $\epsilon$, a specific form 
(e.g., ${\cal H} = {\cal H}({\bf \tilde{p}})$) within the new phase space 
variables $({\bf \tilde{p}},{\bf \tilde{q}})$.

Defining the ``inverse evolution operator'' $T_{\epsilon}^{-1}$ that 
undertakes the associated transformation of phase space distributions
according to
\begin{equation}
\tilde{\cal H}({\bf \tilde{p}},{\bf \tilde{q}}) \equiv 
(T_{\epsilon}^{-1} {\cal H})({\bf \tilde{p}},{\bf \tilde{q}}) :=
{\cal H}({\bf p},{\bf q}) \label{Htilde}
\end{equation}
for all $({\bf p},{\bf q})$, we can straightforwardly verify that
$T_{\epsilon}^{-1}$ satisfies
\begin{equation}
\frac{\partial}{\partial \epsilon} T_{\epsilon}^{-1} = 
L_{\epsilon} T_{\epsilon}^{-1} \label{TL}
\end{equation}
with the ``Lie operator'' $L_{\epsilon}$ defined by
\begin{equation}
L_{\epsilon} := 
\frac{\partial w_{\epsilon}}{\partial {\bf \tilde{q}}}
\frac{\partial}{\partial {\bf \tilde{p}}} -
\frac{\partial w_{\epsilon}}{\partial {\bf \tilde{p}}}
\frac{\partial}{\partial {\bf \tilde{q}}}. \label{L}
\end{equation}
A power series expansion of the original Hamiltonian in the perturbation
parameter $\epsilon$,
\begin{equation}
{\cal H}({\bf p},{\bf q}) = \sum_{n=0}^{\infty} \epsilon^n 
{\cal H}_n({\bf p},{\bf q}), \label{Hn}
\end{equation}
as well as analogous power series expansions of $w_{\epsilon}$, $L_{\epsilon}$,
$T_{\epsilon}^{-1}$, $\tilde{\cal H}$,
\begin{eqnarray}
w_{\epsilon}({\bf p},{\bf q}) & = & 
  \sum_{n=0}^{\infty} \epsilon^n w_{n+1}({\bf p},{\bf q}), \\
L_{\epsilon} & = & \sum_{n=0}^{\infty} \epsilon^n L_{n+1}, \\
T_{\epsilon}^{-1} & = & \sum_{n=0}^{\infty} \epsilon^n T_n^{(-)}
  \quad \mbox{with} \quad T_0^{(-)} = \mbox{Id}, \\
\tilde{\cal H}({\bf p},{\bf q}) & = & 
  \sum_{n=0}^{\infty} \epsilon^n \tilde{\cal H}_n({\bf p},{\bf q})
\label{Htilden}
\end{eqnarray}
can now be used to successively determine $\tilde{\cal H}_n$ as well as the
coefficients $w_n$ of the generating Hamiltonian $w_{\epsilon}$.

In the case of a one-degree-of-freedom system with rapid periodical
driving, we consider a Hamiltonian of the form
\begin{equation}
H = H(p,q,\theta) = H(p,q,\theta + 2 \pi) 
\quad \mbox{with} \quad \theta = \frac{2 \pi}{\tau} t. \label{Ht}
\end{equation}
Here,   the  small   perturbation  parameter   $\epsilon$  essentially
corresponds to the driving period  $\tau$, and the aim of perturbation
theory  consists in eliminating  the explicit  time dependence  of the
Hamiltonian.  In order  to formally obtain a power  series of the type
(\ref{Hn}), we consider  the equivalent, autonomous Hamiltonian system
with two degrees of freedom given by
\begin{equation}
{\cal H} = {\cal H}(p,q,J,\theta) = {\cal H}_0(J) 
  + \tau {\cal H}_1(p,q,\theta) \label{H2}
\end{equation}
with
\begin{eqnarray}
{\cal H}_0(J) & = & J, \\
{\cal H}_1(p,q,\theta) & = & \frac{1}{2 \pi} H(p,q,\theta).
\end{eqnarray}
Here we have additionally introduced the canonically conjugate variables
$(\theta,J)$, corresponding to the scaled time, $\theta = 2 \pi t / \tau$, 
as well as to the inversely scaled energy, $ J = - \tau H / (2 \pi)$, 
respectively.
The new, time-independent Hamiltonian $\tilde{\cal H}$ can then be written
in the form
\begin{equation}
\tilde{\cal H} = \tilde{\cal H}(\tilde{p},\tilde{q},\tilde{J}) = 
  \tilde{J} + \frac{\tau}{2 \pi} \tilde{H}(\tilde{p},\tilde{q}),
\end{equation}
yielding $\tilde{H}(\tilde{p},\tilde{q})$ as integrable
approximation of the rapidly driven system (\ref{Ht}).

Making the power series ansatz
\begin{equation}
\tilde{H}(\tilde{p},\tilde{q}) = \sum_{n=0}^{\infty} \tau^n 
  \tilde{H}_n(\tilde{p},\tilde{q}),
\end{equation}
we obtain, when applying the Lie transformation method 
(\ref{Htilde} -- \ref{Htilden}) to the Hamiltonian (\ref{H2}) with 
${\bf p} \equiv (J,p)$, ${\bf q} \equiv (\theta,q)$ and equating terms of 
equal power in $\tau$,
\begin{equation}
2 \pi \frac{\partial w_n}{\partial \theta} = n \tilde{H}_{n-1} - 
  \sum_{m=1}^{n-1} L_{n-m} \tilde{H}_{m-1} - T_{n-1}^{(-)} H
\end{equation}
from the derivative of (\ref{Htilde}) with respect to $\tau$, and
\begin{equation}
T_n^{(-)} = \frac{1}{n} \sum_{m=0}^{n-1} L_{n-m} T_m^{(-)} \label{Tn}
\end{equation}
with
\begin{equation}
L_n = 
\frac{\partial w_n}{\partial q} \frac{\partial}{\partial p} -
\frac{\partial w_n}{\partial p} \frac{\partial}{\partial q}
\end{equation}
from the relations (\ref{TL}) and (\ref{L}), respectively.
Requiring $w_n$ to be a $2 \pi$ periodic function in $\theta$, it follows
that we have to choose
\begin{equation}
w_n(p,q,\theta) \equiv - \frac{1}{2 \pi} \int_0^{\theta} \left\{ 
  \sum_{m=1}^{n-1} L_{n-m} \tilde{H}_{m-1} + T_{n-1}^{(-)} H \right\}
  (p,q,\theta') \, d \theta'
\end{equation}
in order to obtain a time-independent expression for $\tilde{H}_{n-1}$:
\begin{equation}
\tilde{H}_{n-1}(p,q) = \frac{1}{n} \left\langle \sum_{m=1}^{n-1} 
  L_{n-m} \tilde{H}_{m-1} + T_{n-1}^{(-)} H \right\rangle (p,q).
  \label{Hav}
\end{equation}
Here we have (according to \cite{lichtenberg:book}) introduced the notation
\begin{eqnarray}
\langle H \rangle (p,q) & := & \frac{1}{2 \pi} \int_0^{2 \pi} 
  H(p,q,\theta) \, d \theta, \\
\{H\}(p,q,\theta) & := & H(p,q,\theta) - \langle H \rangle (p,q).
\end{eqnarray}

In zeroth order in $\tau$, we obtain (with $T_0^{(-)} = \mbox{Id}$)
\begin{equation}
w_1(p,q,\theta) = - \frac{1}{2 \pi} \int_0^{\theta} \{ H \}(p,q,\theta') 
  \, d \theta',
\end{equation}
yielding the familiar result that the time-periodic Hamiltonian $H$ 
tends for $\tau \to 0$ to its average within one driving period:
\begin{equation}
\tilde{H}_0(p,q) = \langle H \rangle (p,q).
\end{equation}
The expansion coefficients $w_n$, $T_{n-1}^{(-)}$, $\tilde{H}_{n-1}$ 
with $n > 1$ can be successively determined in ascending order in $n$ 
from the expressions (\ref{Tn} -- \ref{Hav}), as the latter depend 
only on $w_m$, $T_{m-1}^{(-)}$, $\tilde{H}_{m-1}$ with $1 \leq m < n$.

The method shall be exemplified for a Hamiltonian of the form
\begin{equation}
H(p,q,\theta) = H_0(p,q) + K(\theta) V(q)
\end{equation}
which typically appears e.g.\ in the case of electromagnetically driven 
atoms.
Defining
\begin{eqnarray}
K_1(\theta) & := & \frac{1}{2 \pi} \int_0^{\theta} \{K\}(\theta') 
  \, d \theta', \\
K_2(\theta) & := & \frac{1}{2 \pi} \int_0^{\theta} \{K_1\}(\theta') 
  \, d \theta',
\end{eqnarray}
we obtain up to second order in the perturbation $\tau$
\begin{eqnarray}
\tilde{H}(\tilde{p},\tilde{q}) & = & H_0(\tilde{p},\tilde{q}) + \langle K \rangle
  V(\tilde{q}) - \tau \langle K_1 \rangle \frac{\partial V}{\partial \tilde{q}}
  \frac{\partial H_0}{\partial \tilde{p}} + \tau^2 \\
  && \times \left[ \frac{1}{3} \left( \langle K_1 \rangle^2 + \frac{1}{2}
  \langle K_1^2 \rangle - \langle K_2 K \rangle - 2 \langle K_2 \rangle
  \langle K \rangle \right) \left(\frac{\partial V}{\partial \tilde{q}}\right)^2
  \frac{\partial^2 H_0}{\partial \tilde{p}^2} \right. \nonumber \\
  && + \left. \langle K_2 \rangle \left( 
  \frac{\partial^2 V}{\partial \tilde{q}^2}
  \left(\frac{\partial H_0}{\partial \tilde{p}}\right)^2 + 
  \frac{\partial V}{\partial \tilde{q}} \left( 
  \frac{\partial^2 H_0}{\partial \tilde{p} \partial \tilde{q}} 
  \frac{\partial H_0}{\partial \tilde{p}} -
  \frac{\partial^2 H_0}{\partial \tilde{p}^2} 
  \frac{\partial H_0}{\partial \tilde{q}} \right) \right) \right]. \nonumber
\end{eqnarray}
For the particular case of the Kicked Harper Hamiltonian
Eq.~(\ref{eq:kh_H}), we have
\begin{equation}
K(\theta) = \sum_{n=-\infty}^{\infty} \delta 
  ( \theta / (2 \pi) - n ).
\end{equation}
This yields
\begin{eqnarray}
\tilde{H}(\tilde{p},\tilde{q}) & = & \cos\tilde{p} + \cos\tilde{q} 
  - \frac{\tau}{2} \sin\tilde{p} \, \sin\tilde{q}
  - \frac{\tau^2}{12} \left( \cos\tilde{p} \, \sin^2\tilde{q} 
  + \cos\tilde{q} \, \sin^2\tilde{p} \right)\nonumber \\
  && - \frac{\tau^3}{48} \sin(2\tilde{p}) \sin(2\tilde{q}) + {\cal O}(\tau^4)
  \label{Hint}
\end{eqnarray}
as integrable approximation up to third order in $\tau$.
As evaluated by means of the power series coefficients $T_{n}^{(-)}$,
the corresponding transformation to the original phase space variables 
$(p,q)$ reads
\begin{eqnarray}
p & = & \tilde{p} + (\tau - t) \sin\tilde{q} + \frac{1}{2} t (\tau - t)
  \sin\tilde{p} \, \cos\tilde{q} + \frac{1}{12} t (\tau - t) \nonumber \\
  && \times \left[ 2(4\tau - 3 t) \cos\tilde{p}
  \, \cos\tilde{q} \, \sin\tilde{q} + (\tau - 2 t) \sin^2\tilde{p} \, 
  \sin\tilde{q} \right], \\
q & = & \tilde{q} - \frac{1}{2} t (\tau - t) \cos\tilde{p} \, \sin\tilde{q}
  + \frac{1}{12} t (\tau - t) \nonumber \\
  && \times \left[ (5\tau - 4 t) \sin\tilde{p} \, \sin^2\tilde{q}
  + 2 ( \tau - 2 t) \cos\tilde{p} \, \sin\tilde{p} \, \cos\tilde{q} \right]
\label{pqpq}
\end{eqnarray}
for $0 < t  < \tau$.  At times $t_n = n  \tau - \epsilon$ with integer
$n$ and $\epsilon \to 0_{+}$, we have $(p,q) = (\tilde{p},\tilde{q})$.
This implies  that the Kicked Harper  map Eq.~(\ref{eq:kh_map}), which
essentially corresponds  to the stroboscopic  section at $t =  t_n$ of
the dynamics  generated by  $H$, can be  directly approximated  by the
propagation of the integrable approximation (\ref{Hint}).

The  terms beyond  the third  order  in $\tau$  are most  conveniently
derived with symbolic programs  such as {\tt MAPLE}.  The coefficients
$C_{nm}$ that  enter into the corresponding  Fourier series expansions
$\sum_{n,m} C_{nm} \cos  mp \, \cos nq$ (or $\sin mp  \, \sin nq$) can
then be used to propagate the classical equations of motion associated
with $\tilde{H}$ as  well as to create the  quantum Hamiltonian matrix
to  be  diagonalized  in   order  to  determine  the  eigenvalues  and
eigenvectors of $\tilde{H}$.

\section{Analytic continuation of the Kicked Harper tori into the
complex domain}
\label{app:cont}

In contrast to integrable systems, the analytic continuation of the 
Kicked Harper tori into the complex domain cannot be performed by
imaginary time propagation -- which, to a certain extent, can be seen
from the fact that these tori are visualized in phase space by a stroboscopic
map, rather than by a smooth Hamiltonian flow.
The map itself, however, which can be written as
\begin{equation}
\left( \begin{array}{l} p_0 \\ q_0 \end{array} \right) \mapsto
\left( \begin{array}{l} p_\tau \\ q_\tau \end{array} \right) 
  ( p_0, q_0 ) := 
\left( \begin{array}{l} p_0 - \tau \sin q_0 \\ 
  q_0 + \tau \sin ( p_0 - \tau \sin q_0 ) \end{array} \right),
\label{map}
\end{equation}
is amenable to complexification, as it represents an analytic vector field 
in the phase space variables $(p,q)$.
In analogy to the unbroken KAM tori of the real phase space, the analytic
continuations of these tori are then given by invariant manifolds of the
complexified map (\ref{map}).
This implies that the recursive application of (\ref{map}) on any complex
phase space point $(p_0,q_0)$ that is located on such a manifold yields a
closed, quasiperiodic trajectory which, as in the integrable system, is
characterized by the same frequency $\Omega$ (or winding number) as the
real KAM torus from which on the manifold is departing.

This fact can be used to construct the analytic continuation of the KAM tori
by an optimization program that is based on computing complex phase space 
trajectories with the complexified map (\ref{map}).
For this purpose, it is convenient to introduce, for each of such 
trajectories, the ``pseudo angle''
\begin{equation}
\theta_n = n \, \Omega \, \tau \, (\mbox{mod} \, 2\pi)
\end{equation}
of the $n$th iterated point with respect to a given starting point 
$(p_0,q_0)$, where $\Omega$ is the round-trip frequency for which the analytic
continuation of the torus is to be constructed.
Sorting the trajectory with respect to this pseudo angle yields then a 
smooth, closed orbit -- if and only if the trajectory is quasiperiodic with
frequency $\Omega$.
If there is a slight mismatch between the frequency of the trajectory and
$\Omega$ or if the trajectory is not closed at all, then the sorted orbit
exhibits wild fluctuations, which means that huge differences of $(p,q)$ 
may occur between adjacent values of $\theta$.
A phase space point $(p_0,q_0)$ on the ``good'' manifold may then be found
by minimizing a function $f = f(p_0,q_0)$ that measures the ``noise'' of
the sorted trajectory that starts from $(p_0,q_0)$ for a given number $N$ of 
mappings.
Specifically, we chose $f$ as the average squared difference between
adjacent slopes in the sorted orbit -- i.e.,
\begin{eqnarray}
f & = & \frac{1}{N} \sum_{i=1}^{N-1} \left[ \left|
\frac{p_{p(i+1)} - p_{p(i)}}{\theta_{p(i+1)} - \theta_{p(i)}} -
\frac{p_{p(i)} - p_{p(i-1)}}{\theta_{p(i)} - \theta_{p(i-1)}} \right|^2 
+ \right. \nonumber \\
&& \left. \left| 
\frac{q_{p(i+1)} - q_{p(i)}}{\theta_{p(i+1)} - \theta_{p(i)}} -
\frac{q_{p(i)} - q_{p(i-1)}}{\theta_{p(i)} - \theta_{p(i-1)}} \right|^2 \right]
\label{noise}
\end{eqnarray}
where $p(i)$ represents, for given $N$, the index permutation that sorts the
pseudo angles $\theta_i$ (i.e., $\theta_{p(j)} < \theta_{p(i)}$ for all 
$j < i$).
Clearly, this noise function $f$ will become minimal (and will 
actually vanish in the limit $N \to \infty$) if all phase space points lie 
on a smooth complex curve -- in which case the slope between adjacent points 
is locally almost constant.

Since  $f$ contains  many local  minima in  general,  the optimization
method has to be implemented in a stepwise way.  We first start with a
rather small  number $N$  of mappings, corresponding  to two  or three
round-trips  of the trajectory.   Calculating the  noise (\ref{noise})
for a given starting point $(p_0,q_0)$ and for points in its immediate
environment yields then, by means  of the Newton method, a rough (and,
at this  stage of the  algorithm, rather imprecise) estimation  of the
correct phase space  point on the KAM manifold.   Taking this point as
new starting point, the procedure is repeated with a higher number $N$
of  mappings, which  makes  the local  ``valley''  around the  minimum
steeper  and thereby  allows  for  a more  precise  estimation of  the
correct  point.  The  iteration is  continued until  the point  on the
manifold  is known with  the desired  precision: At  each step  of the
iteration, the starting  point $(p_0,q_0)$ is improved by  one step of
the Newton  method, and the value  of the noise at  the improved point
gives  an indication  by which  amount  $N$ may  be increased  without
leaving the  ``good'' minimization valley.  At a  given maximum number
of mappings,  convergence of  the method is  checked by  comparing the
noise $f(p_0,q_0)$ of the  optimized starting point $(p_0,q_0)$ with a
suitably chosen reference value  $f_{crit}$ (recall that $f$ should be
zero on the  KAM manifold for $N \to \infty$).  Cases  of doubt may be
resolved  by directly  looking  at the  sorted  trajectory plotted  in
pseudo angle space (as shown in Fig.~\ref{fig:orbit}).

The optimization  method is  additionally simplified for  the specific
case of  the Kicked Harper.   Due to the  symmetry of the  phase space
with respect to the  diagonals (see e.g.~Fig.~\ref{fig:khmap}), we can
infer that  the analytic continuations of  the real KAM  tori pass all
through the complex hyperplane defined by $p = q^*$.  We may therefore
restrict  our starting  points  to that  hyperplane.  By  additionally
fixing the  imaginary part  of $p$ (which,  in some sense,  defines at
which distance from the real phase  space we want to calculate a point
on the manifold), only one parameter,  namely the real part of $p$ and
$q$, needs to be varied for the Newton method.

\begin{figure}[h]
\begin{center}
\leavevmode
\epsfxsize15cm
\raisebox{8cm}{\epsfbox{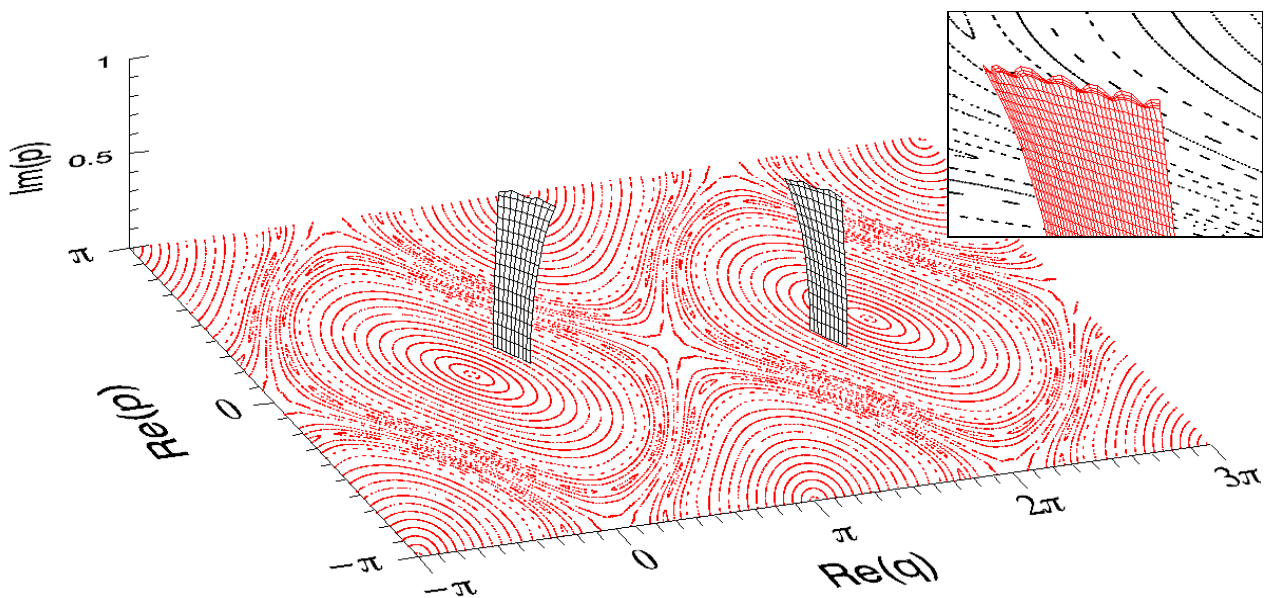}}
\end{center}
\vspace*{2cm}
\caption{Part  of the  analytic continuation  of a  KAM torus  and its
symmetrically shifted counterpart in the near-integrable Kicked Harper
at $\tau  = 1$.   The imaginary  part of momentum  of the  manifold is
plotted versus the real parts of  position and momentum, on top of the
real  phase  space.   In  contrast  to integrable  dynamics,  the  two
manifolds on each side of the separatrix do {\em not} join each other,
but are interrupted  far inside the regular phase  space domain.  Note
the fluctuations  that appear  at the outermost  end of  the converged
manifold (as  shown in the  inset for the  manifold on the  right hand
side).
\label{fig:khcont}}
\end{figure}

\begin{figure}
\begin{center}
\leavevmode
\epsfxsize15cm
\raisebox{1cm}{\epsfbox{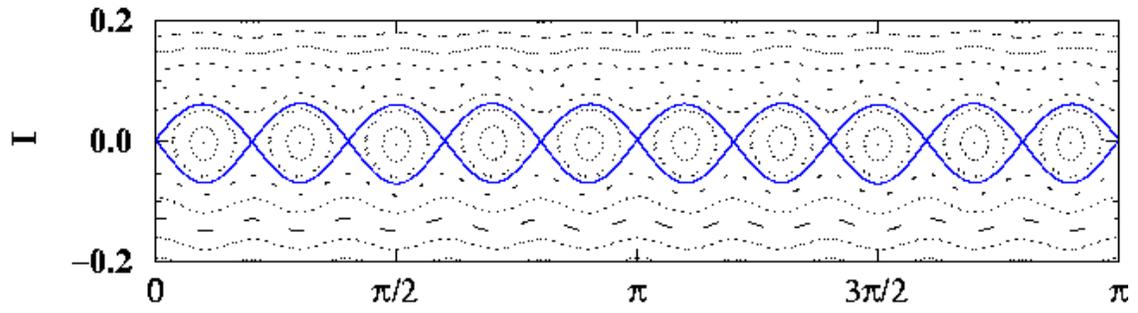}}
\end{center}
\caption{Poincar{\'e} section in the vicinity of the $10$:$1$ resonance
of the kicked Harper, in the action angle variables of the integrable
approximation $\widetilde{H}^{(n)}$ ($n=6$).
Note the similarity of the phase space structure with the one of a generalized
mathematical pendulum, with $10$ islands instead of one.
\label{fig:res_10}}
\end{figure}

\begin{figure}[h]
\begin{center}
\leavevmode
\epsfxsize15cm
\epsfbox{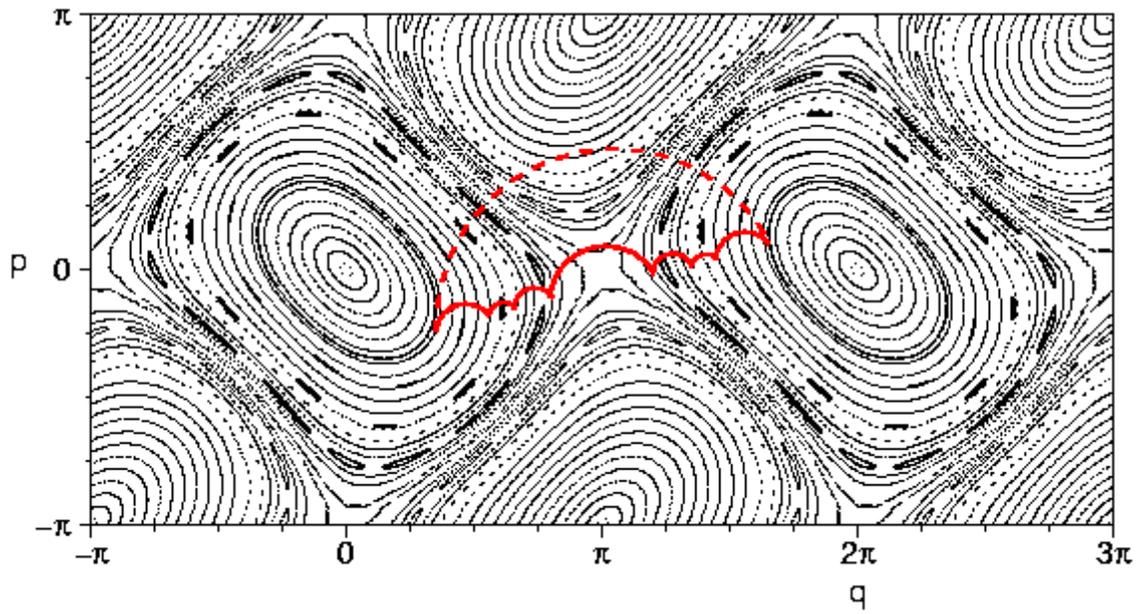}
\end{center}
\caption{Visual representation of resonance-assisted tunneling:
Plotted is, for the Kicked Harper system with $\tau = 1$, the succession 
of complex paths that are used to accomplish the tunneling process
for the 10th excited state at $2\pi/\hbar = 126$.
Instead of a direct transition (dashed curve) to the symmetry related torus,
the quasi-mode is coupled via the $8$:$1$ (hardly visible) the $10$:$1$, and
the $14$:$1$ resonance to the 50th excited state, which lies sufficiently
close to the separatrix to tunnel with a large rate to its counterpart.
\label{fig:saute_moutons}
}
\end{figure}

\begin{figure}[h]
\begin{center}
\leavevmode
\epsfxsize8cm
\epsfbox{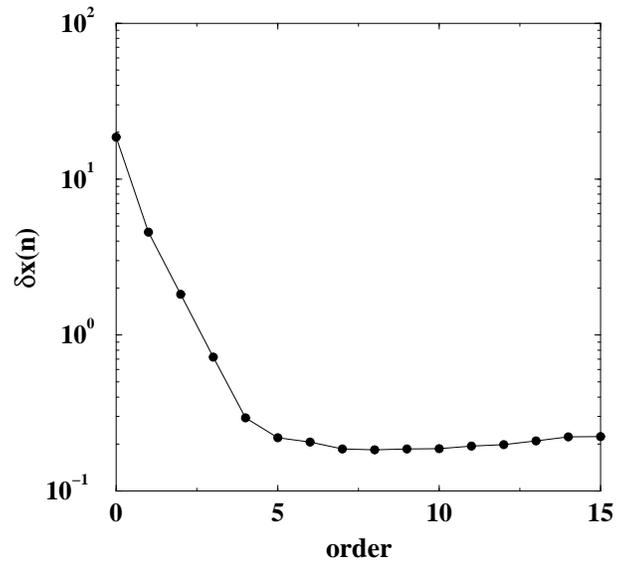}
\end{center}
\caption{ For the tenth  iterate of a set of randomly distributed
points in phase space, distance, as a function  of the order
of  the approximation $n$,  between the sets of resulting points  depending on
whether the kicked Harper  or its integrable approximation $\widetilde
H^{(n)}$  is  used. The perturbation parameter is $\tau=1.0$.
\label{fig:asympH_n}}
\end{figure}

\begin{figure}[h]
\begin{center}
\leavevmode
\epsfxsize8cm
\epsfbox{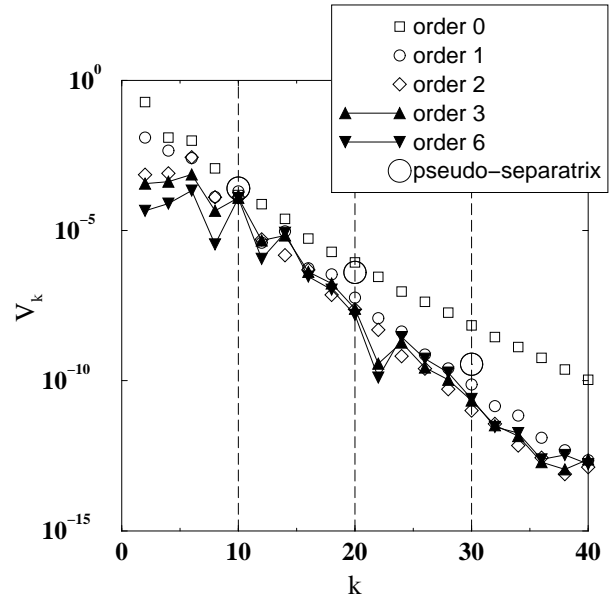}
\end{center}
\caption{\label{fig:Vk_prl_tps}   $k$-dependence  of  the  resonance
coefficients  $V^{10:1}_k$ for  the  resonance $10$:$1$  of the  kicked
Harper with  $\tau = 1$.   The various symbols correspond  to different
orders $n$ of  the integrable approximation $\widetilde{H}^{(n)}$ used
for the calculation, demonstrating the independence of the $V^{r:s}_k$
on $n$  for $n \simeq n_0=6$.   The large circle  symbols  at $k=10,20,30$ show
the  result   obtained  for   these  Fourier  coefficients   with  the
pseudo-separatrix method used in {\protect \cite{prlbsu}}.
}
\end{figure}

\begin{figure}[t]
\begin{center}
\leavevmode
\epsfxsize15cm
\epsfbox{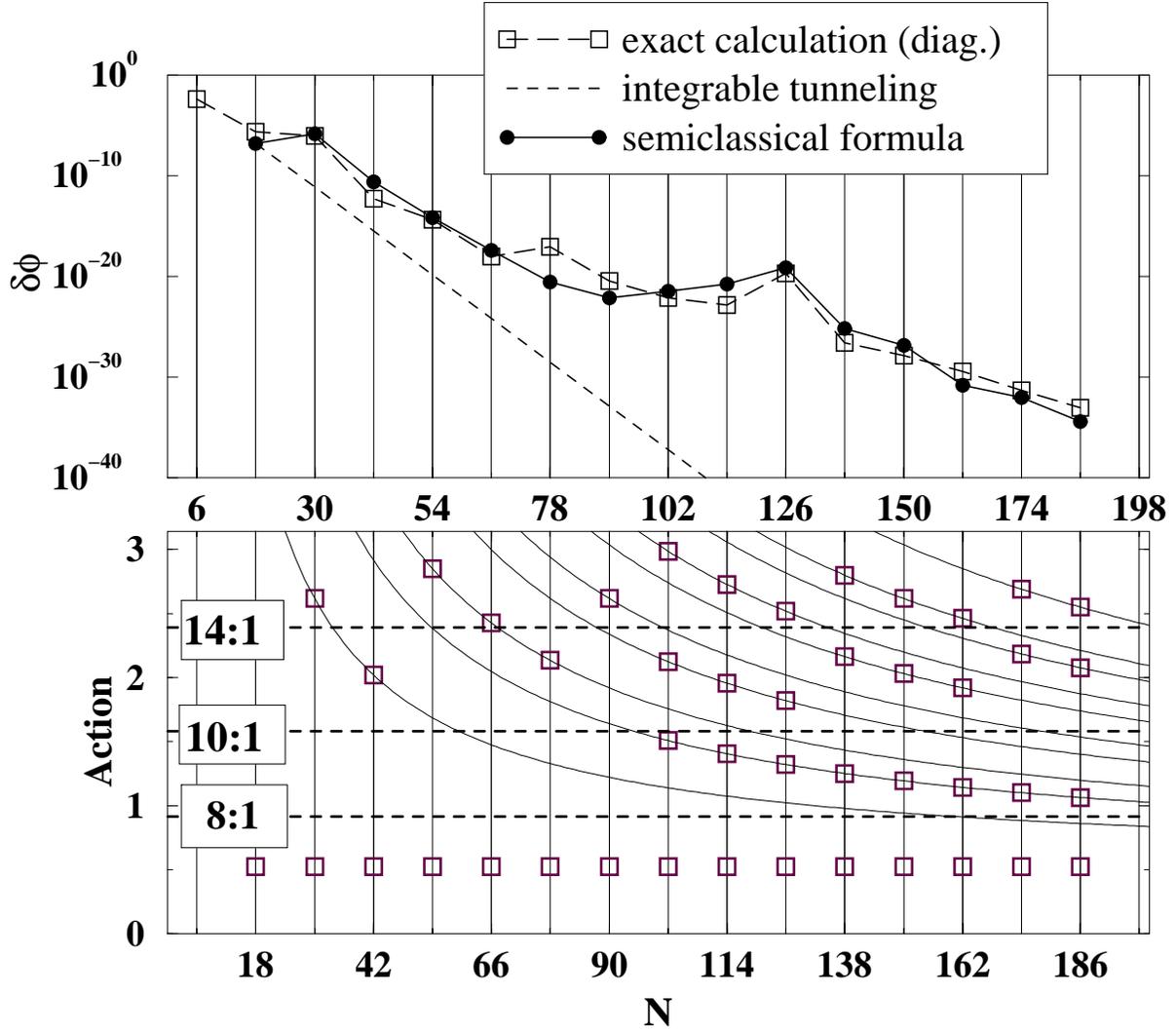}
\end{center}
\caption{Top~: same  as Fig.~{\protect \ref{fig:split1}},  but for the
Floquet mode whose WKB  approximation quantizes on the classical torus
with  action $I_{\pi/6}=\pi/6$.   Bottom~: the  classical  action, for
each value  of $N$,  of the modes  involved in the  dominant tunneling
path.   The horizontal doted  line corresponds  to the  initial action
$I_{\pi/6}$.  The horizontal solid  lines correspond to the actions of
the resonances $8$:$1$, $10$:$1$ and $14$:$1$.  The curved solid lines
are such that $I-I_{\pi/6} = {\rm const.}  \times \hbar$ (that is, the
distance from  $I_{\pi/6}$ is constance  in mode number  unit). Notice
that as $N = 2\pi/\hbar$ increases, more resonances are involved.
\label{fig:split2}
}
\end{figure}

\begin{figure}[h]
\begin{center}
\leavevmode
\vspace*{4cm}
\epsfxsize15cm
\epsfbox{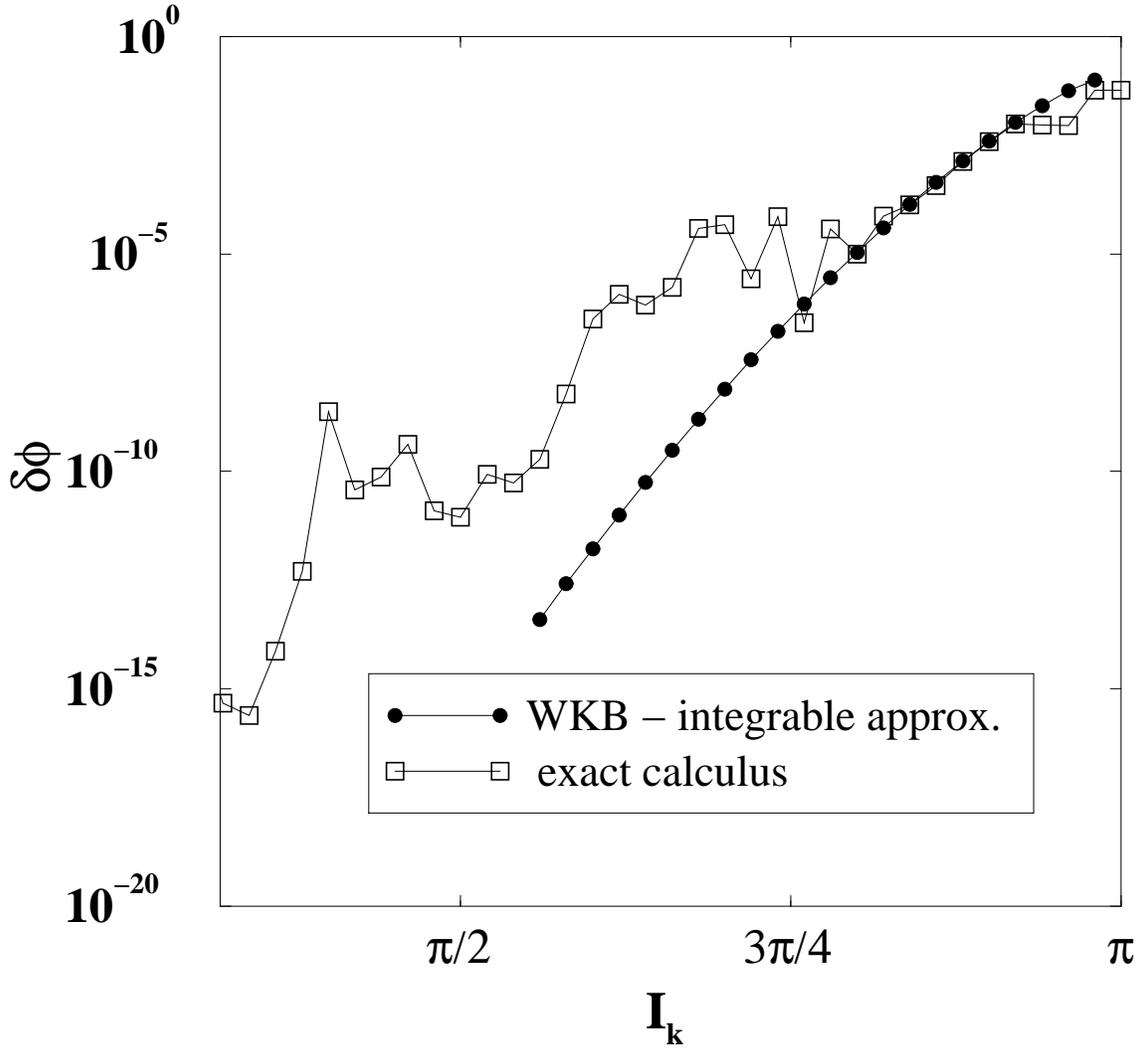}
\end{center}
\caption{  Eigenphase splitting for  the kicked  Harper system  in the
nearly integrable  regime $(\tau = 1)$,  for a fixed  value of $\hbar$
($=2\pi/100$),  and   a  variable  initial   torus.  Squares:  exact
numerical calculations.  Dots: integrable semiclassical approximation,
without including the effect of  the resonances.  As the initial torus
get  close  enough  from  the  separatrix $I_{\rm  max}  =  \pi$,  the
integrable  approximation for the  tunneling reproduces  faithfully the
exact  quantum  result,  showing  that  the  chaotic  layer  near  the
separatrix plays no role here.
\label{fig:split-ki}}
\end{figure}

\begin{figure}[h]
\begin{center}
\leavevmode
\epsfxsize15cm
\epsfbox{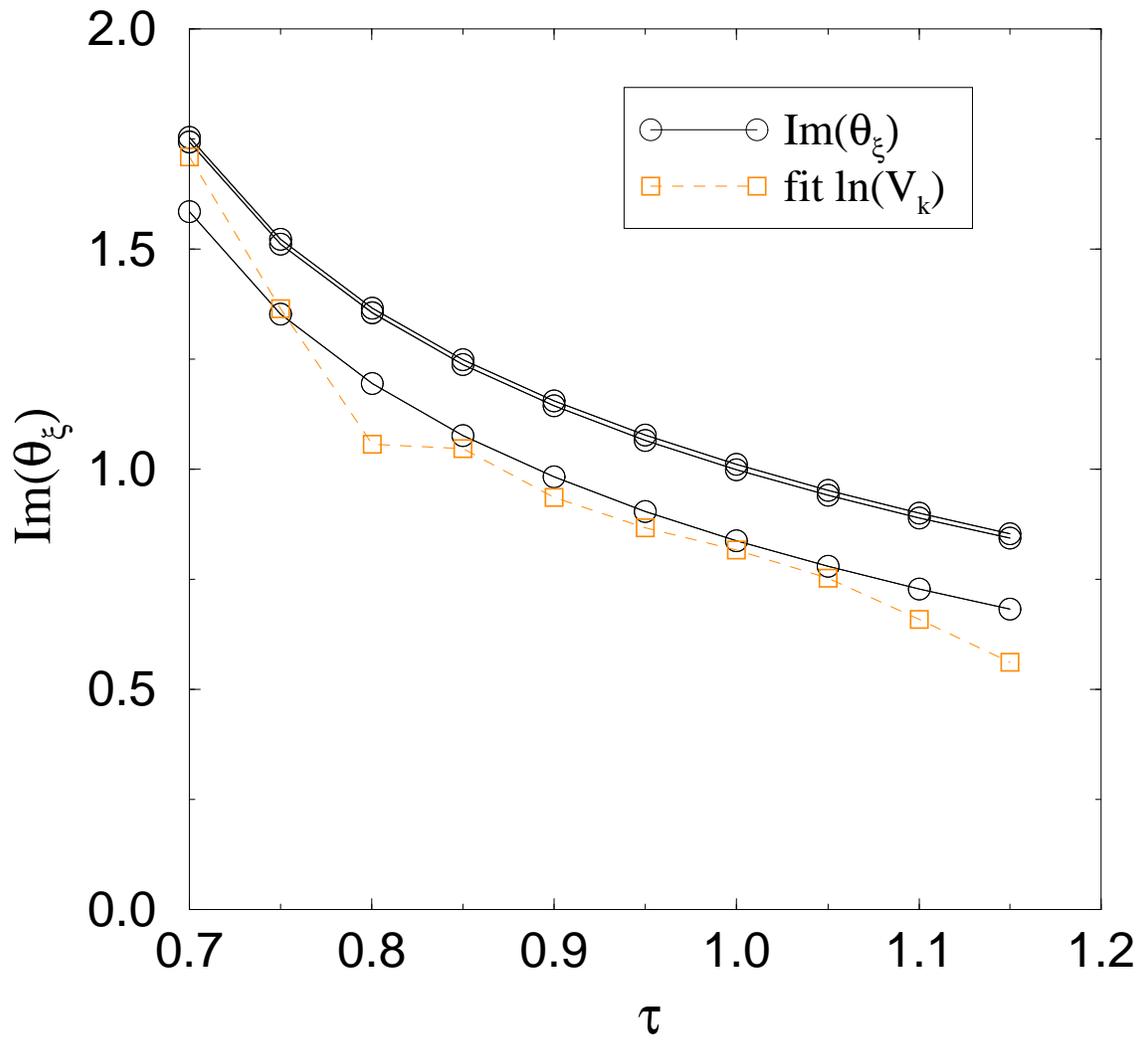}
\end{center}
\caption{For     the    order    $n=2$     integrable    approximation
$\widetilde{H}^{(n)}$: comparison,  as a function  of the perturbation
parameter $\tau$, of the fitted  value $\vartheta_\xi$ of the slope of
the log  of the coefficient  $V^{10:1}_k$ (squares) with  the imaginary
part of the angle coordinate of the visible singularities on the torus
$I_{10:1}$ (circles) The lines are just guides to the eye.
\label{fig:txi_vs_tau:2}}
\end{figure}

\begin{figure}[h]
\begin{center}
\leavevmode
\epsfxsize15cm
\epsfbox{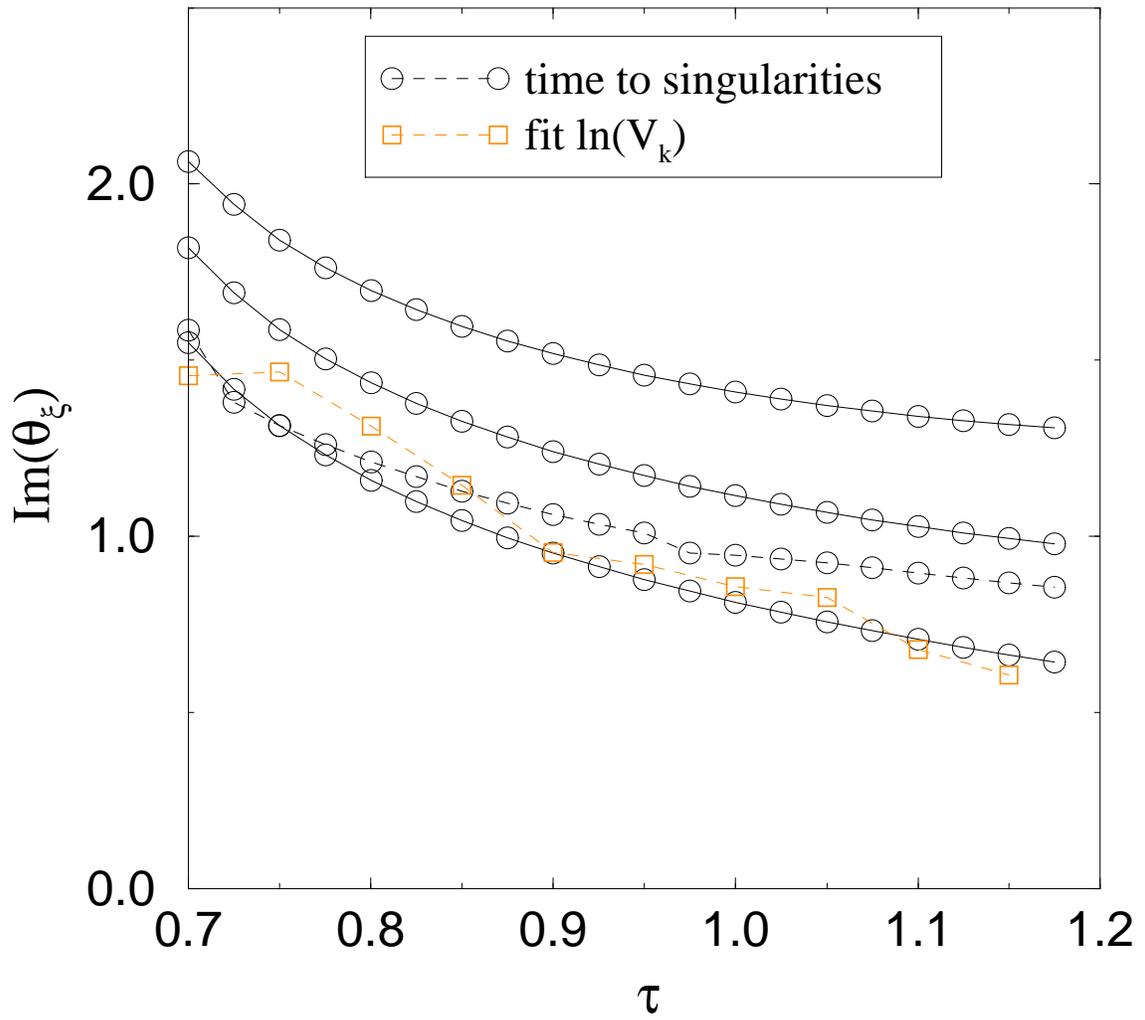}
\end{center}
\caption{Same as  Fig.~{\protect \ref{fig:txi_vs_tau:2}}, but  for the
order $n=3$ integrable approximation $\widetilde{H}^{(n)}$.
\label{fig:txi_vs_tau:3}}
\end{figure}

\begin{figure}[h]
\begin{center}
\leavevmode
\epsfxsize12cm
\epsfbox{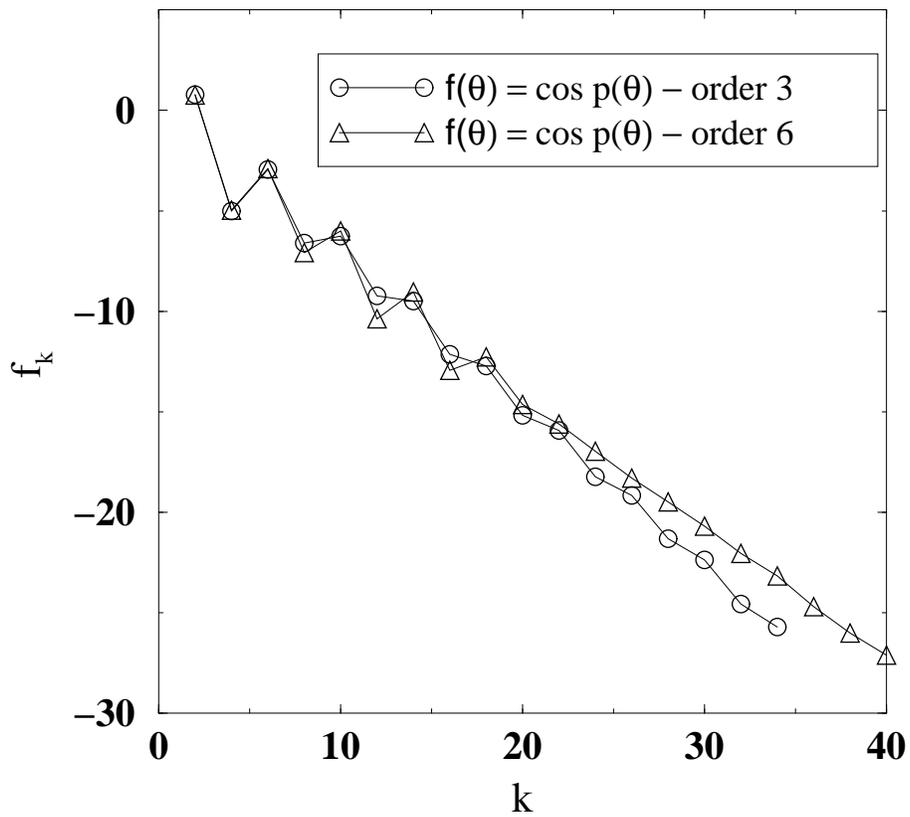}
\end{center}
\caption{Fourier  coefficients  $f_k$  of  the functions  $f(\theta)  =
\cos\left[   p(\theta,I_{10:1})   \right]$.    The  different   curves
correspond  to different  orders $n$  of the  integrable approximation
used. We observe that although the $f_k$'s remain insensitive to the
choice of the integrable approximation for not too large $k$, their
asymptotic behavior differ depending on whether the integrable
approximation with $n=3$ or $n=6$ is chosen.
\label{fig:fk}}
\end{figure}

\begin{figure}[h]
\begin{center}
\leavevmode
\epsfxsize12cm
\epsfbox{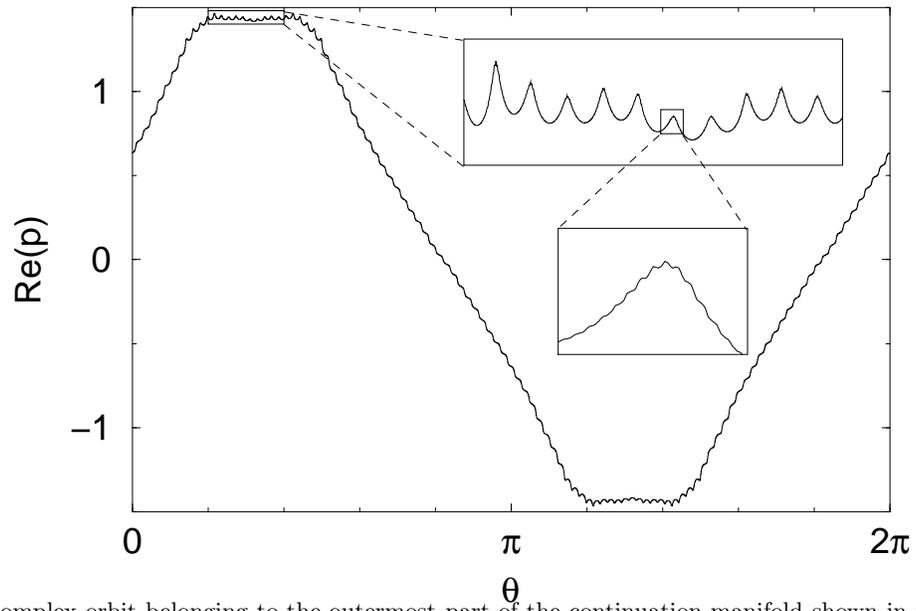}
\end{center}
\caption{Sorted complex  orbit belonging to the outermost  part of the
continuation manifold shown  in fig.~\ref{fig:khcont}.  Plotted is the
real part of momentum as a function of the pseudo angle $\theta$.  One
clearly recognizes significant periodic modulations of the orbit which
appear on  different scales,  as can be  seen from  the magnifications
shown in the insets.
\label{fig:orbit}}
\end{figure}

A typical  example for the analytic continuation  of a near-integrable
kicked  Harper  torus  was  shown in  Fig.~\ref{fig:khcont}.   In  the
corresponding integrable system, the two manifolds shown in the figure
would join each  other across the separatrix to  form a smooth complex
surface.  For the near-integrable Kicked Harper, however, the analytic
continuations can  be constructed only  up to a certain  distance from
the real phase space (namely  up to $\mbox{Imag}(p) \simeq 0.8$ within
the $p =  q^*$ subspace); beyond that value,  the optimization program
does not converge any longer.  A  closer look on the outermost part of
the  converged manifold  reveals significant  fluctuations, as  can be
seen in the inset of  Fig.~\ref{fig:khcont}.  Plotting one of the last
converged   trajectories   as  a   function   of   the  pseudo   angle
(Fig.~\ref{fig:orbit}) reveals that  this trajectory exhibits periodic
modulations  which seem  to  appear on  many  different scales.   This
observation  is in  fact consistent  with the  findings of  Greene and
Percival in the standard map,  where a fractal structure was found for
the outermost part of the  complex torus (compare in particular Fig.~3
of  \cite{greene81}   with  our  Fig.~\ref{fig:orbit}).    It  thereby
supports  our conviction  that  the  reason why  our  method fails  to
converge is indeed the fact that  the natural boundary of the torus is
approached.


\section{Tunneling in the pendulum}
\label{app:tunnpend}

In this appendix, we give a semiclassical expression for tunneling
in the generalized quantum pendulum
\begin{equation}
  \widehat{H}=\frac{(\widehat{I}-I_0)^2}{2\mu}+V_1\cos{r\Theta}
  \; .
\end{equation}
We shall moreover assume antiperiodic boundary condition for the angle
$\Theta$  as a  remnant  of the  existence  of Maslov  indices in  the
original $(p,q)$ variables.

\subsection{Invariant torus geometry} 

\begin{figure}[h]
\begin{center}
\leavevmode
\epsfxsize15cm
\epsfbox{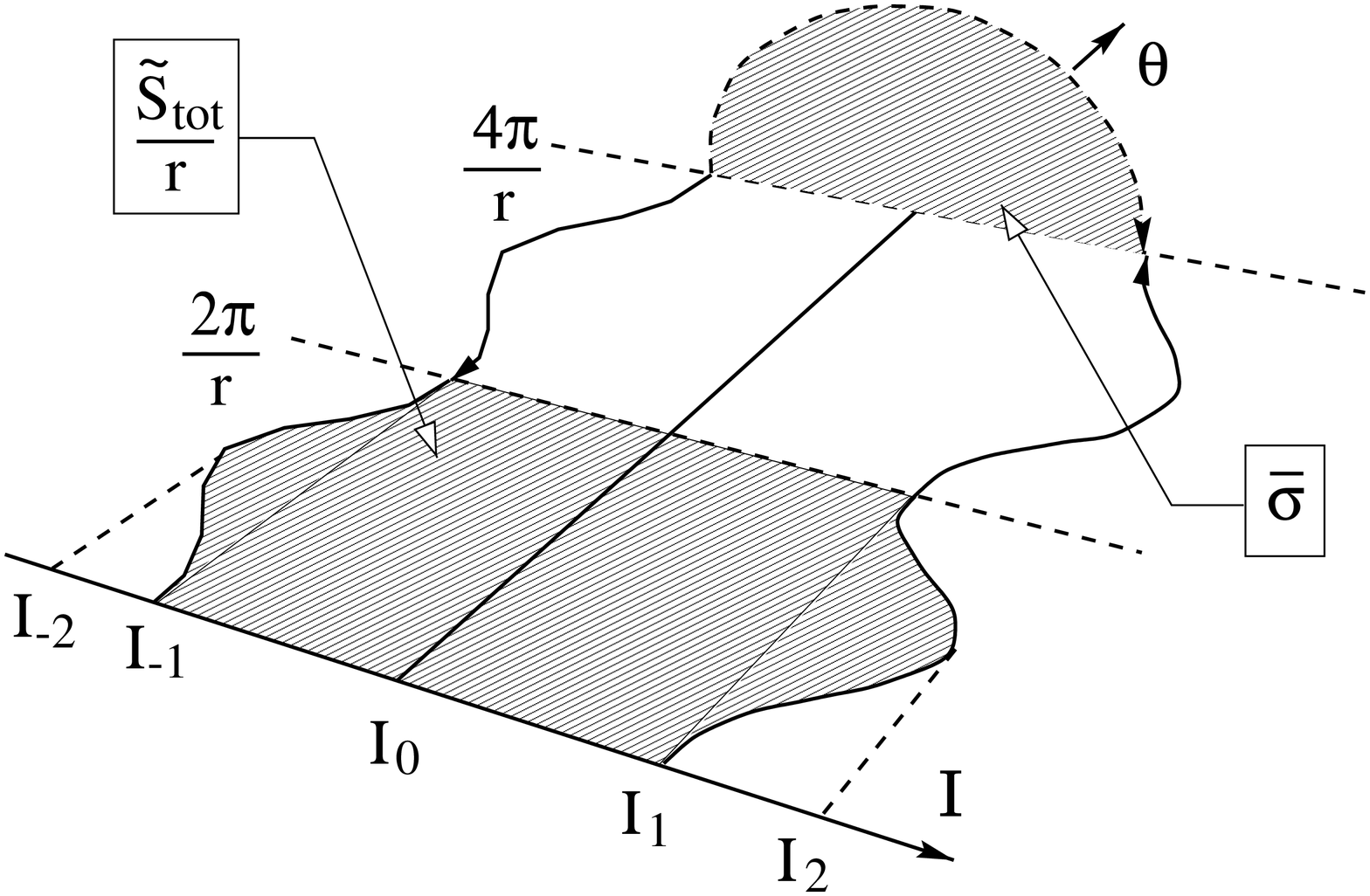}
\end{center}
\caption{Sketch    of    the    invariant   manifold    Eq.~({\protect
\ref{eq:manifold}}) in  complex phase space.  The  solid and interrupted
lines belong respectively to  the classically authorized and forbidden
part of this manifold.  The  shaded area represent the action integral
that parametrize the tunneling process.
\label{fig:Ipend}}
\end{figure}

It turns out to be  more convenient to adopt an action representation,
that  is,  to work  in  the Fourier  space  of  the pendulum  problem.
Considering  an   eigenmode  of   energy  $E_k$,  this   implies  that
semiclassically, the  manifold $\Gamma$ we are  interested in  is the
solution of
\begin{equation} \label{eq:manifold}
H(I,\Theta)  = E_k
\end{equation}
for real action $I$, but possibly complex angle $\Theta$.  Introducing
$I_{\pm  1} =I_0 \pm  \sqrt{2\mu(E_k-V_1)}$, and  $I_{\pm 2}  =I_0 \pm
\sqrt{2\mu(E_k+V_1)}$, the abscissa of  the turning points, on can, as
schemed  on Fig.~\ref{fig:Ipend},  distinguish two  parts  in this
manifold.  For $I  \in [I_1,I_2]$ for instance, the  $2r$ solutions of
Eq.~(\ref{eq:manifold})   are   real,   and   we   shall   denote   by
${\Gamma}^{l\pm}_{[1,2]}$ the graph of
\begin{equation} \label{eq:manifold1}
    \Theta^{l\pm}_{[1,2]}(I)  = \frac{2l\pi}{r} \pm
    \frac{1}{r}\arccos{\left[ (E-(I-I_0)^2/2\mu)/V_1 \right]}
  \qquad (l=0,\ldots,r-1) \; . 
\end{equation}
This,  together with  the ${\Gamma}^{l\pm}_{[-2,-1]}$ defined  in the
same way in $[I_{-2},I_{-1}]$,  corresponds to the classically allowed
region  (solid line  in Fig.~\ref{fig:Ipend}).   Outside  these intervals,
$\Theta$ has a  non zero imaginary part, and the  $2r$ sheets of $\Gamma$ come by pairs corresponding to complex conjugated functions
\begin{equation} \label{eq:manifold2}
    \Theta^{l\pm}_{[-1,1]}(I)  = \frac{2l\pi}{r} \pm
    \frac{i}{r}{\rm argch}\left[(E_k-(I-I_0)^2/2\mu)/V_1\right] 
  \qquad (l=0,\ldots,r-1) \; . 
\end{equation}
for $I \in [I_{-1},I_1]$,
\begin{equation} \label{eq:manifold3}
    \Theta^{l\pm}_{[2,\infty]}(I)  = \frac{(2l+1)\pi}{r} \pm
    \frac{i}{r}{\rm argch}\left[(E-(I-I_0)^2/2\mu)/V_1\right] 
  \qquad (l=0,\ldots,r-1) \; . 
\end{equation}
for $I>I_2$, and $\Theta^{l\pm}_{[-\infty,-2]}(I)$ defined in the same
way  for  $I<I_{-2}$.  The  interrupted  line in  Fig.~\ref{fig:Ipend}
represent a  complex path $\Theta(I)$  ($I$ real) in  this classically
forbidden  region.   In  the  following,  we  will  denote  by  ${\Gamma}^{l\pm}_{[a,b]}$  ($[a,b] =  [-\infty,-2],  [-2,-1], [-1,1],  [1,2],
[2,\infty]$)  the  various  parts  of  ${\Gamma}$  described  by  the
$\Theta^{l\pm}_{[a,b]}$ above.  We shall also make use of the notation
$I_{[a,b]}(\Theta)$ for the inverse of $\Theta^{l-}_{[a,b]}(I)$ on the
interval         $\Theta\in[(2l-1)\pi/r,2l\pi/r[$        and        of
$\Theta^{l+}_{[a,b]}(I)$           on           the           interval
$\Theta \in [2 l \pi /r , (2l+1) \pi / r]$.

\subsection{Local  solutions and connection formula}

Following  standard  WKB  analysis,   a  local  approximation  to  the
Schr\"odinger equation  $\widehat{H} \Psi = E_k \Psi$  can be obtained
between       two        turning       points, ie.\ under each of the 
${\Gamma}^{l\pm}_{[a,b]}$ as
\begin{equation}
  \widetilde{\psi}^{l\pm}_{[a,b]}(I)=
  \frac{{\mathcal A}^{l\pm}_{[ab]}}{\sqrt{\tilde Q^{l\pm}_{[ab]}(I)}}
  \left(\exp{\frac{i\widetilde{S}^{l\pm}_{[a,b]}(I)}{\hbar}} \right) \; ,
\label{eq:local}
\end{equation}
with the action integral
\begin{equation} \label{eq:tldS}
  \widetilde{S}^{l\pm}_{[a,b]} = - \int^{\Theta^{l\pm}_{[a,b]}(I)} 
  \Theta \frac{dI_{[a,b]}}{d\Theta}(\Theta)d\Theta \; ,
\end{equation}
$\tilde    Q^{l\pm}_{[ab]}(I)    =    |    T_k    V_1    r    \sin\left(
r\Theta^{l\pm}_{[a,b]}(I)  \right)|$,  and  $T_k$  the period  of  the
classical motion at energy $E_k$.

This analysis can be legitimated a priori when the interval spanned by
the manifold on the $I$ coordinate is large as compared to the typical
domain where the WKB construction diverges at the caustics. 
Here it means
that $|I_2 - I_1| \gg \hbar^{2/3}$, which can be written
\be
\frac{\mu V_1}{\hbar k} \gg \hbar^{2/3}
\label{validsemic}
\ee
if $V_1$ is small as compared to energy $E_k$.

The  first step  in the  construction of  a global  wavefunction is  then 
to patch
together these  local solutions at  caustics.  A convenient way  to do
this here is to use Langer \cite{langer37} connection formula which, for
a given  choice of  the origin of  the action integrals,  imposes some
relations  between the  various coefficients  ${\mathcal A}$.   Let us
denote for instance by ${\mathcal A}_{\rm out}$ and ${\mathcal A}_{\rm
in}$  the  amplitude  of  the   outgoing  and  incoming  wave  on  the
classically  authorized side  of a  turning point,  and  by ${\mathcal
A}_{\rm >}$ and  ${\mathcal A}_{\rm <}$ the ones  of the exponentially
decreasing and increasing functions on the classically forbidden side.
Then, except for a  global multiplicative constant ${\mathcal A}$, the
relation between these various amplitudes can be expressed in term of a
single parameter $\eta$.  Assuming the origin of the phase integral is
taken at the turning point this relation can be expressed as
\begin{eqnarray}
   {\mathcal A}_{\rm out} & = & {\mathcal A} \exp[- i(\pi/4 - \eta)] \\
   {\mathcal A}_{\rm in}  & = & {\mathcal A} \exp[+ i(\pi/4 - \eta)] \\
   {\mathcal A}_{\rm >} & = & {\mathcal A} \cos[\eta] \\
   {\mathcal A}_{\rm <} & = & {\mathcal A} 2 \sin [\eta] \; .
\end{eqnarray}
Neglecting  tunneling amount  to assume  that only  the exponentially
decreasing  function  should be  used,  in  which  case the  dephasing
between the incoming  and outcoming  wave on  the classically
allowed  side  of  the  caustic  is $\pi/2$.   The  above  expressions
generalize this to the case where  there is some non zero amplitude on
the exponentially  increasing wave.

For the pendulum  the parameter $\eta$ is obviously  zero for caustics
at  $I=I_{\pm  2}$ since  there,  any  amplitude  on an  exponentially
increasing  function  would  lead  to  a  non  normalisable  solution.
Moreover, by  symmetry, the $r$  caustics at $I =  I_1$, (respectively
$I_{-1}$), must have the  same parameter $\eta = \eta_1$ (respectively
$\eta_{-1}$).


\subsection{Paths within the classically allowed regions}

To obtain a global approximation  to the Schr\"odinger equation, it is
now  necessary to  choose the  manifold $\Gamma$, that  is  the energy
$E_k$, and the  caustics parameter $\eta_1$ and $\eta_{-1}$,  in such a
way that the  various connections are consistent one  with each other,
or in  other words that,  once a choice  of the origins of  the action
integrals in  Eq.~(\ref{eq:tldS}) is made,  a given set  of parameters
${\mathcal  A}^{l\pm}_{[ab]}$ fulfill  all the  constraints due  to the
connection  formula.   This  is  equivalent  to  the  fact  that  such
condition apply to any independent loop of $\Gamma$.

The first closed path of the manifold that we are going to consider is
the one belonging to the classically allowed region $[I_1,I_2]$.
Moving in the upward direction on this loop, one gains a phase $- \pi/2 +
\eta_1$ at each of the $r$ turning points with abscissa $I_1$, and
$+\pi/2$ at each of the one at $I=I_2$.  Consistency should therefore
imposes that
\begin{equation} \label{eq:QantCond1}
 \widetilde{S}_{\rm  tot}/\hbar  + 2 r \eta_1 = \pi 
  \quad [{\rm mod} \;  2\pi] \; .
\end{equation}
where $\widetilde{S}^{[12]}_{\rm  tot}$ is defined as in Eq.~(\ref{eq:tldS}),
but between $(I,\Theta)$ and $(I,\Theta+2\pi)$.

One should be a little bit  more careful here because the geometry of
the  phase  space of  the  pendulum is  a  torus,  which implies  that
$\widetilde{S}^{[12]}_{\rm  tot}$  is not  independent  of the  origin
$(I,\Theta)$  of the  integration.  Indeed,  the fact  that  the final
point is really $(I,\Theta+2\pi)$,  and not $(I,\Theta)$, implies that
$d\widetilde{S}^{[12]}_{\rm   tot}/dI=-2\pi$,   and   that   therefore
$\widetilde{S}^{[12]}_{\rm    tot}(I)    =   \widetilde{S}^{[12]}_{\rm
tot}(I_0) - 2\pi (I-I_0)$.
 
It    is    therefore    impossible    to   impose    the    condition
Eq.~(\ref{eq:QantCond1})  for  all $I$.   However,  because the  phase
space is  a cylinder,  the only admissible  values of the  action are,
assuming antiperiodicity  in $\Theta$, the  discrete set $I_n  = \hbar
(n+\frac{1}{2})$, with $n$ a positive or negative integer.  For all of
those,  Eq.~(\ref{eq:QantCond1})  remains   unchanged  if  we  replace
$\widetilde{S}_{\rm tot}$ by  ${S}_{\rm tot} = \widetilde{S}_{\rm tot}
+ 2 \pi I$ (since $2 \pi I_n/\hbar$ is equal to zero modulo $2\pi$ for
all  $I_n$).   One therefore  recovers  in  this  way the  ``natural''
quantization condition
\begin{equation} \label{eq:QantCond2}
  S^{[12]}_{\rm   tot}/\hbar + 2 r \eta_1 = 
  \frac{1}{\hbar} \oint I_{[12]}(\Theta) d\Theta + 2 r \eta_1 
  = 2 \pi \left(k +  \frac{1}{2} \right) \; .
\end{equation}
Using the same reasoning, we obtain also that 
\begin{equation} \label{eq:QantCond3}
   S^{[-1-2]}_{\rm   tot}/\hbar - 2 r \eta_{-1} = 
  \frac{1}{\hbar} \oint I_{[-1-2]}(\Theta) d\Theta - 2 r
  \eta_{-1} 
  = 2 \pi \left(k' + \frac{1}{2} \right) \; .
\end{equation}

\subsection{Classically forbidden paths}

The second kind of closed loop that we shall consider is, as usual in
tunneling problems, the one starting from some turning point
$(I_1,\Theta^0_l)$, with $\Theta^0_l = \frac{2l\pi}{r}$ for some given
$l=0,\dots r-1$, following one of the complex branch of $\Gamma$ up to 
$(I_{-1},\Theta^0_l)$, switch to the complex conjugated branch and
move back to $(I_1,\Theta^0_l)$.  Consistency then impose that
\begin{equation}
\tan\eta_1\tan\eta_{-1}=\frac{1}{4}\exp\left(-2\frac{\bar \sigma}{\hbar}\right)
\; ,
\label{eqetazeta}
\end{equation}
with
\begin{equation}
\bar \sigma \equiv \int_{I_{-1}}^{I_1} |\Im\left({\Theta_{[-1,1]}(I)}\right)| dI
= \frac{1}{r}\int_{I_{-1}}^{I_1}
{\rm argch}\left(\frac{E_k}{V_1}-\frac{(I-I_0)^2}{2\mu V_1}\right)dI \;
.
\end{equation}

Finally the  last kind of loop  that we need to  consider is the
one which starts from some turning point $(I_1,\Theta^0_l)$, follows one
of the complex branch of $\Gamma$ up to $(I_{-1},\Theta^0_l)$, moves
upward on the classically allowed branch up to
$(I_{-1},\Theta^0_{l+1})$, then backward on a complex branch to 
$(I_{1},\Theta^0_{l+1})$, and finally down to $(I_1,\Theta^0_l)$.
Assuming Eq.~(\ref{eqetazeta}) is verified, which make unimportant
which of the two possible complex branches is used at each traversal of
the classically forbidden region, this still imposes a constraint on the
phase of the wavefunction, namely, after an integration by parts of
the action integral
\begin{equation} \label{eq:lastpath}
  \frac{1}{\hbar} 
  \int_{\Theta^0_l}^{\Theta^0_{l+1}} d\Theta 
  \left[I_{[1,2]}(\Theta)  - I_{[-1,-2]}(\Theta) \right]
    - 2 (\eta_1+\eta_{-1})
  = 0   \quad [{\rm mod} \;  2\pi] \; .
\end{equation}
Any other loop on $\Gamma$ can be
decomposed onto the ones considered above.

\subsection{Quantization conditions}

The left hand side  of Eq.~(\ref{eqetazeta}) being exponentially small
in  $\hbar$, this  implies  that either  $\eta_1$  or $\eta_{-1}$  are
exponentially small, but, assuming
\begin{equation}
  I_0 = \hbar \left(\kappa_0 + \frac{1}{2} \right) \; ,
\end{equation}
where  $\kappa_0$ is  {\em not}  an integer,  not both.  We  assume for
instance that $\eta_1 \ll \eta_{-1} \sim 1$.

This means  that the real  torus ${\Gamma}_{[12]}$ almost  quantizes in
the absence of the corrections due to tunneling, ie.\ that
\begin{equation} \label{eq:QantCond2:approx}
  S^{[12]}_{\rm   tot}/\hbar = 
  \frac{1}{\hbar} \oint I_{[12]}(\Theta) d\Theta  
  \simeq 2 \pi \left(k + \frac{1}{2}\right) \; .
\end{equation}
The  normalization of  the wavefunction  furthermore implies  that the
modulus of all the ${\mathcal A}^{l\pm}_{[12]}$ is one.

Our goal here is not to evaluate the shift in energy $E_k$ associated
with the modification of the quantization condition between
Eqs.~(\ref{eq:QantCond2}) and (\ref{eq:QantCond2:approx}).  Indeed this is
presumably much smaller than polynomial corrections in $\hbar$ that we
have neglected.  Rather, what we are interested in are the amplitudes
${\mathcal A}^{l\pm}_{[-2-1]}$ of the part of the wave function which
has tunneled on the other side of the resonance.  We thus assume
that the energy $E_k$  fulfills Eqs.~(\ref{eq:QantCond2:approx}), and
therefore that the manifold  $\Gamma$ is fixed. Neglecting again
$\eta_1$, the parameter $\eta_{-1}$ is then  determined by
Eq.~(\ref{eq:lastpath}) as
\begin{equation} 
  \eta_{-1} \equiv \frac{1}{2\hbar} 
  \int_{\Theta^0_l}^{\Theta^0_{l+1}} d\Theta 
  \left[I_{[1,2]}(\Theta)  - I_{[-1,-2]}(\Theta) \right]
    \quad [{\rm mod} \;  2\pi] \; .
\end{equation}
Note that   this    choice     of    $\eta_{-1}$    necessarily    fulfills
Eq.~(\ref{eq:QantCond3}),  but  that   the  reciprocal  is  not  true.
Indeed,  as we  shall see  below, Eq.~(\ref{eq:lastpath})  imposes the
``selection rule'', which stands that only W.K.B.  modes whose quantum
numbers  differ  in  a  multiple   of  $r$  are  coupled.   Using  the
quantization     condition    Eq.~(\ref{eq:QantCond2})     and    that
$I_{[1,2]}(\Theta)$  and   $I_{[-1,-2]}(\Theta)$  are  symmetric  with
respect to $I_0 = \hbar (\kappa_0 + \frac{1}{2})$, we obtain
\begin{equation} \label{explicitzeta}
  \eta_{-1} \equiv \frac{2\pi(k - \kappa_0)}{r} \; .
\end{equation}

Applying then Eq.~(\ref{eqetazeta}) fixes the value of $\eta_{1}$ as 
\begin{equation} 
  \eta_{1}\simeq\frac{1}{4\tan\eta_{-1}}
  \exp\left(-2\frac{\bar \sigma}{\hbar}\right) 
=\frac{1}{\disp    4\tan\left(\frac{2\pi(k-\kappa_0)}{r}\right)}    
\exp\left( -2\frac{\bar \sigma}{\hbar} \right),
\label{finvaleta}
\end{equation}
and, taking the origin of  the phase integral at $(I_{-1},0)$ for all
the  $\widetilde{S}^{l\pm}_{[-2,-1]}$  (and  at  $(I_{1},0)$  for  the
$\widetilde{S}^{l\pm}_{[1,2]}$) the transmitted amplitude as
\begin{eqnarray}
{\mathcal A}^{l\pm}_{[-2-1]} & \simeq & {\mathcal A}_T
\exp\left(- i (2l\pm1) \eta_{-1} \pm i \pi/4         \right) \\
{\mathcal A}_T & =  &
\frac{1}{\disp 2\sin\left(\frac{2\pi(k-\kappa_0)}{r}\right)}
\exp\left(-\frac{\bar \sigma}{\hbar}\right) 
\end{eqnarray}

\subsection{Selection rules}

Going now back to the $\Theta$ representation, and evaluating the
corresponding Fourier transformation at the stationary phase
approximation, we obtain
\begin{equation}
 \Psi(\Theta) = \frac{1}{\sqrt{T_k |\dot \Theta_k|}}
\left[ \exp \left( \frac{i}{\hbar} S_{[1,2]}(\Theta) \right) + 
{\mathcal A}_T \exp \left(\frac{i}{\hbar} S_{[-2,-1]}(\Theta) - 
 i(2\nu(\Theta)+1)\eta_{-1} \right)
\right]
\label{sol_globale_theta}
\end{equation}
with $\nu(\Theta) = {\rm Int} [\Theta/(2\pi/r)]$.
 It is straightforward to generalize this expression to a more general
potential, like in Eq. (\ref{localapprox}), as long as the general
structure of branches, the periodicity and the symmetry are preserved.

Now we introduce  the quasimodes 
\begin{equation}
u_{k'}(\Theta) = \frac{1}{\sqrt{T_{k'} |\dot \Theta_{k'}|}}
 \exp ({i S_{k'}(\Theta)/\hbar}) \; ,
\end{equation}
which are built on real invariant tori $k'$ such that
\begin{equation}
  S_{k'}^{\rm tot} \equiv \int_0^{2\pi} I_{k'}(\Theta) d\Theta 
  = 2\pi \left(k' + \frac{1}{2} \right) \; .
\end{equation}
$\Psi$ is essentially equal to $u_k$, plus some exponentially small
tunneling contribution $\Psi_T$ localized on the other side of the
resonance. If we write the scalar product of $\Psi_T$ with some
$u_{k'}$, assuming $-(k'-\kappa_0) \simeq (k-\kappa_0)$ (and thus $T_{k'}
\simeq T_k \equiv T$, and $|\dot \Theta_k| \simeq |\dot \Theta_{k'}|
\equiv \dot \Theta$)  we obtain
\begin{eqnarray}
\langle u_{k'} | \Psi_T \rangle & = & \frac{{\mathcal A}_T }{T} \int_0^{2\pi}
\frac{d\Theta}{\dot \Theta} 
\exp i \left( S_{[-2,-1]}(\Theta)/\hbar -(2l(\Theta)+1)\eta_{-1}
- S_{k'}(\Theta)/\hbar  \right) \\
& = & {\mathcal A}_T \times
\frac{r}{T} \int_0^{2\pi/r}
\frac{d\Theta}{\dot \Theta} 
\exp i \left( S_{[-2,-1]}(\Theta)/\hbar - \eta_{-1}
- S_{k'}(\Theta)/\hbar  \right) \nonumber \\
& \times &  \frac{1}{r} \sum_{l=0}^{r-1} \exp i l
\left( S_{[-2,-1]}(2\pi/r)/\hbar - 2\eta_{-1}
- S_{k'}(2\pi/r)/\hbar  \right) \; .
\end{eqnarray}
The integral on $\Theta$ can be evaluated using that 
$S_{[-2,-1]}(\Theta) - S_{k'}(\Theta) \simeq t(\Theta) (E_k-E_{k'})$. 
For the remaining sum on $l$, one should note that
$S_{k'}(2\pi/r)/\hbar = 2 \pi k'/r$, and that in the same way,
Eq.~(\ref{eq:lastpath}) imposes that $S_{[-2,-1]}(2\pi/r) - 2 \eta_{-1} =
S_{[12]}(2\pi/r) \;  [{\rm mod} \;  2\pi]  \equiv 2k\pi/r \;
[{\rm mod} \;  2\pi]$.  We therefore obtain that, up to a global phase
\begin{equation} \label{eq:selection}
 \langle u_{k'} | \Psi_T \rangle  = {\mathcal A}_T 
\frac{\sin(\pi \delta E/ r \Delta)}{\pi \delta E/ r \Delta}
\sum_m \delta_{k,{k'+rm}} \; ,
\end{equation}
with $\delta E \equiv E_k - E_{k'}$ and $\Delta = \hbar \omega$ is the
spacing between  the quasimodes energy  at $E\simeq E_k  \simeq E_k'$.
Eq.~(\ref{eq:selection})  contains the  selection rule,  stated above,
that a resonance with  $r$ islands couples through tunneling quasimodes
with quantum numbers differing in a integer multiple of $r$.

\subsection{Comparison with the result of perturbation theory}

The  semiclassical   treatment  described  can be justified a priori in the
domain  $\hbar^{2/3}\ll  \sqrt{2\mu}  (\sqrt{E_k+V_1}-\sqrt{E_k-V_1})$,
with $E_k\simeq  \hbar^2 (k+\frac{1}{2})^2/(2\mu)$ which is disjoint from the
regime of validity of the quantum
theory of perturbation. However we can show that for 
\be
\begin{array}{ccc}
V_1 & \ll & (I_k-I_0)^2 \\
|I_k-I_0| & \gg & \hbar r \\
\eta_{-1} & \ll & 1\footnote{In the case $\eta\ll 1$ the intervals on which the main semiclassical function and the transmitted one dominate are quite well 
separated}.
\end{array}
\ee
 the two formulas coincide. Indeed in this case the transmitted WKB function
has an overlap with the plane wave $|k-rm\rangle$ almost equal to $1$,
and then we show in the following that  ${\mathcal A}^T \simeq 
\langle k-rm|k^{(m)}\rangle$, with $k^{(m)}$ the estimation 
of the $k^{th}$ eigenmode of the system given by the theory
of perturbation at the order $m$.
Whether the coincidence holds for every $\eta$ is not obvious for now.

 The $m^{th}$ order perturbation theory gives
\begin{equation}
\langle k-rm|k^{(m)}\rangle = \left(\frac{\mu V_1}{\hbar^2}\right)^m
\prod_{l=1}^m \frac{1}{(k-\kappa_0)^2-(k-\kappa_0-lr)^2},
\end{equation}
which, in the limit of large $m$, is equivalent to
\begin{equation}
\langle k-rm|k^{(m)}\rangle \simeq 
\left(\frac{\mu V_1}{\hbar^2r^2}\right)^m
\frac{\disp\Gamma \left( \frac{2(k-\kappa_0)}{r}-m \right)}
{\disp m!\Gamma \left(\frac{2(k-\kappa_0)}{r} \right)}
= \left(\frac{\mu V_1}{\hbar^2r^2}\right)^m
\frac{\disp\Gamma\left(\frac{\eta_{-1}}{\pi}\right)}{\disp m!\Gamma
\left(m+\frac{\eta_{-1}}{\pi} \right)}. 
\end{equation}
Now we use the Stirling formula and the equivalent $x\Gamma(x)\thicksim 1$
for $x\rightarrow 0$, and we get
\begin{equation}
\langle k-rm|k^{(m)}\rangle \simeq \frac{1}{2\eta_{-1}}
\left(\frac{\mu V_1 e^2}{\hbar^2 r^2 m^2}\right)^m.
\label{couplpert}
\end{equation}

On the other hand, the classical action $\bar \sigma$ along the imaginary path
relating the $k^{th}$ K.A.M. torus to its symmetric is equivalent to
\begin{equation}
\bar \sigma\simeq\frac{2(I_k-I_0)}{r}
\left(-\log\left(\frac{\mu V_1}{4(I_k-I_0)^2}\right) -2+
{\mathcal O}\left(\frac{\mu V_1}{(I_k-I_0)^2}\right)\right)
\end{equation}
in the perturbative regime. This can be rewritten,
by using (\ref{explicitzeta}) and $\eta_{-1} \ll 1$,
\begin{equation}
\frac{\bar \sigma}{\hbar}\simeq \left(m+\frac{\eta_{-1}}{\pi}\right)
\left(-\log\left(\frac{\mu V_1}{\hbar^2 r^2m^2}\right) -2+
{\mathcal O}\left(\frac{\mu V_1}{\hbar^2 r^2m^2}\right)\right).
\end{equation}
Then
\begin{equation}
\exp\left(-\frac{\bar \sigma}{\hbar}\right)\simeq
\left(\frac{\mu V_1 e^2}{\hbar^2 r^2m^2}\right)^m,
\end{equation}
and the expression for ${\mathcal A}_T$ is
\begin{equation}
{\mathcal A}_T\simeq \frac{1}{2\eta_{-1}}
\left(\frac{\mu V_1 e^2}{\hbar^2 r^2m^2}\right)^m,
\end{equation}
which is identical to (\ref{couplpert}).


\newpage

\newpage

\newpage

\newpage


\newpage

\vspace{2cm}

\newpage

\newpage

\newpage

\newpage

\newpage

\newpage

\newpage

\newpage

\newpage

\newpage

\begin{references}




\bibitem{davi81} M.~J.~Davis and E.~J.~Heller, J.~Chem.~Phys. {\bf 75}
(1981) 246.

\bibitem{barnas98} J.~Barnas and A.~Fert Phys.~Rev.~Lett.\ {\bf 80},
1058-1061 (1998). 

\bibitem{creagh96_99} S.~C. Creagh and N.~D. Whelan,
Phys. Rev. Lett. {\bf 77}, 4975 (1996); and {\bf 82}, 5237 (1999).

\bibitem{lin90}
W.~A.~Lin and L.E.~Ballentine, Phys.~Rev.~Lett.\ {\bf 65},  2927 (1990).

\bibitem{boh93}
O. Bohigas, S. Tomsovic, D. Ullmo, Phys.~Rep.~{\bf 223}, (1993) 43.
O.~Bohigas, D.~Boos{\'e}, R.~Egydio de Carvalho,
and V.~Marvulle, Nucl.~Phys. {\bf A560} (1993) 197.


\bibitem{tom94}
S. Tomsovic and D. Ullmo, Phys. Rev. E. {\bf 50} (1994) 145. 


\bibitem{doron} 
E.~Doron and S.D.~Frischat, Phys.~Rev.~Lett. {\bf 75}, 3661 (1995).  
S.D.~Frischat and E.~Doron, Phys.~Rev.~E {\bf 57}, 1421 (1998).

\bibitem{leyvraz96}
F. Leyvraz and D. Ullmo, J. Phys.{\bf A 29} (1996) 2529-2551. 

\bibitem{zakrzewski98}
J.~Zakrzewski, D.~Delande, A.~Buchleitner Phys.~Rev.~E {\bf 57}, 1458 (1998).

\bibitem{hensinger01}
W.~.K.~Hensinger, et al., Nature {\bf 412}, 52 (2001).

\bibitem{steck01}
D.~A.~Steck, W.~H.~Oskay and M.~G.~Raizen, Science {\bf 293} 274 (2001); 
Phys. Rev. Lett. {\bf 88}, 120406 (2002).

\bibitem{mouchet01}
A.~Mouchet, C.~Miniatura, R.~Kaiser, B.~Gr{\'e}maud and D.~Delande,
Phys.~Rev.~E {\bf 64}, 016221 (2001).


\bibitem{dembowski} C. Dembowski et al., Phys. Rev. Lett. {\bf 84}, 867 (2000).

\bibitem{noeckel} J. U. N{\"o}ckel and A. D. Stone, Nature {\bf 385}, 45 (1997).

\bibitem{greene81} J.M.~Greene and I.C.~Percival, Physica {\bf 3D},
530-548 (1981).


\bibitem{creagh_in_steve} S.~Creagh, in {\em Tunneling in Complex Systems},
  edited by S. Tomsovic (World Scientific, Singapore, 1998), p. 35.

\bibitem{shudo} A. Shudo and K. S. Ikeda, Phys.~Rev.~Lett. {\bf 74}, 682;
Phys.~Rev.~Lett. {\bf 76}, 4151.
T. Onishi, A. Shudo, K. S. Ikeda, and K. Takahashi, Phys. Rev. E {\bf 64},
025201 (2001).  A.~Shudo, Y.~Ishii and K.~S.~Ikeda,
J.~Phys.~A:Math.~Gen. {\bf 35} (2002) L1-L7.

\bibitem{bonci98}
L.~Bonci, F.~Farusi, P.~Grigolini and R.~Roncaglia,  
Phys.~Rev.~E {\bf 58} 5689 (1998).

\bibitem{UzeNoiMar83JCP}
T. Uzer, D.~W. Noid, and R.~A. Marcus, J. Chem. Phys. {\bf 79},  4412  (1983).

\bibitem{Ozo84JPC}
A.~M. {Ozorio de Almeida}, J. Phys. Chem. {\bf 88},  6139  (1984).

\bibitem{FarUze86JCP}
D. Farrely and T. Uzer, J. Chem. Phys. {\bf 85}, 308 (1986).

\bibitem{BreHol91AP}
H.~P. Breuer and M. Holthaus, Ann. Phys. (N.J.) {\bf 211},  249  (1991).

\bibitem{prlbsu}
O. Brodier, P. Schlagheck, and D. Ullmo, Phys. Rev. Lett. {\bf 87},  064101
  (2001).

\bibitem{Leboeuf90} 
P.~Leboeuf, J.~Kurchan, M.~Feingold, and D.~P.~Arovas,
Phys.~Rev.~Lett. {\bf 65}, 3076 (1990).

\bibitem{LimShe91PRL}
R. Lima and D. Shepelyansky, Phys. Rev. Lett. 67, 1377 (1991).

\bibitem{GeiKetPet91PRL}
T. Geisel, R. Ketzmerick, and G. Petschel, Phys. Rev. Lett. 67, 3635 (1991).

\bibitem{RonO94PRL}
R.~Roncaglia, L.~Bonci, F.~Izrailev, B.J.~West and P.~Grigolini
Phys.~Rev.~Lett.\ {\bf 73}, 802 (1994).

\bibitem{Sok86SJTMP}
V.~V. Sokolov, Sov. J. Theor. Math. Phys. {\bf 67},  223  (1986).
V.~V. Sokolov, O.~V. Zhirov, D. Alonso, and G. Casati, Phys. Rev. Lett. {\bf
  84},  3566  (2000).

\bibitem{lichtenberg:book} 
A.~J. Lichtenberg and M.~A. Lieberman, {\em
Regular and Stochastic Motion} (Springer-Verlag, New York, 1983).


\bibitem{GroO91PRL}
F.~Grossmann, T.~Dittrich, P.~Jung, and P.~H{\"a}nggi, 
Phys.~Rev.~Lett.\ {\bf 67} (1991) 516.



\bibitem{moser:lecture} Memoirs of the American Mathematical Society
{\bf 81}: Lectures on Hamiltonian systems by J{\"u}rgen Moser.

\bibitem{maslov:book}  V.P.~Maslov and M.V.~Fedoriuk, {\em Semiclassical
approximation in quantum mechanics}, (Reidel, Dordrecht,1981)
[original Russian edition 1965].

\bibitem{arnold:book} See e.g.\ appendices 10 and 11 of : V.I.~Arnold,
{\em Mathematical methods of classical mechanics}, Springer Verlag, 1978.

\bibitem{landau:book} L.~Landau et E.~Lifchitz, {\em M{\'e}canique
Quantique}, \'Editions Mir 1966.

\bibitem{Cre94JPA} S.~C. Creagh, J. Phys. A {\bf 27},  4969  (1994).

\bibitem{chirikov} B.~V. Chirikov, Phys. Rep. {\bf 52},  265  (1979).

\bibitem{wilkinson86}
M. Wilkinson, Physica {\bf 21D}, 341 (1986);  J. Phys. A {\bf 20}, 635 (1987).


\bibitem{langer37}
R.~E. Langer, Phys. Rev. {\bf 51}, 669 (1937).

\bibitem{validity} We should note that that we cannot rigourously show
Eq.~(\ref{eq:from_appendix4_2})  to hold  in the  extreme perturbative
regime.   The  matching  between  the semiclassical  and  the  quantum
transition amplitudes is therefore not necessarily expected.

\bibitem{bogomolny84}
E.~B.~Bogomolny, Physica {\bf 13D}, 281-301 (1984).





\end{references}
\end{document}